\RequirePackage{lineno}
\documentclass[useAMS,usenatbib,usegraphicx]{mn2e}
\usepackage{color}
\usepackage{epsf}
\usepackage{ulem}
\usepackage{epsfig}
\usepackage{amsmath}
\usepackage{amssymb}
\usepackage{subfigure}
\newcommand{\hMpc}{h^{-1}\rm{Mpc}}
\newcommand{\hGpc}{h^{-1}\rm{Gpc}}
\newcommand{\snl}{\Sigma_{\rm nl}}

\newcommand{\ddda}{\frac{\partial D_A}{\partial \alpha}}
\newcommand{\dhda}{\frac{\partial H}{\partial \alpha}}
\newcommand{\ddde}{\frac{\partial D_A}{\partial \epsilon}}
\newcommand{\dhde}{\frac{\partial H}{\partial \epsilon}}
\newcommand{\sas}{\sigma_\alpha^2}
\newcommand{\ses}{\sigma_\epsilon^2}
\newcommand{\sae}{\sigma_{\alpha\epsilon}}
\newcommand{\sds}{\sigma_{D_A}^2}
\newcommand{\shs}{\sigma_H^2}
\newcommand{\sdh}{\sigma_{D_A H}}
\newcommand{\dadd}{\frac{\partial\alpha}{\partial D_A}}
\newcommand{\dadh}{\frac{\partial\alpha}{\partial H}}
\newcommand{\dedd}{\frac{\partial\epsilon}{\partial D_A}}
\newcommand{\dedh}{\frac{\partial\epsilon}{\partial H}}

\title[Measuring $D_A$ and $H$ using BAO]
{Measuring $D_A$ and $H$ at $z=0.35$ from the SDSS DR7 LRGs using baryon
acoustic oscillations}

\author[X. Xu et al.]
{Xiaoying Xu$^{1}$, Antonio J. Cuesta$^{2}$, Nikhil Padmanabhan$^{2}$,
Daniel J. Eisenstein$^{3}$, \newauthor Cameron K. McBride$^{3}$ \\
$^{1}$ Steward Observatory, University of Arizona, 933 N. Cherry Ave., 
Tucson, AZ 85721; xxu@as.arizona.edu \\
$^{2}$ Department of Physics, Yale University, 260 Whitney Ave., New Haven,
CT 06511 \\
$^{3}$ Harvard-Smithsonian Center for Astrophysics, Harvard University, 60 Garden St., Cambridge,
MA 02138}

\begin{document}
\maketitle
\label{firstpage}
\begin{abstract}
We present measurements of the angular diameter distance $D_A(z)$ and the
Hubble parameter $H(z)$ at $z=0.35$ using the anisotropy of the baryon
acoustic oscillation (BAO) signal measured in the galaxy clustering
distribution of the Sloan Digital Sky Survey (SDSS) Data Release 7
(DR7) Luminous Red Galaxies (LRG) sample. Our work is the first to apply
density-field reconstruction to an anisotropic analysis of the acoustic
peak. Reconstruction partially removes the effects of non-linear evolution
and redshift-space distortions in order to sharpen the acoustic signal. We
present the theoretical framework behind the anisotropic BAO signal and
give a detailed account of the fitting model we use to extract this signal
from the data. Our method focuses only on the acoustic peak anisotropy,
rather than the more model-dependent anisotropic information from the
broadband power. We test the robustness of our analysis methods on 160
LasDamas DR7 mock catalogues and find that our models are unbiased at the
$\sim0.2\%$ level in measuring the BAO anisotropy. After reconstruction
we measure $D_A(z=0.35)=1050\pm38$ Mpc and $H(z=0.35)=84.4\pm7.0$
km/s/Mpc assuming a sound horizon of $r_s=152.76$ Mpc. Note that these
measurements are correlated with a correlation coefficient of 0.57. This
represents a factor of 1.4 improvement in the error on $D_A$ relative to
the pre-reconstruction case; a factor of 1.2 improvement is seen for $H$.
\end{abstract}



\section{Introduction}\label{sec:intro}

One of the major challenges in modern cosmology is to understand cosmic
acceleration, a crucial but perplexing discovery made through observations
of distant supernovae \citep{Rea98, Pea99}. This expansion is typically
attributed to ``dark energy'' which makes up nearly 75\% of the universal
energy budget (e.g. \citealt{Mehta12}). Despite its ubiquitous nature,
our understanding of its physical properties is poor. It is possible
to gain leverage on the dark energy equation of state $w$, as well as
other cosmological parameters such as $\Omega_{\rm K}$ through studying
the cosmic expansion history. We will use an extension of the baryon
acoustic oscillations method, namely its anisotropic signature, to probe
cosmic expansion in this paper.

Baryon acoustic oscillations (BAO) arise from the interactions between
matter and radiation in the early universe. The story begins with the
seeding of primordial overdensities composed of dark matter, baryons,
photons and neutrinos. Prior to recombination, the large density
of free electrons locks the photons and baryons together via Thomson
scattering. As the overdensities grow gravitationally, they also heat up
and the photons become very energetic. The radiation pressure becomes
large enough to push the photon-baryon fluid outwards in a spherical
soundwave. The dark matter, which remains at the central overdensity,
exerts a gravitational restoring force on the fluid. The competition
between the outward push by radiation and the inward pull of gravity gives
rise to a system of standing sound waves within the plasma \citep{PY70,
SZ70, BE84, H89, HS96, HW96, EH98}. The period of these waves corresponds
to a characteristic spatial scale known as the acoustic scale or the
sound horizon. This scale, $\sim150$ comoving Mpc, is the distance
traveled by the sound wave in the plasma prior to recombination. When
recombination occurs, the free electrons are quickly captured which ends
the coupling between the photons and baryons. The photons stream away,
forming the cosmic microwave background (CMB), and the baryons are left
behind at characteristic separations corresponding to the acoustic
scale. This scale is still measurable in the clustering distribution
of galaxies today, which makes it an ideal standard ruler \citep{EH98,
EHT98, BG03, E03, HH03, L03, SE03, M04, AR05, AQG05, Angulo05, GB05,
DJT06}. We can use this ruler to measure the distance to various redshifts
using the clustering distribution of galaxies measured from large galaxy
surveys. This distance-redshift relation depends on the values of the
cosmological parameters, including those governing dark energy, so it
can be used to infer the cosmic expansion history and the cosmology of
the universe.

To use the BAO method, we must first measure the acoustic scale from
the clustering of galaxies. This is typically done statistically using
the 2-point correlation function of galaxy separations or its Fourier
transform, the power spectrum. Past BAO studies have primarily been
focused on the spherically-averaged (monopole) statistics \citep{Cea05,
Eea05, H06, Tea06, Pea07, Kea10, Pea10, Beutler11, Bea11a, Pea12,
Xea12, Mehta12}, which only allow us to measure the spherically-averaged
distance $D_V(z) \propto D_A(z)^2/H(z)$, where $z$ is the median redshift
of the galaxy sample. This effectively assumes that the clustering of
galaxies is isotropic. Most importantly, the Hubble parameter $H(z)$
is degenerate with $D_A(z)$ in this measure and hence we cannot directly
probe the cosmic expansion history encoded by $H(z)$.

The clustering of galaxies, however, is not truly isotropic. Anisotropies
arise from large-scale redshift-space distortions caused by the
line-of-sight velocity of galaxies \citep{Kaiser87} and from assuming
the wrong cosmology when calculating the 2-point statistics. This second
point can be used to break the degeneracy between $H(z)$ and $D_A(z)$. One
can imagine that if we assume the wrong cosmology, then the BAO will
appear at slightly different locations along the line-of-sight and
transverse directions because the line-of-sight distances are measured
from redshifts and $H(z)$, whereas the transverse distances are measured
using the angular separation and $D_A(z)$.

The use of such anisotropies to measure the cosmic expansion history was
first proposed by \citet{AP79}. In the case of the BAO, the anisotropy
can be measured using the clustering signal along different directions
\citep{Oea08,GCH11,Bea11b,CW11,KSB12} or by looking at higher order
multipoles of the 2-point statistics \citep{PW08,TSN11, CW12}. If the
clustering were isotropic, then all higher order multipoles should be
zero. However, anisotropies introduce power into the even multipoles (the
odd multipoles remain zero due to symmetry). This fact can be exploited
to infer the values of $H(z)$ and $D_A(z)$ (see \S\ref{sec:theory}).

In this paper, we will focus on the multipole method and apply the
technique described in \citet{PW08} to the 7th data release (DR7;
\citealt{Aea09}) of the Sloan Digital Sky Survey (SDSS). We present
the first application of this method to a galaxy redshift survey and
demonstrate its feasibility. We calibrate our methods on 160 LasDamas
mocks and perform detailed tests to ensure their robustness. We also
apply reconstruction, a technique for partially removing the effects of
non-linear structure growth on the BAO \citep{Eea07}. This technique
has been tested extensively on the monopole through simulations
\citep{Sea08,PWC09,Nea09,Sea10,Mea11} and has recently been applied
to SDSS DR7 data \citep{Pea12,Xea12,Mehta12}. The current work takes a
first look at how reconstruction affects the anisotropic BAO signal and
uses this to infer $H(z)$ and $D_A(z)$ from the DR7 data.

In \S\ref{sec:theory} we present the theoretical background for the
multipole method. In \S\ref{sec:analysis} we present our analysis
techniques with emphasis on the intricacies of the fitting model. In
\S\ref{sec:datasets} we introduce the mock catalogues used and the
SDSS DR7 data. \S\ref{sec:mockres} and \S\ref{sec:datares} present our
fitting results and detailed tests of our fitting model and reconstruction
technique on the mocks and data respectively. We present the cosmological
implications of our $D_A(z)$ and $H(z)$ measurements in \S\ref{sec:cosmo}
and conclude in \S\ref{sec:theend}.

\section{Theory}\label{sec:theory}

\subsection{Background, Basics and Definitions} \label{sec:bkgrd}

There are two main effects that give rise to anisotropic galaxy
clustering. The first are redshift-space distortions, which arise due
to the line-of-sight velocities of galaxies such as their peculiar
motions within clusters (the Finger-of-God effect) or their coherent
infall towards overdense regions \citep{Kaiser87}. These distort our
measurements of cosmological redshifts and hence the line-of-sight
separation between galaxies. Such effects change the shape of the
correlation function smoothly with scale (i.e. they have no features
and are purely broadband in nature).

Anisotropic clustering can also arise if we assume the wrong cosmology
when calculating the separations between galaxies. Since the distribution
of matter is mostly isotropic at large scales, artificial anisotropies
are introduced by calculating distances assuming the wrong cosmology
as each cosmology predicts a unique distance scale. Specifically,
we calculate line-of-sight separations using redshifts and the Hubble
parameter $H(z)$ while transverse separations are calculated using the
angular separation between pairs of galaxies and the angular diameter
distance $D_A(z)$. Both $H(z)$ and $D_A(z)$ are predicted given a set
of cosmological parameters; however, if these do not match the true
cosmology of the universe, we will measure different clustering signals
along the line-of-sight and transverse directions. This implies that
the BAO signal in the line-of-sight direction will be slightly offset
from its location in the transverse direction. This is a manifestation
of the Alcock-Paczynski technique \citep{AP79}, which uses the measured
anisotropy in an object thought to be isotropic to constrain the true
cosmology of the universe.

In the past, limited survey volume has made it difficult to analyze
the differential clustering along the line-of-sight and transverse
directions. As a result, most BAO analyses have been based on
the spherically-averaged (i.e. monopole) clustering statistics
(e.g. \citealt{Pea10} and \citealt{Pea12}), which only allow us to measure
$D_V(z)$, the spherically-averaged distance to redshift $z$. This quantity
is defined as
\begin{equation}
D_V(z) = \bigg[(1+z)^2D_A^2(z)\frac{cz}{H(z)}\bigg]^{1/3}
\end{equation}
which corresponds to two powers of $D_A(z)$ from our transverse distance
measure along the two transverse directions on the sky and one power of
$H(z)$ from our line-of-sight distance measure. However, by measuring
the anisotropy, we will be able to break this degeneracy between $D_A(z)$
and $H(z)$, and measure these two quantities separately.

To measure the anisotropy, we must construct a clustering model that
includes a parameterization of the anisotropic signal. We can then
fit this model to the data and measure this parameter. Essentially
one is presented with two choices. The first is to perform fits to
the transverse and radial correlation functions (e.g. \citealt{Oea08})
and the second is to simultaneously fit the monopole and higher order
multipoles of the clustering statistics \citep{PW08}. 

In this work, we will follow the formalism for the multipole method
based on the work of \citet{PW08}. We note that the BAO can additionally
be shifted in an isotropic manner if the assumed cosmology is not the
true cosmology. Isotropic shifts also occur due to non-linear structure
growth. We define the isotropic shift in BAO position as
\begin{eqnarray}
\alpha &=& \frac{D_V(z)/r_s}{D_{V,f}(z)/r_{s,f}}\\
&=& \left[
\frac{D^2_A(z)}{D^2_{A,f}(z)} \frac{H_f(z)}{H(z)} \right]^{1/3}
\frac{r_{s,f}}{r_s}
\label{eqn:ahd}
\end{eqnarray}
where $D_V$ is defined as above and $r_s$ is the sound horizon (BAO
scale). The $f$ subscripts correspond to a fiducial or reference cosmology
on which we anchor our measurements: all isotropic shifts and anisotropic
signals in the BAO are measured relative to this fiducial cosmology (see
\S\ref{sec:fitting}). This parameterization has been used extensively
in past BAO studies focusing on the monopole.

We parameterize the anisotropic BAO signal as $\epsilon$
\begin{equation}
1+\epsilon = \left[
\frac{H_f(z)}{H(z)} \frac{D_{A,f}(z)}{D_A(z)}
\right]^{1/3}.
\label{eqn:ehd}
\end{equation}
These parameterizations of $\alpha$ and $\epsilon$ are derived
from isotropic coordinate dilations and anisotropic coordinate
warpings between the true and fiducial cosmology spaces (see Equations
\ref{eqn:adef} \& \ref{eqn:edef}). Note that if there is no isotropic
shift, then $\alpha=1$. Similarly, the lack of anisotropy implies
$\epsilon=0$. Combining Equations (\ref{eqn:ahd}) \& (\ref{eqn:ehd}),
we arrive at
\begin{eqnarray}
\frac{D_A(z)}{r_s} &=& \frac{\alpha}{1+\epsilon} \frac{D_{A,f}(z)}{r_{s,f}} 
\label{eqn:daz} \\
H(z) r_s &=& \frac{1}{\alpha(1+\epsilon)^2} H_f(z) r_{s,f} 
\label{eqn:hz}.
\end{eqnarray}
By measuring both the isotropic and anisotropic BAO shifts, we
can separately constrain the angular diameter distance $D_A(z)$
and the Hubble parameter $H(z)$ at the median redshift of our galaxy
sample $z$. If we denote the errorbars on $\alpha$ and $\epsilon$ as
$\sigma_\alpha$ and $\sigma_\epsilon$, and the covariance between them
as $\sigma_{\alpha\epsilon}$, then the errorbars on $D_A(z)$ and $H(z)$
can be calculated as
\begin{equation}
\Bigg(
\begin{matrix}
\sds & \sdh \\
\sdh & \shs
\end{matrix}
\Bigg)
= \Bigg(
\begin{matrix}
\ddda & \ddde \\
\dhda & \dhde 
\end{matrix}
\Bigg)
\Bigg(
\begin{matrix}
\sas & \sae \\
\sae & \ses
\end{matrix}
\Bigg)
\Bigg(
\begin{matrix}
\ddda & \ddde \\
\dhda & \dhde 
\end{matrix}
\Bigg)^T.
\label{eqn:ermat}
\end{equation}
This yields,
\begin{eqnarray}
\frac{\sigma_{D_A}^2}{D_A^2} &=& \alpha^{-2} \sigma_\alpha^2 +
(1+\epsilon)^{-2} \sigma_\epsilon^2 - 2\alpha^{-1}(1+\epsilon)^{-1}
\sigma_{\alpha \epsilon} \label{eqn:edaz}\\
\frac{\sigma_H^2}{H^2} &=& \alpha^{-2} \sigma_\alpha^2 +
4(1+\epsilon)^{-2} \sigma_\epsilon^2 + 4\alpha^{-1}(1+\epsilon)^{-1}
\sigma_{\alpha\epsilon} \label{eqn:ehz}\\
\frac{\sigma_{D_AH}}{D_AH} &=& -\alpha^{-2}\sigma_\alpha^2 +
2(1+\epsilon)^{-2}\sigma_\epsilon^2  -
\alpha^{-1}(1+\epsilon)^{-1}\sigma_{\alpha\epsilon}. \nonumber \\
&&\label{eqn:cdah}
\end{eqnarray}

We will also need a method for distinguishing between anisotropies
introduced by redshift-space distortions into the broadband shape
of our clustering statistic and those introduced through assuming
the wrong cosmology (i.e. the Alcock-Paczynski signal). There exist
simple redshift-space distortion models that can be used if we are
only interested in analyzing the anisotropy in the BAO signal and
not the details of the redshift-space distortions themselves. Any
residual inadequate matching between these models and the actual
broadband shape of the data can be mostly compensated by including a few
additional marginalization terms \citep{Sea08,Xea12} such as Equation
(\ref{eqn:fida}) described in \S\ref{sec:fitting}.

\subsection{Formalism for the Correlation Function} \label{sec:xiform}

The clustering of galaxies can be measured using the correlation function
$\xi(r)$ or the power spectrum $P(k)$. Since our analysis will focus on
the correlation function, we will present the formalism for configuration
space here and state the analogue for the power spectrum in Fourier space,
which can also be found in \citet{PW08}.

We begin with a few basic coordinate definitions,
\begin{eqnarray}
r^2 &=& r^2_\parallel+r^2_\perp \label{eqn:rdef} \\
\mu^2 &=& \cos^2\theta = \frac{r^2_\parallel}{r^2} \label{eqn:mudef}
\end{eqnarray}
where $r$ is the separation between two galaxies and $\theta$ is the angle
between a galaxy pair and the line-of-sight direction. $r_\parallel$
is the separation of the galaxies in the line-of-sight direction and
$r_\perp$ is their transverse separation. In the following, unprimed
coordinates will denote the fiducial cosmology space and primed
coordinates will denote the true cosmology space.

The isotropic dilation ($\alpha$) and anisotropic warping ($\epsilon$)
parameters are then defined by
\begin{eqnarray}
r'_\parallel &=& \alpha (1+\epsilon)^2 r_\parallel \label{eqn:adef} \\
r'_\perp &=& \alpha (1+\epsilon)^{-1} r_\perp. \label{eqn:edef}
\end{eqnarray}

Substituting Equations (\ref{eqn:adef}) \& (\ref{eqn:edef}) into the
definitions of $r'$ and $\mu'$ as in Equations (\ref{eqn:rdef}) \&
(\ref{eqn:mudef}), we see that
\begin{eqnarray}
r' &=& \alpha \sqrt{(1+\epsilon)^4 r^2_\parallel 
+ (1+\epsilon)^{-2} r^2_\perp} \nonumber \\
&=& \alpha r [1+2\epsilon L_2(\mu)] + \mathcal{O}(\epsilon^2)
\label{eqn:rrdef}
\end{eqnarray}
where in the last line we have substituted the second order Legendre
polynomial
$L_2(\mu) = (3\mu^2-1)/2$. Also,
\begin{eqnarray}
\mu'^2 &=& \frac{\alpha^2 (1+\epsilon)^4 r^2_\parallel}{\alpha^2
(1+\epsilon)^4r^2_\parallel + \alpha^2(1+\epsilon)^{-2}r^2_\perp} 
\nonumber \\
&=& \mu^2 + 6\epsilon(\mu^2-\mu^4) + \mathcal{O}(\epsilon^2).
\label{eqn:mumudef}
\end{eqnarray}

The true 2D correlation function $\xi(\vec{r'})$ can be Legendre
decomposed into multipole moments as
\begin{equation}
\xi(\vec{r'}) = \sum^{\infty}_{\ell'=0} \xi_{\ell'}(r') L_{\ell'}(\mu')
\end{equation}
where the $L_{\ell'}(\mu')$ are Legendre polynomials of order
$\ell'$. Again, if clustering were perfectly isotropic the $\ell>0$
moments would all be zero. Anisotropy introduces power into the
even-order multipoles, however the odd-order multipoles are always
zero due to symmetry. We can substitute Equation (\ref{eqn:rrdef}) into
$\xi_{\ell'}(r')$ and Equation (\ref{eqn:mumudef}) into $L_{\ell'}(\mu')$
and write
\begin{eqnarray}
\xi(\vec{r'}) &=& \sum^{\infty}_{\ell'=0} 
\left[ \xi_{\ell'}(\alpha r) + 2\epsilon L_2(\mu)
\frac{d \xi_{\ell'}(\alpha r)}{d\log(r)} \right] \nonumber \\
&& \cdot
\left[ L_{\ell'}(\mu) + 3\epsilon\mu(1-\mu^2)
\frac{d L_{\ell'}(\mu)}{d\mu} \right],
\end{eqnarray}
where we have made a Taylor expansion in both large brackets. 

Finally, in the fiducial cosmology space, we measure the multipole moments
\begin{eqnarray}
\xi_\ell(r) &=& \frac{2\ell+1}{2} \int^{1}_{-1} \xi(\vec{r'})
L_\ell(\mu) d\mu \\
&=& \xi_\ell(\alpha r) \nonumber \\
&& + 3\epsilon \bigg[ \frac{-\ell(\ell-1)(\ell-2)}{(2\ell-3)(2\ell-1)}
\xi_{\ell-2}(\alpha r) \nonumber \\
&& + \frac{\ell(\ell+1)}{(2\ell-1)(2\ell+3)}
\xi_{\ell}(\alpha r) \nonumber \\
&& + \frac{(\ell+1)(\ell+2)(\ell+3)}{(2\ell+3)(2\ell+5)}
\xi_{\ell+2}(\alpha r) \bigg] \nonumber \\
&& + 2\epsilon \bigg[ \frac{3\ell(\ell-1)}{2(2\ell-3)(2\ell-1)}
\frac{d \xi_{\ell-2}(\alpha r)}{d \log(r)} \nonumber \\
&& + \frac{\ell(\ell+1)}{(2\ell-1)(2\ell+3)}
\frac{d \xi_{\ell}(\alpha r)}{d \log(r)} \nonumber \\
&& + \frac{3(\ell+1)(\ell+2)}{2(2\ell+3)(2\ell+5)}
\frac{d\xi_{\ell+2}(\alpha r)}{d \log(r)} \bigg],
\label{eqn:xidef}
\end{eqnarray}
where we have used the recursion relation for Legendre polynomials
\begin{equation}
(\ell+1)L_{\ell+1}(\mu) = (2\ell+1)\mu L_{\ell}(\mu) - \ell L_{\ell-1}(\mu),
\end{equation}
the derivative relation 
\begin{equation}
\frac{d L_\ell(\mu)}{d\mu} = 
\frac{(\ell+1)[\mu L_\ell(\mu) - L_{\ell+1}(\mu)]}{1-\mu^2}
\end{equation}
and the orthogonality of Legendre polynomials
\begin{equation}
\int_{-1}^{1} L_\ell(\mu)L_{\ell'}(\mu) d\mu = \frac{2}{2\ell+1}
\delta_{\ell\ell'}.
\end{equation}
Here, $\delta_{\ell\ell'}$ is the delta function. Since we measure the
correlation function using our choice of fiducial cosmology, Equation
(\ref{eqn:xidef}) will form the basis of our fitting template for
extracting the isotropic dilation ($\alpha$) and the anisotropic warping
($\epsilon$) signal from the data.

The same relations can be derived in Fourier space from the definitions
\citep{PW08}
\begin{eqnarray}
k'_\parallel &=& \alpha^{-1} (1+\epsilon)^{-2} k_\parallel \\
k'_\perp &=& \alpha^{-1} (1+\epsilon) k_\perp.
\end{eqnarray}
The final equation for the measured multipole moments is identical to
the configuration space case except $\xi_\ell(\alpha r) \rightarrow
P_\ell(k/\alpha)$ and the sign on each occurrence of $\epsilon$
is flipped.

\subsection{The Anisotropic Signal} \label{sec:anisig}

\begin{figure*}
\vspace{0.4cm}
\centering
\subfigure[The monopole (left) and quadrupole (right) we expect to
measure in the presense of anisotropic clustering according to Equations
(\ref{eqn:mono}) \& (\ref{eqn:quad}) for a linear theory based model
including the Kaiser effect. The Kaiser effect gives rise to the BAO
bump near $110\hMpc$ in the quadrupole. One can see that the monopole
is insensitive to $\epsilon$. However, $\epsilon$ works to change the
position of the line-of-sight and transverse BAO features in opposing
directions with the line-of-sight having a more prominent shift. The
differential nature of these shifts moves the quadrupole BAO, compounding
with any isotropic shifts. Hence, we expect the quadrupole to be sensitive
to an anisotropic BAO signal.]
{
\epsfig{file=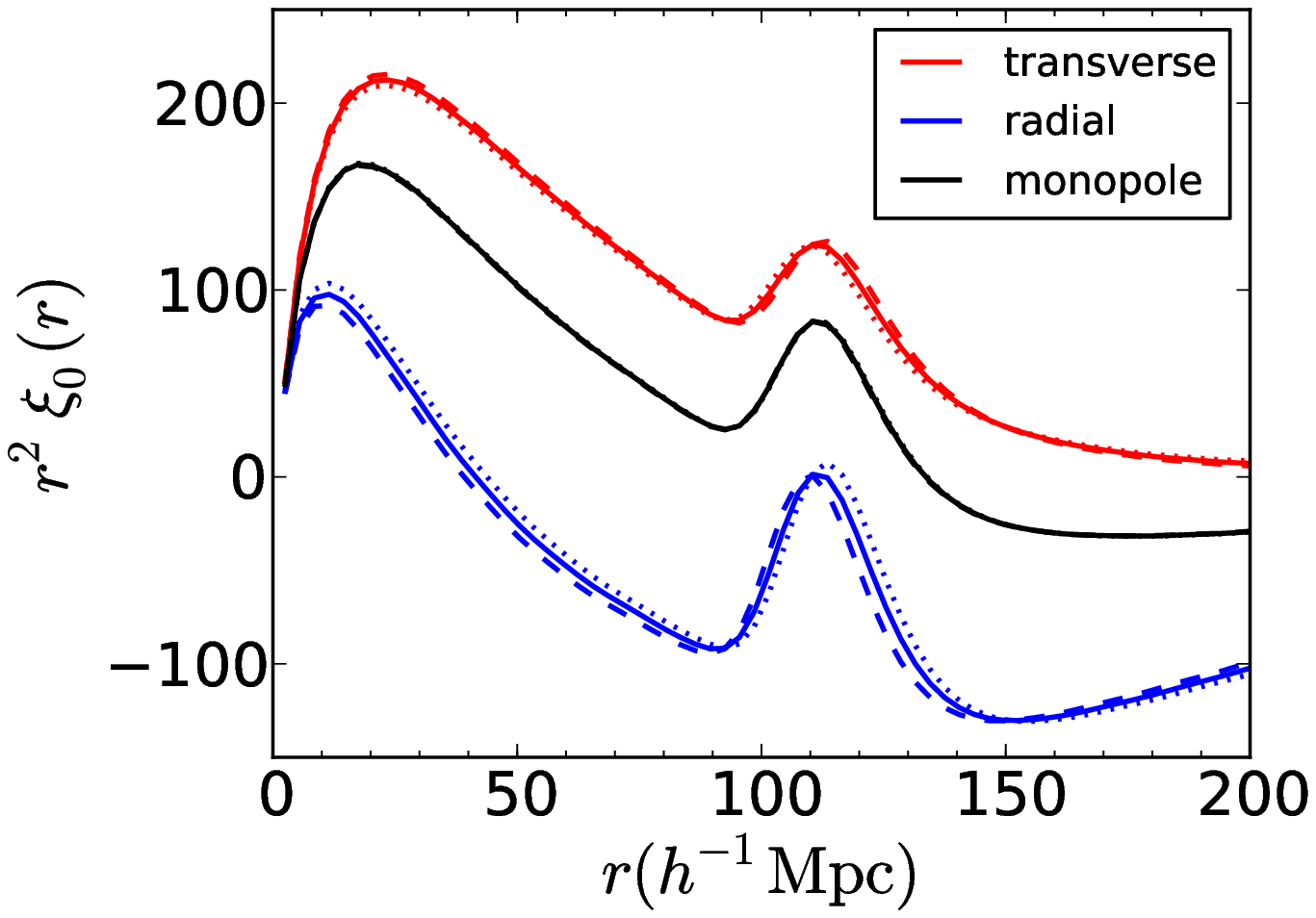, width=0.4\linewidth, clip=}
\hspace{0.5cm} 
\epsfig{file=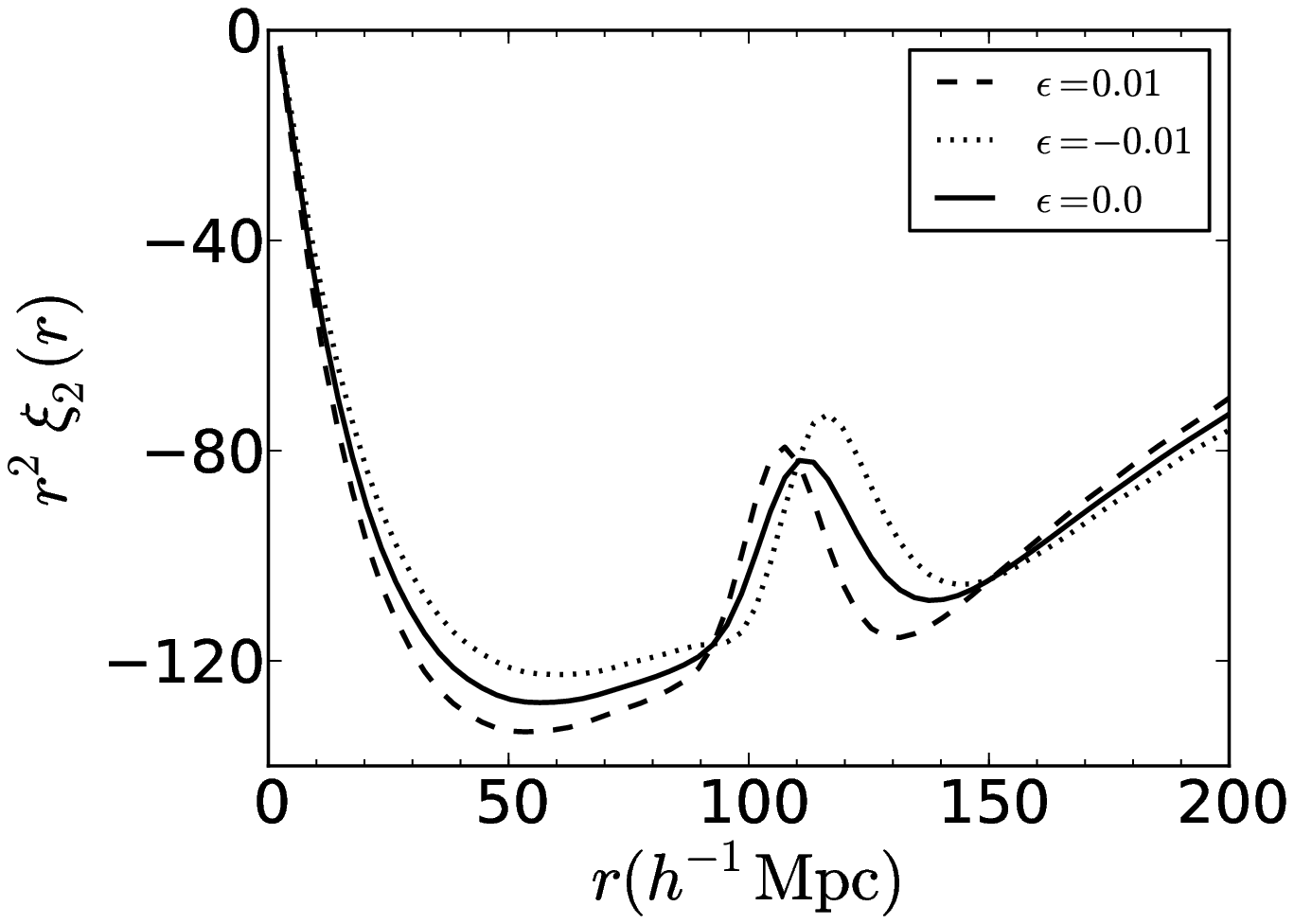, width=0.4\linewidth, clip=}
\label{fig:ep_linfig}
}
\subfigure[Variation of monopole (left) and quadrupole (right) models
including the Kaiser effect and a full non-linear treatment of FoG
and anisotropic $\snl$. The solid black line in this and the similar
plots following always corresponds to the fiducial model parameters
$\Sigma_\perp = 6\hMpc$, $\Sigma_\parallel = 10\hMpc$ and $\Sigma_s =
4\hMpc$ with $\beta = 0.35$ (center of the $\beta$ prior in our fits). The
monopole is again affected very little by $\epsilon$. The fiducial
quadrupole model picks up a crest-trough-crest structure at the BAO scale
due to the differential broadening of the line-of-sight and transverse
BAO signals by Kaiser, FoG and anisotropic $\snl$. We again see that
the anisotropic warping parameterized by $\epsilon$ works to shift the
location of the quadrupole BAO. In addition, it can adjust the relative
amplitude of the crests. $\epsilon$ is the only parameter that can shift
the BAO in the quadrupole while leaving the monopole BAO unaffected;
the isotropic shift $\alpha$ changes the BAO position equally in both.]
{
\epsfig{file=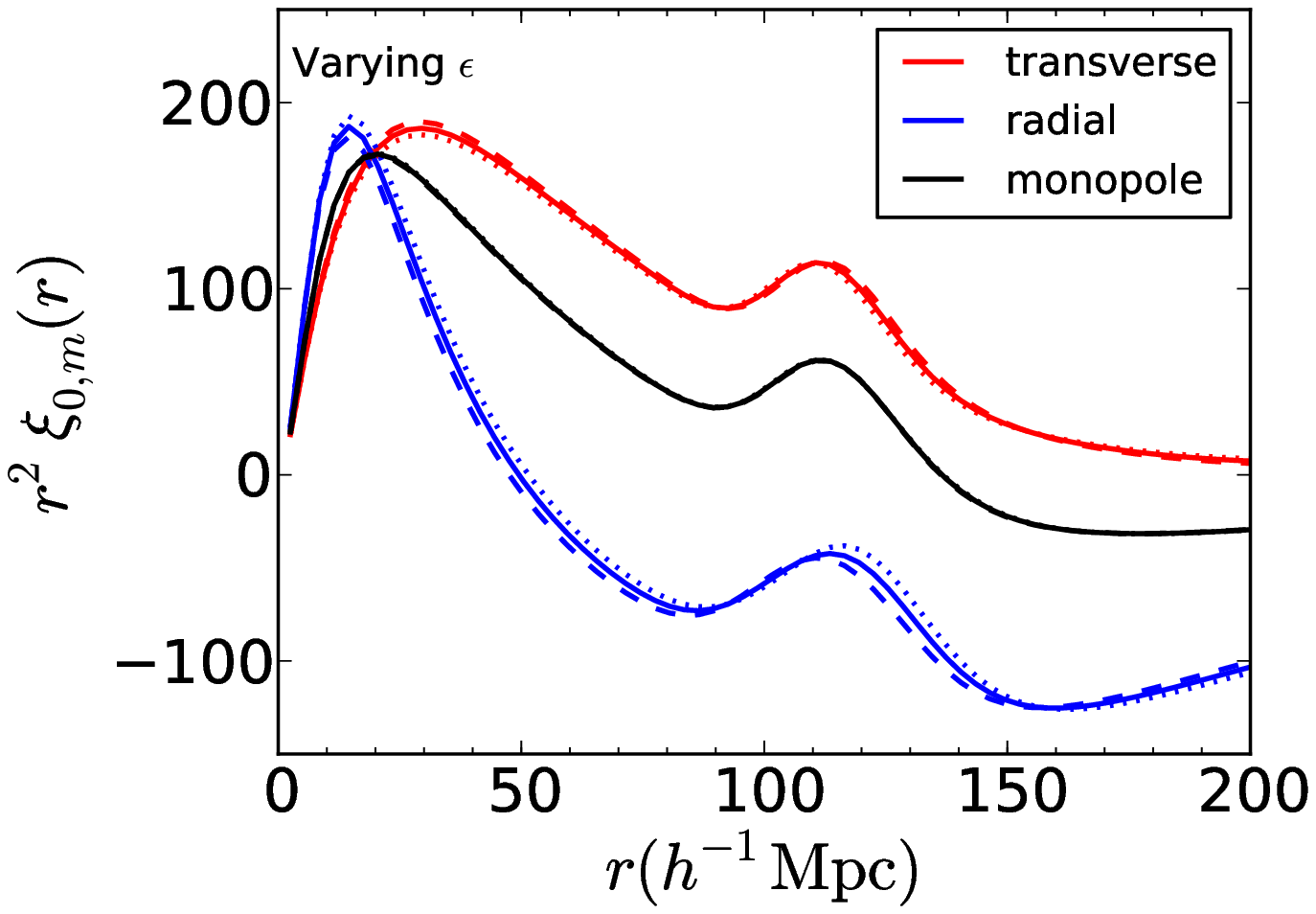, width=0.4\linewidth, clip=}
\hspace{0.5cm}
\epsfig{file=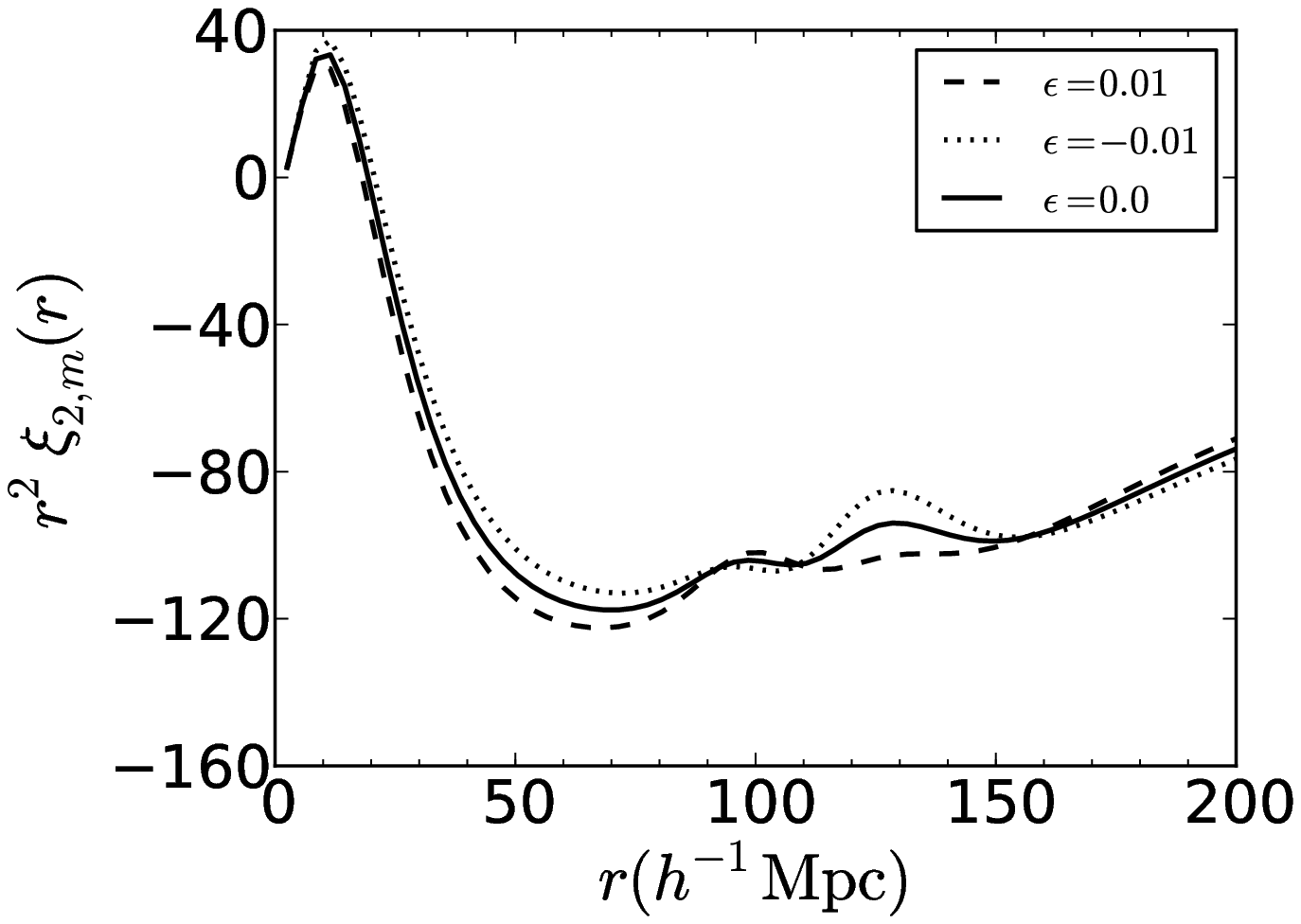, width=0.4\linewidth, clip=}
\label{fig:epfig}
}
\caption{Variation of our models with $\epsilon$ for a linear theory
based model including the Kaiser redshift-space distortion (a) and a full
non-linear model including FoG and Kaiser redshift-space distortions as
well as anisotropic $\snl$ (b). In these and the following two figures,
we have assumed a cosmology of $\Omega_b = 0.04$, $\Omega_m = 0.25$,
$h=0.7$, $n_s=1.0$ and $\sigma_8=0.8$. $\epsilon$ parameterizes the
amount of Alcock-Paczynski anisotropy, which, if there was none, would
be equal to 0. The left panel shows the monopole (black), the transverse
correlation function (red) and the radial correlation function (blue),
where the difference between these latter two yields a measurement
of the quadrupole. Solid, dashed and dotted lines are defined as in
the plot legend of the right panel which shows the quadrupole. Note
that the quadrupole BAO feature is much weaker in the more realistic
non-linear model.}
\label{fig:epfig_full}
\end{figure*}

Although anisotropic BAO information exists in all higher order
multipoles, its magnitude decreases considerably. Hence, for the purposes
of this study, we will only focus on the monopole ($\ell=0$) and the
quadrupole ($\ell=2$). The monopole and quadrupole we expect to measure
according to Equation (\ref{eqn:xidef}) are
\begin{eqnarray}
\xi_0(r) &=& \xi _0(\alpha r) + \frac{2}{5}\epsilon
\left[ 3\xi_2(\alpha r) + \frac{d\xi_2(\alpha r)}{d\log(r)} \right] 
\label{eqn:mono} \\
\xi_2(r) &=& 2\epsilon \frac{d\xi_0(\alpha r)}{d\log(r)}
+\bigg( 1 + \frac{6}{7}\epsilon \bigg)\xi_2(\alpha r)
+\frac{4}{7}\epsilon \frac{d\xi_2(\alpha r)}{d\log(r)} \nonumber \\
&& + \frac{4}{7}\epsilon \bigg[ 5\xi_4(\alpha r) + 
\frac{d\xi_4(\alpha r)}{d\log(r)} \bigg],
\label{eqn:quad}
\end{eqnarray}
and will form the basis of our analysis. Here, $\xi_4(r)$ is the
hexadecapole ($\ell=4$).

Figure \ref{fig:ep_linfig} shows variations of the expected monopole
(Equation \ref{eqn:mono}), the transverse and radial components
of the full 2D correlation function, and the quadrupole (Equation
\ref{eqn:quad}) with $\epsilon$ for a linear theory based model. For
Figures \ref{fig:epfig_full}, \ref{fig:varyfig} and \ref{fig:derfig},
we have assumed a cosmology with $\Omega_b = 0.04$, $\Omega_m = 0.25$,
$h=0.7$, $n_s=1.0$ and $\sigma_8=0.8$ (the LasDamas cosmology described in
\S\ref{sec:sims}). We have included large-scale redshift-space distortions
(the Kaiser effect) so that the quadrupole becomes non-zero, but not
the Finger-of-God (FoG) effect. In Fourier space, this model is given
by Equation (\ref{eqn:tdp}) without the $F(k,\mu,\Sigma_s)$ term. The
transverse and radial correlation functions were calculated as $\xi_0
+ L_2(\mu)\xi_2$ for $\mu=0$ and 1 respectively. Note that taking the
difference between these yields the quadrupole.

One can see that the sensitivity of the monopole to $\epsilon$ is quite
low. However $\epsilon$ does cause significant shifts in the BAO position
in the line-of-sight and transverse components of the 2D correlation
function. The line-of-sight shift is larger than and opposite in direction
to the transverse shift. This causes the BAO feature in the quadrupole at
$\sim110\hMpc$ to move with $\epsilon$, in addition to the shift caused by
the isotropic dilation $\alpha$. Hence we see that the quadrupole can be
used to obtain a measurement of the anisotropic BAO signal via $\epsilon$.

\subsection{A Non-linear Model}\label{sec:nonlin}

In Figure \ref{fig:ep_linfig} discussed in the previous section,
we assumed linear theory including the Kaiser effect for $\xi_0(r)$
and $\xi_2(r)$. However, in order to model actual observations with
fidelity, we must also account for the FoG effect and non-linear structure
growth. This section details a plausible model that includes all of
these effects and will be used in our fitting procedure described in
\S\ref{sec:fitting} to measure $\alpha$ and $\epsilon$.

In Fourier space we can write the following template for the 2D non-linear
power spectrum
\begin{equation}
P_t(k,\mu) = (1+\beta\mu^2)^2 F(k,\mu,\Sigma_s)P_{\rm dw}(k,\mu)
\label{eqn:tdp}
\end{equation}
\citep{Fea94} where
\begin{equation}
F(k,\mu,\Sigma_s) = \frac{1}{(1+k^2\mu^2\Sigma_s^2)^2}
\end{equation}
\citep{Pea94} corresponds to a streaming model for the FoG effect and the
$(1+\beta\mu^2)^2$ term corresponds to the Kaiser model for large-scale
redshift-space distortions. Here $\Sigma_s$ is the streaming scale and
is typically $\sim3-4\hMpc$.

The de-wiggled power spectrum $P_{\rm dw}(k,\mu)$ is defined as
\begin{eqnarray}
P_{\rm dw}(k,\mu) &=& [P_{\rm lin}(k) - P_{\rm nw}(k)] \nonumber \\
&& \cdot \exp \bigg[
-\frac{k^2\mu^2\Sigma_\parallel^2 +k^2(1-\mu^2)\Sigma_\perp^2}{2} \bigg]
+P_{\rm nw}(k) \nonumber \\
\end{eqnarray}
\citep{ESW07} where $P_{\rm lin}(k)$ is the linear theory power spectrum
and $P_{\rm nw}(k)$ is a power spectrum without an acoustic signature
\citep{EH98}. $\Sigma_\parallel$ and $\Sigma_\perp$ are the line-of-sight
and transverse components of $\Sigma_{\rm nl}$, i.e.  $\Sigma_{\rm nl}^2
= (\Sigma_\parallel^2 + \Sigma_\perp^2)/2$, where $\Sigma_{\rm nl}$ is
the standard Gaussian damping of the BAO used to model the degradation
of the signal due to non-linear structure growth \citep{ESW07}. Here,
the damping is anisotropic due to the Kaiser effect.

The multipoles of this template are then
\begin{equation}
P_{\ell,t}(k) = \frac{2\ell+1}{2}\int^1_{-1} P_t(k,\mu) L_\ell(\mu) d\mu,
\end{equation}
which can be transformed to configuration space using
\begin{equation}
\xi_{\ell,t}(r) = i^{\ell} \int \frac{k^3 d\log(k)}{2\pi^2}
P_{\ell,t}(k) j_\ell(kr).
\end{equation}

\begin{figure*}
\vspace{0.4cm}
\centering
\subfigure[Variation of the monopole (left) and quadrupole (right) models
with $\Sigma_\perp$ and $\Sigma_\parallel$. The BAO peak in the monopole
can be affected by $\snl$ as expected since this parameter is used to
model the smearing of the BAO due to non-linear structure growth. In these
plots we have demonstrated the effects of going to an isotropic $\snl$
of roughly the same magnitude as the fiducial case (dotted line) and a
smaller $\snl$ (dashed line). We see that going to an isotropic $\snl$
has little effect on the monopole, however, it completely eliminates the
trough feature at the BAO scale in the quadrupole. Going to a smaller
$\snl$ makes the peak appear sharper in the monopole as expected. In
the quadrupole, it alters the structure of the peaks and reduces the
crest-trough contrast.]
{
\epsfig{file=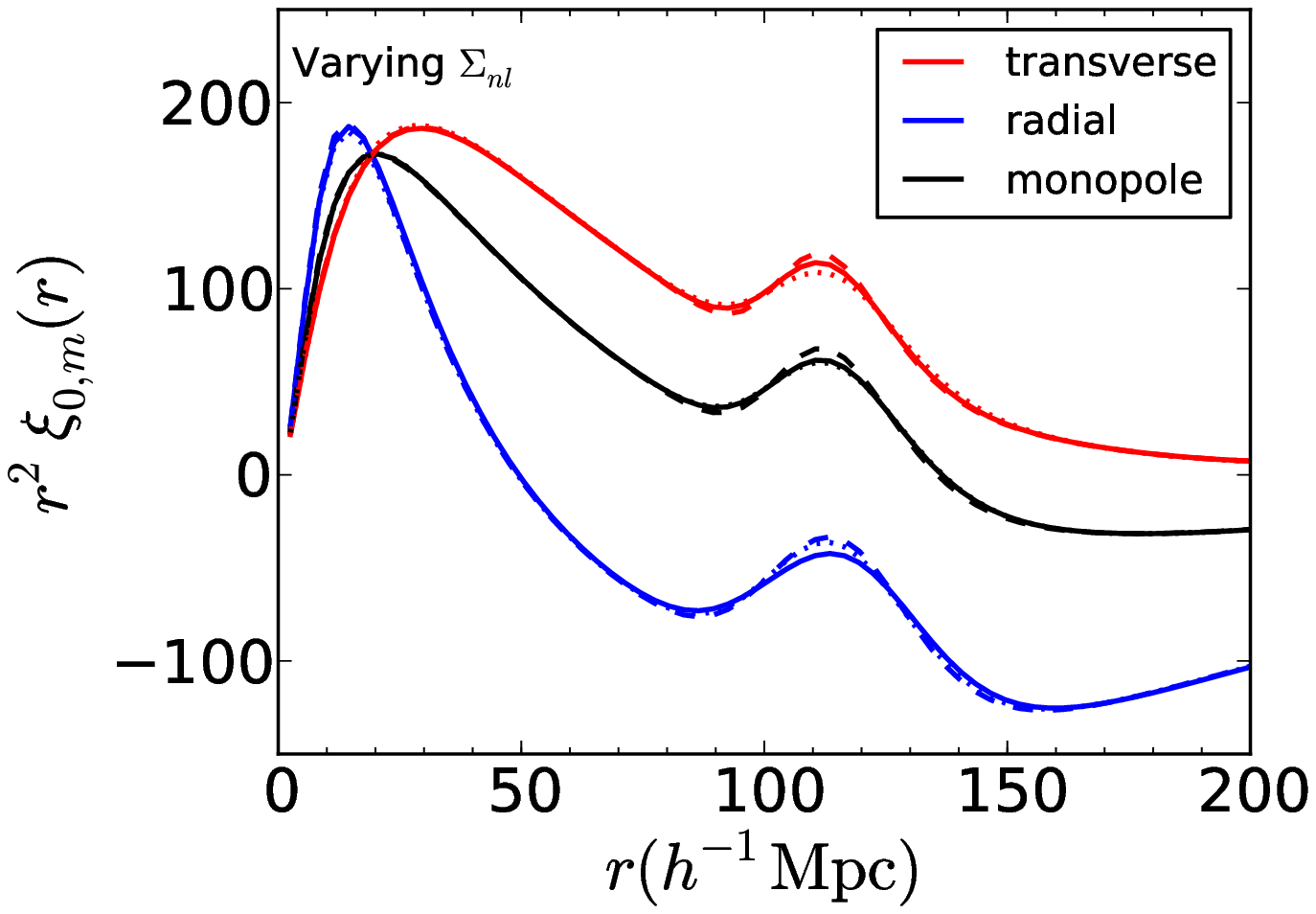, width=0.4\linewidth, clip=}
\hspace{0.5cm}
\epsfig{file=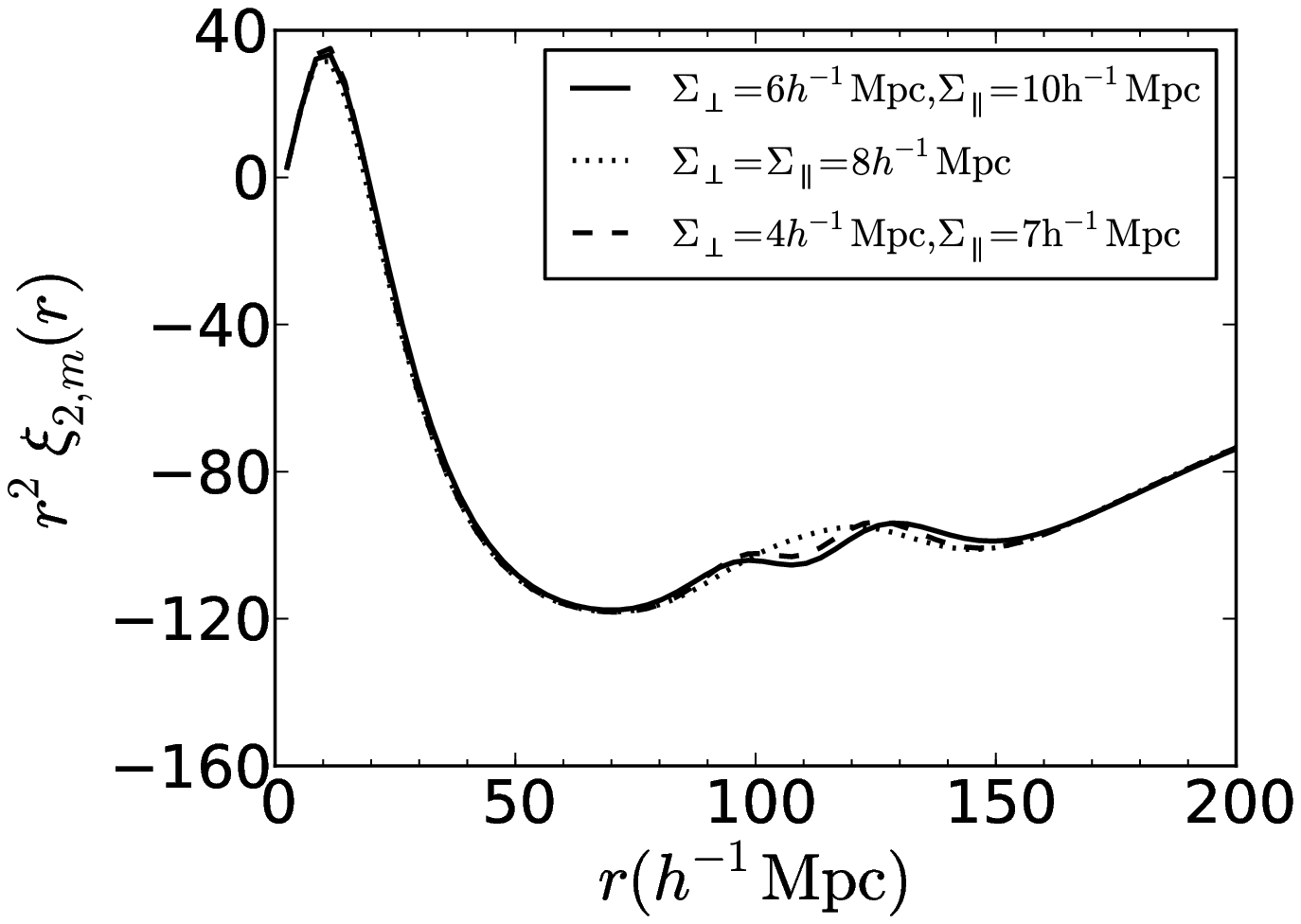, width=0.4\linewidth, clip=}
\label{fig:snlfig}
}
\subfigure[Variation of the monopole (left) and quadrupole (right)
models with $\Sigma_s$. The BAO feature in the monopole can be slightly
broadened by a large $\Sigma_s$. In the quadrupole, the effects of
$\Sigma_s$ are partially degenerate with $\snl$ in that it can alter
the crest-trough contrast and can also completely eliminate the trough
($\Sigma_s=0\hMpc$). However, the effects of $\Sigma_s$ are much stronger
at small scales due to its large influence on the radial and transverse
correlation functions, giving us leverage on this parameter. These
variations are not surprising since the FoG effect is most pronounced
at smaller scales.]
{
\epsfig{file=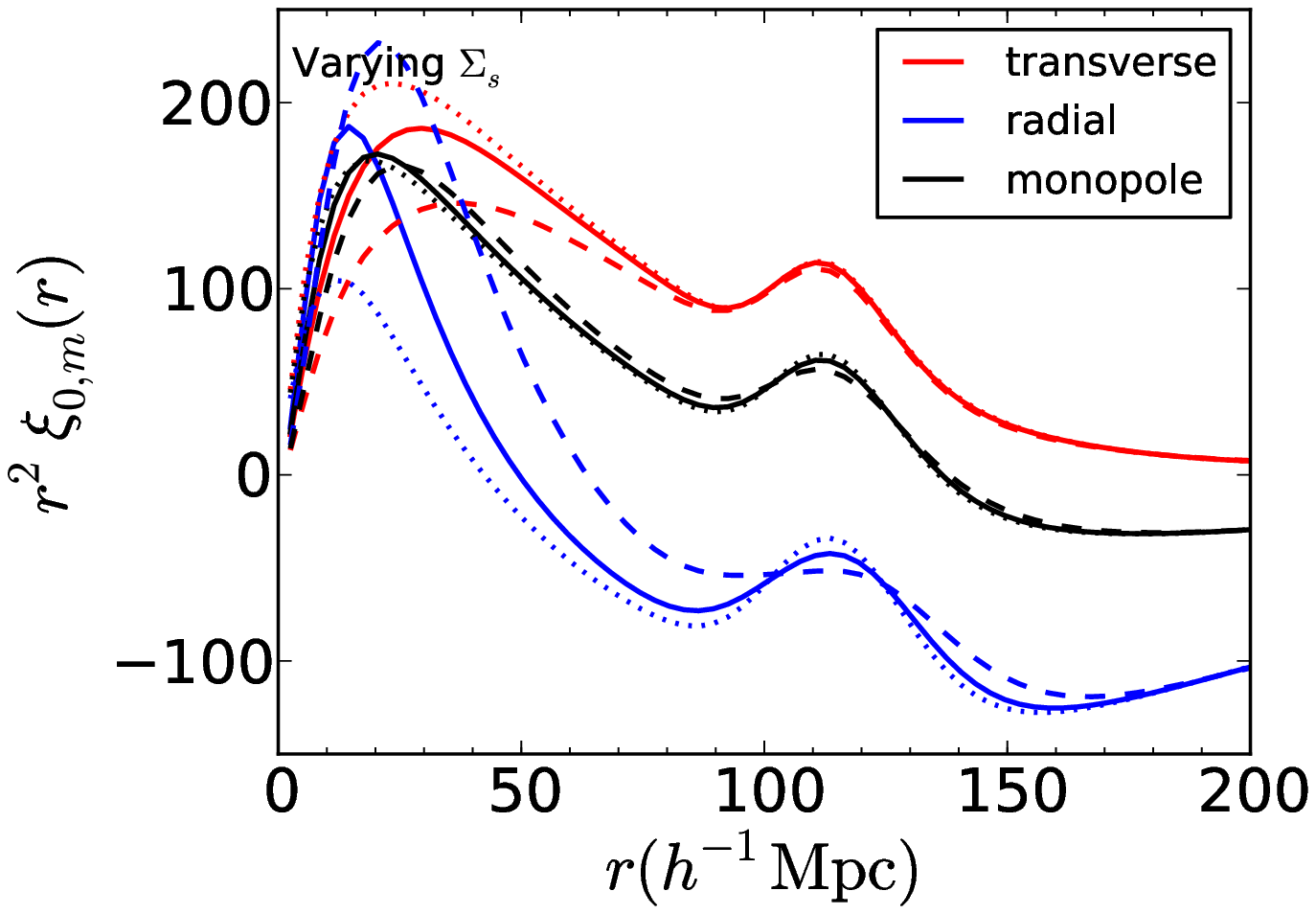, width=0.4\linewidth, clip=}
\hspace{0.5cm}
\epsfig{file=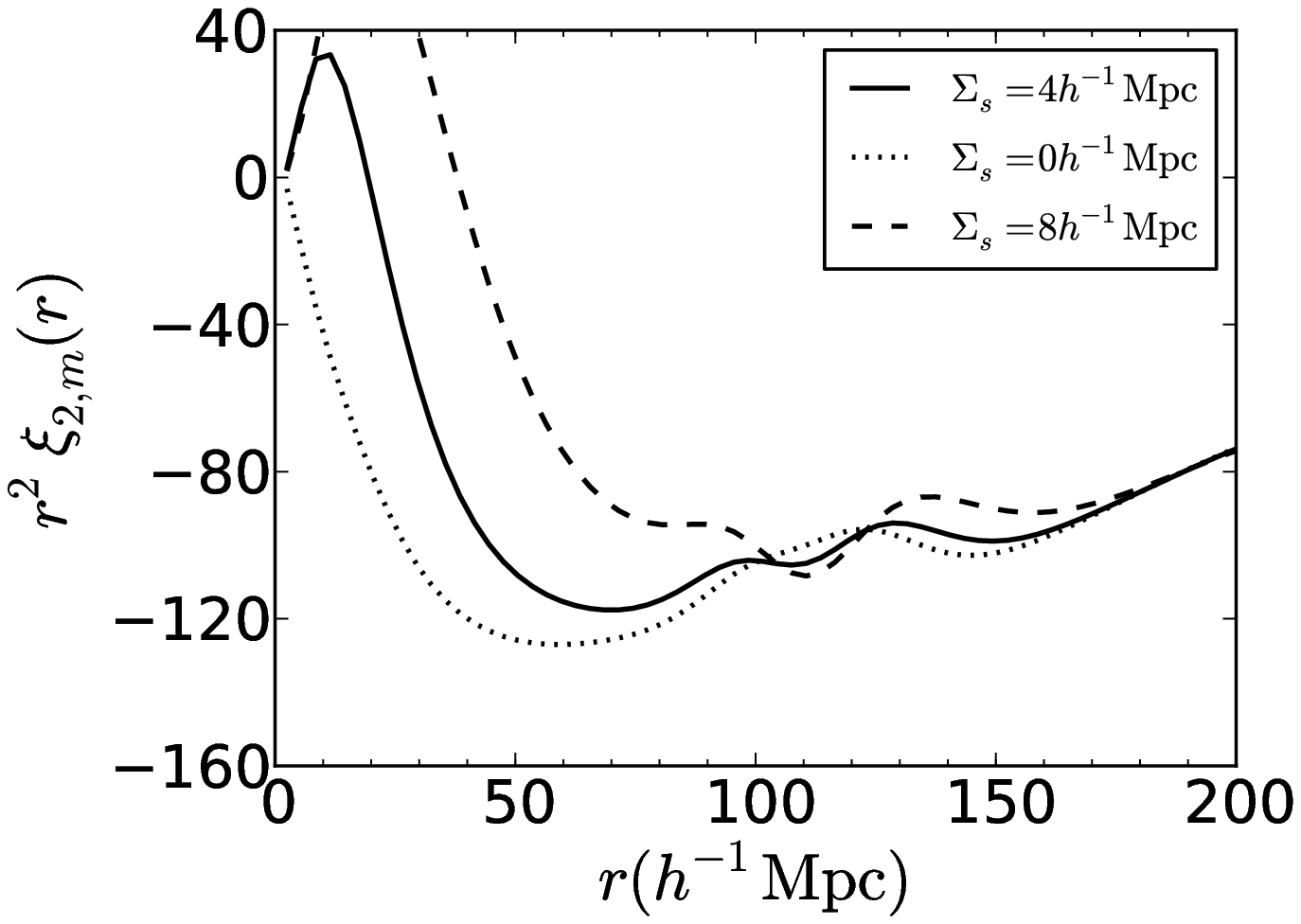, width=0.4\linewidth, clip=}
\label{fig:ssfig}
}
\subfigure[Variation of the monopole (left) and quadrupole (right)
models with $\beta$. Again we see that the monopole is not significantly
affected by $\beta$. The line-of-sight and transverse correlation
functions experience large changes in amplitude with $\beta$, leading
to amplitude differences in the quadrupole at large scales.]
{
\epsfig{file=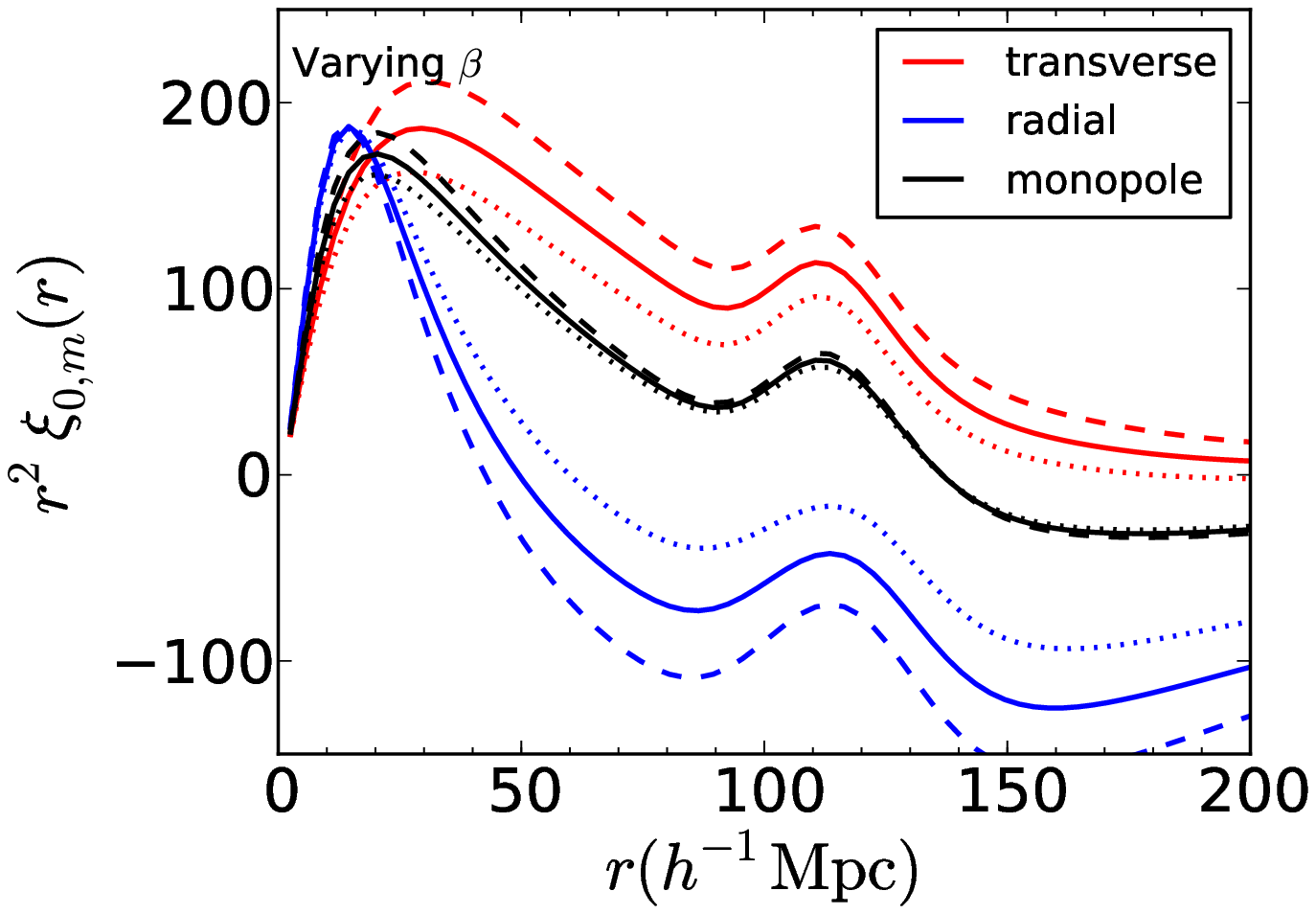, width=0.4\linewidth, clip=}
\hspace{0.5cm}
\epsfig{file=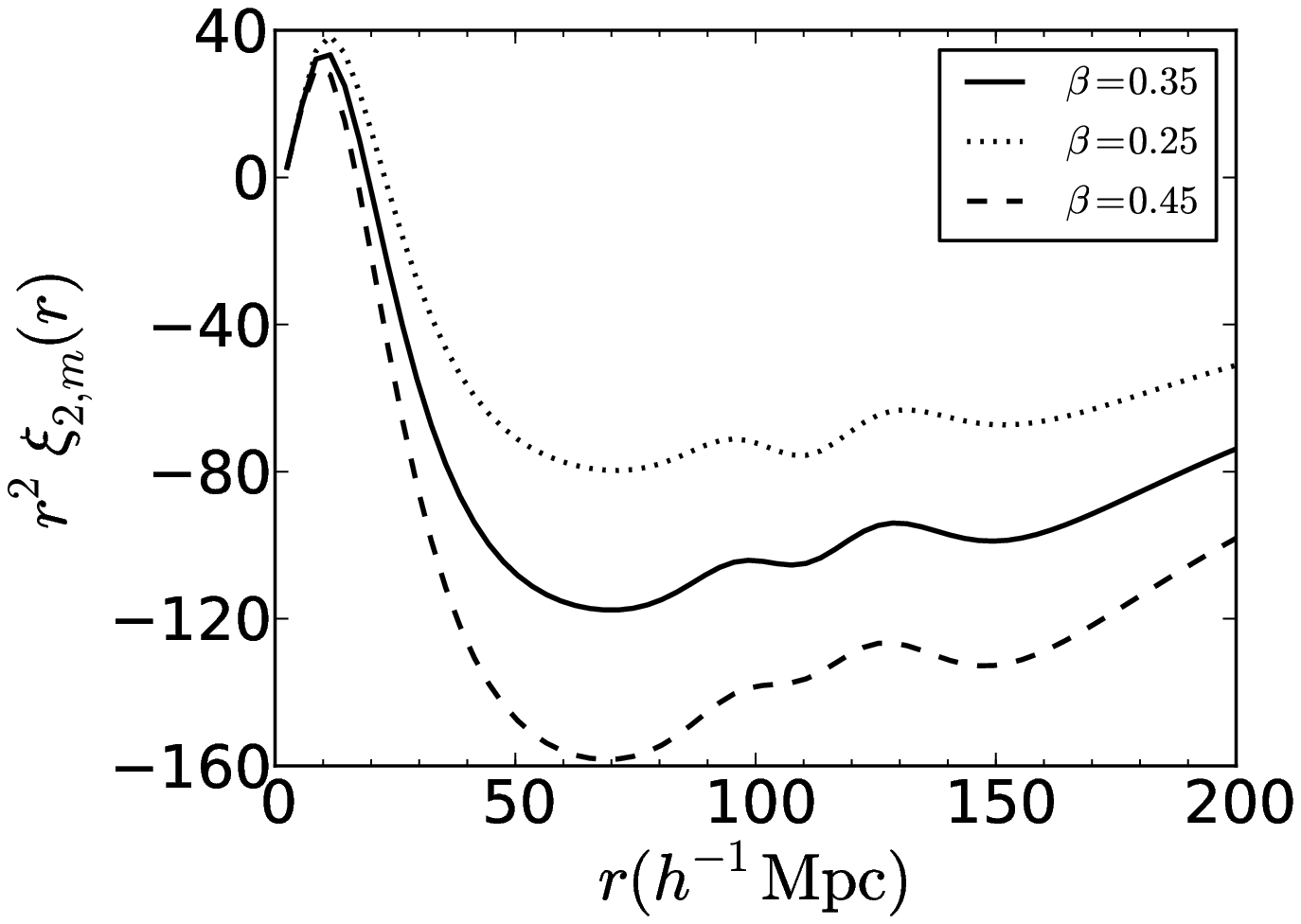, width=0.4\linewidth, clip=}
\label{fig:betafig}
}
\caption{Variation of the non-linear monopole and quadrupole models
with different model parameters: $\snl$ (a), $\Sigma_s$ (b) and $\beta$
(c). Comparing the behaviour of these parameters to $\epsilon$ (Figure
\ref{fig:epfig}) indicates that the various model parameters have mostly
different effects on the quadrupole. While all of these parameters
can affect the shape of the quadrupole, only $\epsilon$ can change
the quadrupole BAO position separately from the monopole BAO ($\alpha$
changes both in lock-step). Hence we expect that the effects of $\epsilon$
should be detectable.}
\label{fig:varyfig}
\end{figure*}

Figure \ref{fig:epfig} shows the variation of our non-linear monopole
and quadrupole models with $\epsilon$ while Figure \ref{fig:varyfig}
shows the variations with other parameters. The variation with $\alpha$
is not shown as its role is well understood: $\alpha$ works to shift the
BAO feature around equally in both the monopole and quadrupole. Figure
\ref{fig:snlfig} shows the variations with $\Sigma_\perp$ and
$\Sigma_\parallel$. Figure \ref{fig:ssfig} shows the variations with
$\Sigma_s$ and Figure \ref{fig:betafig} shows the variations with
$\beta$. In these plots, the solid black line always corresponds to the
fiducial model parameters $\Sigma_\perp = 6\hMpc$, $\Sigma_\parallel =
10\hMpc$ and $\Sigma_s = 4\hMpc$. $\beta$ is set to the center of the
prior, 0.35, unless indicated otherwise. These fiducial parameters will
be discussed in more detail in \S\ref{sec:fitting}.

We see that the monopole model is only weakly affected by varying these
parameters. $\snl$ has an immediately obvious effect on the BAO peak
since it is designed to damp the BAO to model non-linear evolution. The
change in the peak is only significant with a large change in $\snl$. The
$\Sigma_\perp = \Sigma_\parallel = 8\hMpc$ case and the fiducial case
have very similar $\snl$ values. Hence, we see little difference between
the monopole models in these two cases. However, the $\Sigma_\perp =
4\hMpc$, $\Sigma_\parallel = 7\hMpc$ case corresponds to $\snl\sim6\hMpc$
which affords a weaker smearing of the peak. Large $\Sigma_s$ values
can also weakly damp the monopole BAO, causing slight modifications
to its shape. $\beta$ appears to have little effect on the monople and
$\epsilon$ has no effect. The only parameter that can shift the monopole
BAO position is $\alpha$, which also shifts the quadrupole BAO equally.

Looking at the quadrupole model in the fiducial case, we see that the
BAO (at $r\sim110\hMpc$) looks different from the case presented in
Figure \ref{fig:ep_linfig}. There we saw a single bump at the acoustic
scale due to the Kaiser anisotropy $(1+\beta\mu^2)^2$. Including FoG
and anisotropic $\snl$ introduces more structure at the BAO scale. We
see that in the fiducial model, a dip has appeared at the center of the
linear-theory peak, creating a crest-trough-crest structure. This is
the result of anisotropic $\snl$, $\Sigma_s$ and $\beta$ differentially
broadening the BAO peak in the radial and line-of-sight directions. We
see that the radial BAO is wider than the transverse BAO, but the radial
BAO peak has more contrast. Subtracting the two therefore yields the
observed crest-trough-crest shape of the quadrupole near the BAO. 

Figure \ref{fig:snlfig} shows that changing $\Sigma_\perp$ and
$\Sigma_\parallel$ gives rise to crests and troughs near the BAO
scale in the quadrupole. Taking $\snl$ to be isotropic and non-zero
completely removes the trough, leaving a single peak that is broader
than the linear-theory case. Changing the values of $\Sigma_\perp$
and $\Sigma_\parallel$ (but keeping their ratios roughly the same)
alters the structure of the peaks and changes the crest-trough contrast.

Changing $\Sigma_s$ has the most prominent effects at small scales
as expected, since the FoG is strongest there. It also causes a
noticeable change in the quadrupole BAO signal which is partially
degenerate with the effects of anisotropic $\snl$, i.e. it can also
adjust the crest-trough contrast and can eliminate the trough entirely
($\Sigma_s=0\hMpc$). Leverage on this parameter mostly comes from small
scales, where $\Sigma_s$ has a significant effect on the quadrupole shape.

Changing $\beta$ shifts the overall magnitude of the quadrupole at large
scales. Hence, we see that varying $\Sigma_\perp$, $\Sigma_\parallel$,
$\Sigma_s$ and $\beta$ only affects the shape of the quadrupole, not
the BAO position.

Only $\epsilon$ and $\alpha$ can shift the location of the BAO. While
$\alpha$ shifts the BAO position equally along all directions,
Figure \ref{fig:epfig} shows that $\epsilon$ shifts the radial BAO
position more than and in the opposite direction to the transverse BAO
position. Therefore changing $\epsilon$ will cause the quadrupole BAO
position to change in addition to the shift induced by $\alpha$. We
see from \ref{fig:epfig} that $\epsilon$ can also adjust the BAO shape
so it is not completely non-degenerate with the other parameters. We
emphasize that $\Sigma_\perp$, $\Sigma_\parallel$, $\Sigma_s$ and $\beta$
cannot shift the quadrupole BAO, they merely work to change the BAO
shape. $\epsilon$ is the only parameter that can change the quadrupole
BAO position without changing the monopole BAO position, so its effects
should be detectable.

\begin{figure*}
\vspace{0.4cm}
\centering
\begin{tabular}{cc}
\hspace{0.07cm}
\epsfig{file=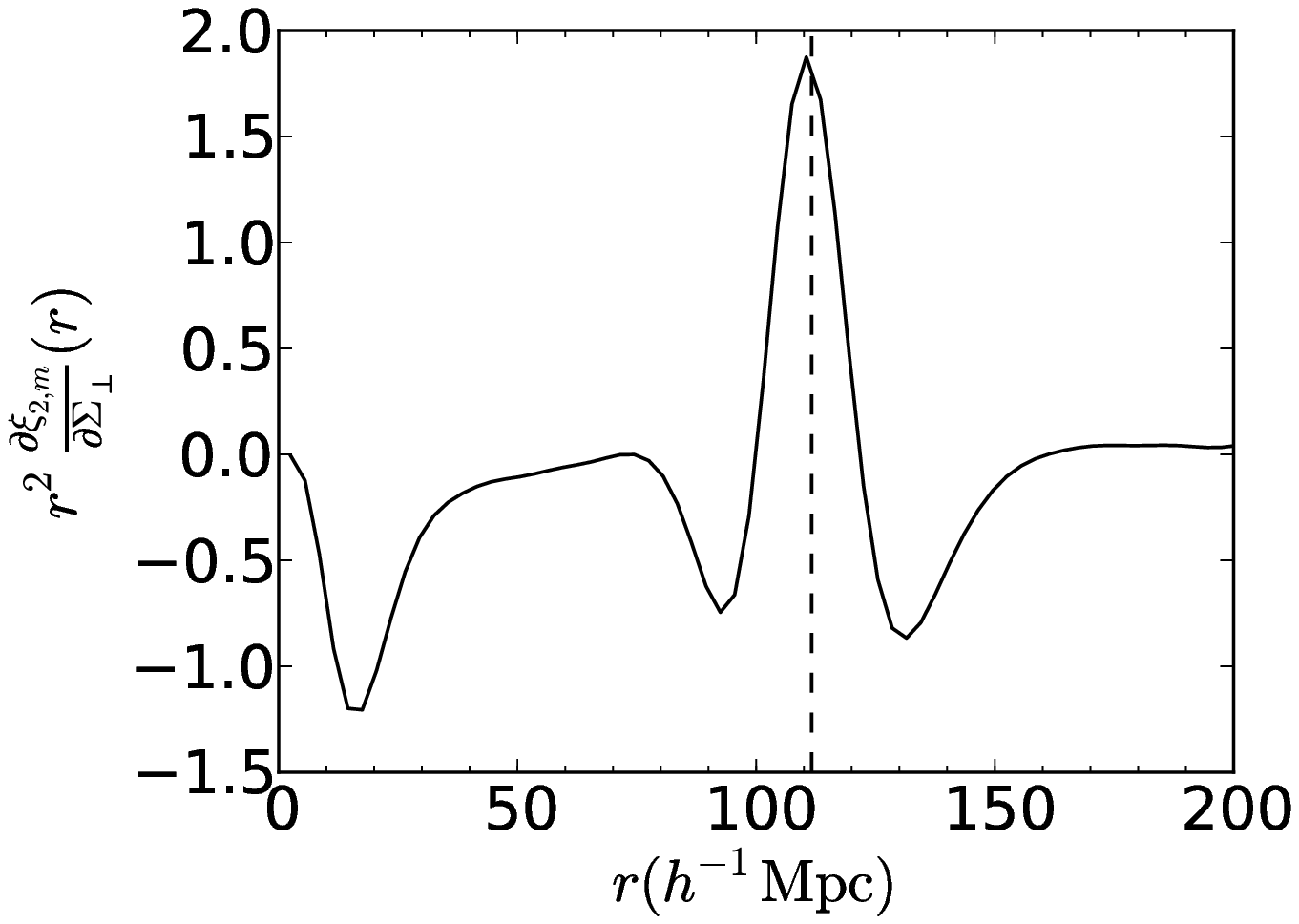, width=0.45\linewidth, clip=}&
\hspace{0.45cm}
\epsfig{file=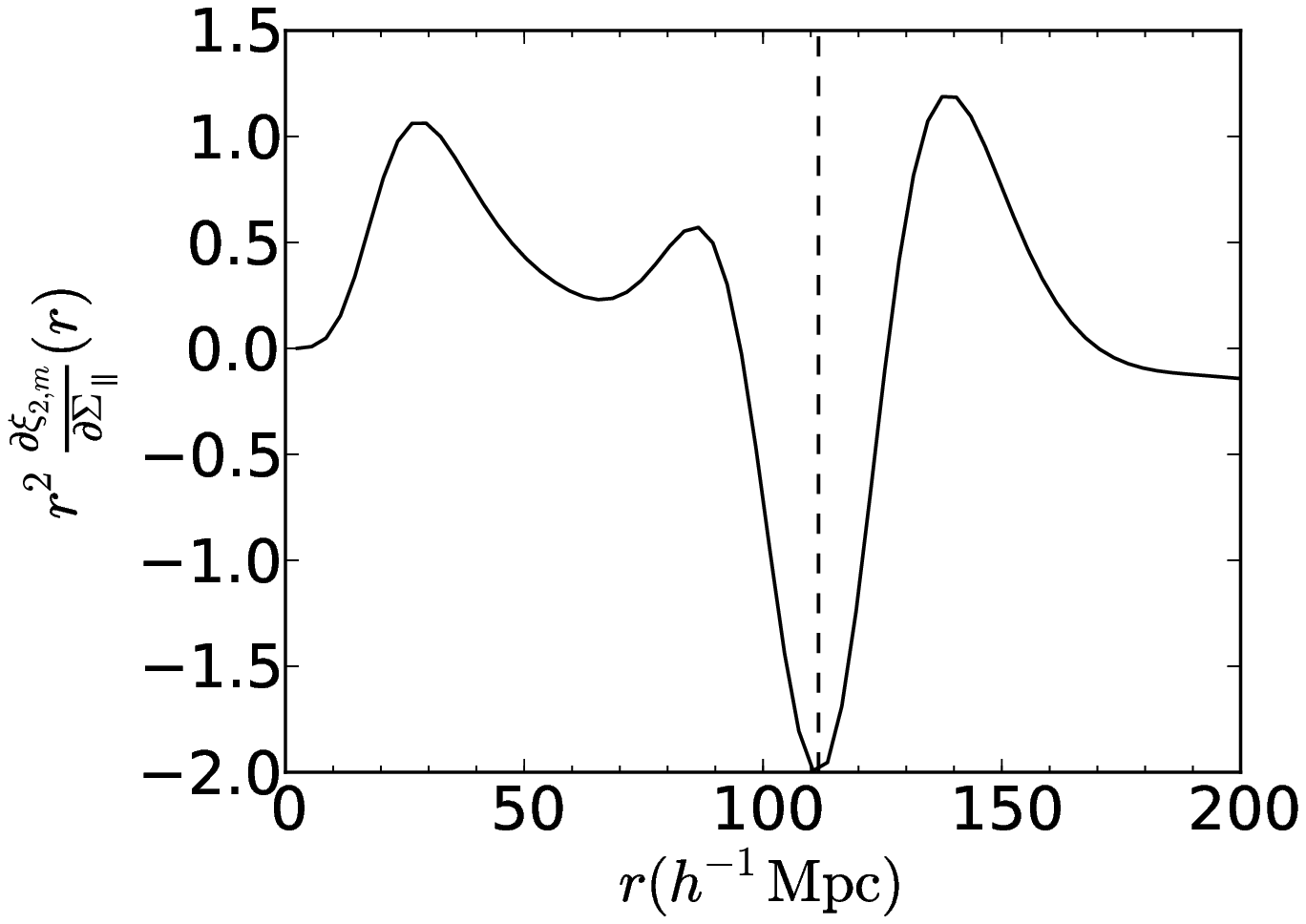, width=0.45\linewidth, clip=}\\
\hspace{0.438cm}
\epsfig{file=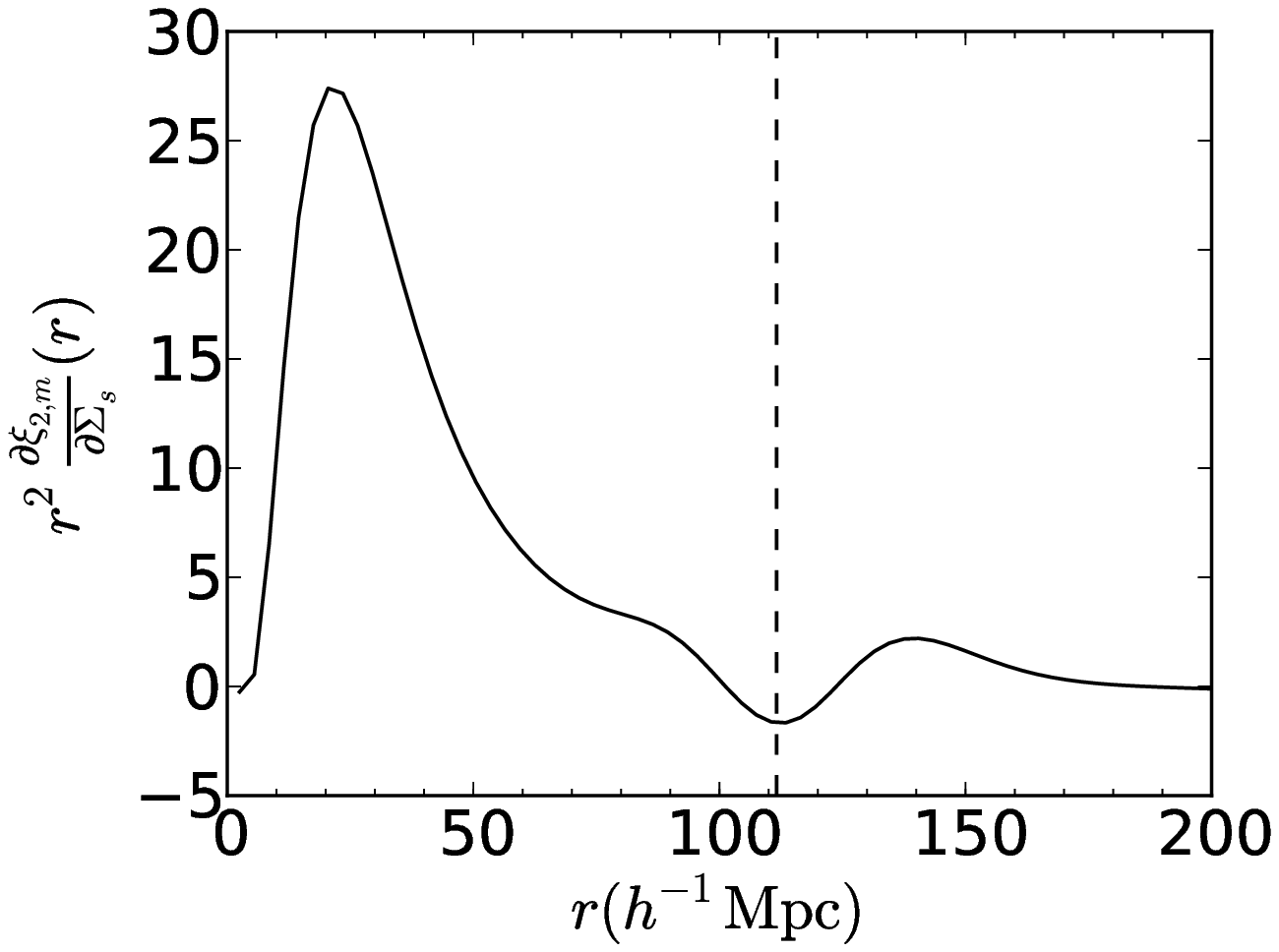, width=0.43\linewidth, height=0.31\linewidth, clip=}&
\hspace{0.435cm}
\epsfig{file=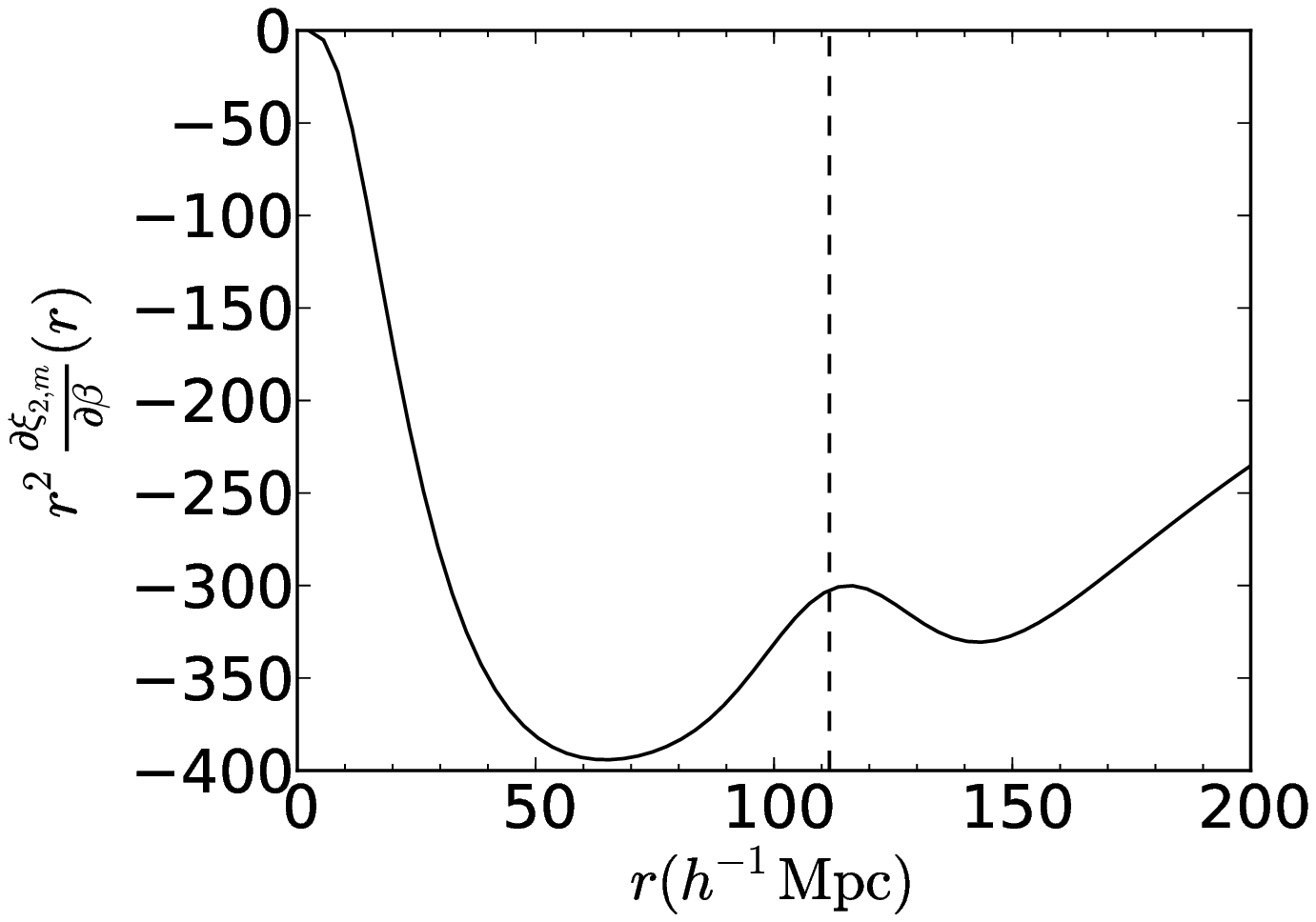, width=0.45\linewidth, clip=}\\
\epsfig{file=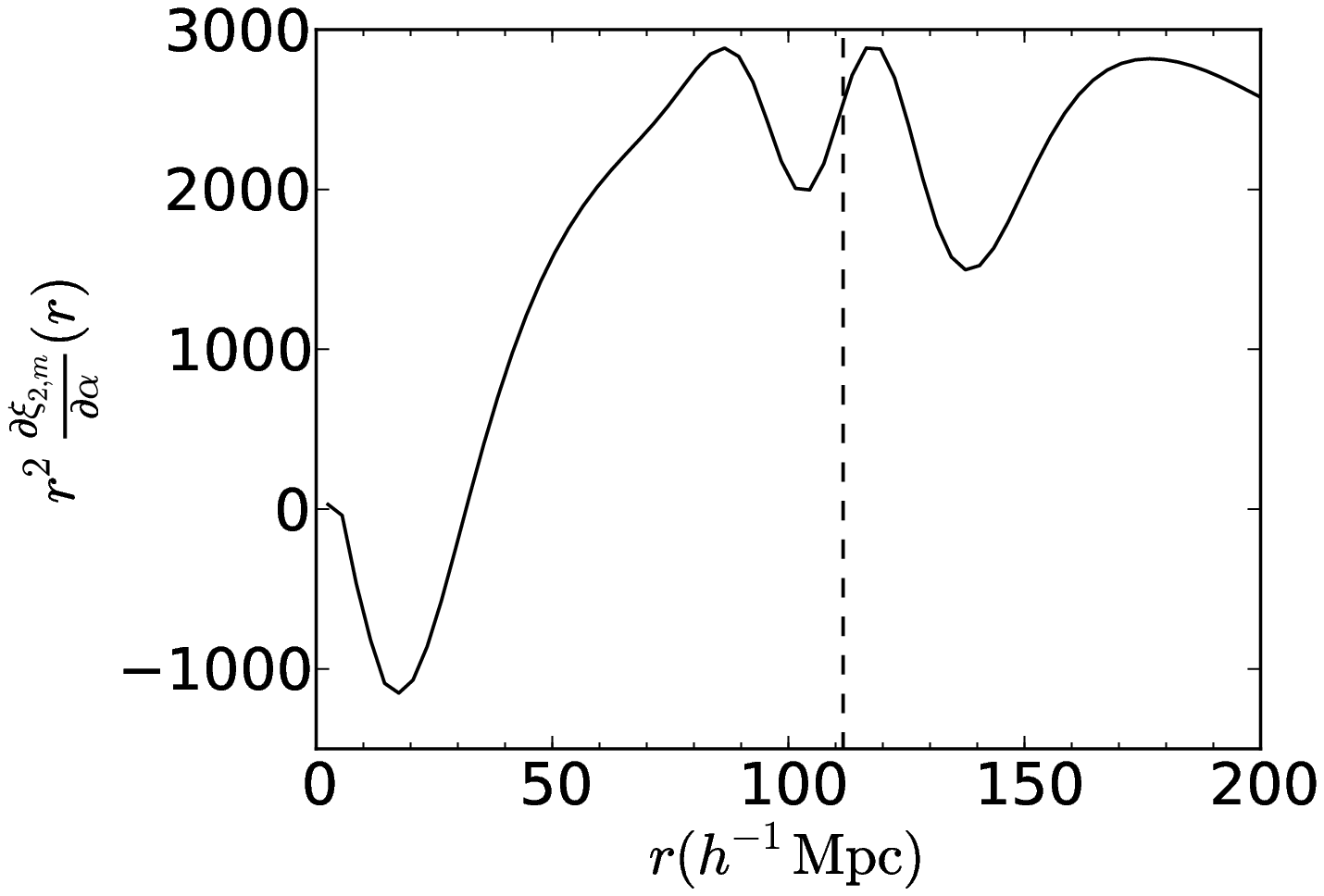, width=0.465\linewidth, clip=}&
\hspace{0.442cm}
\epsfig{file=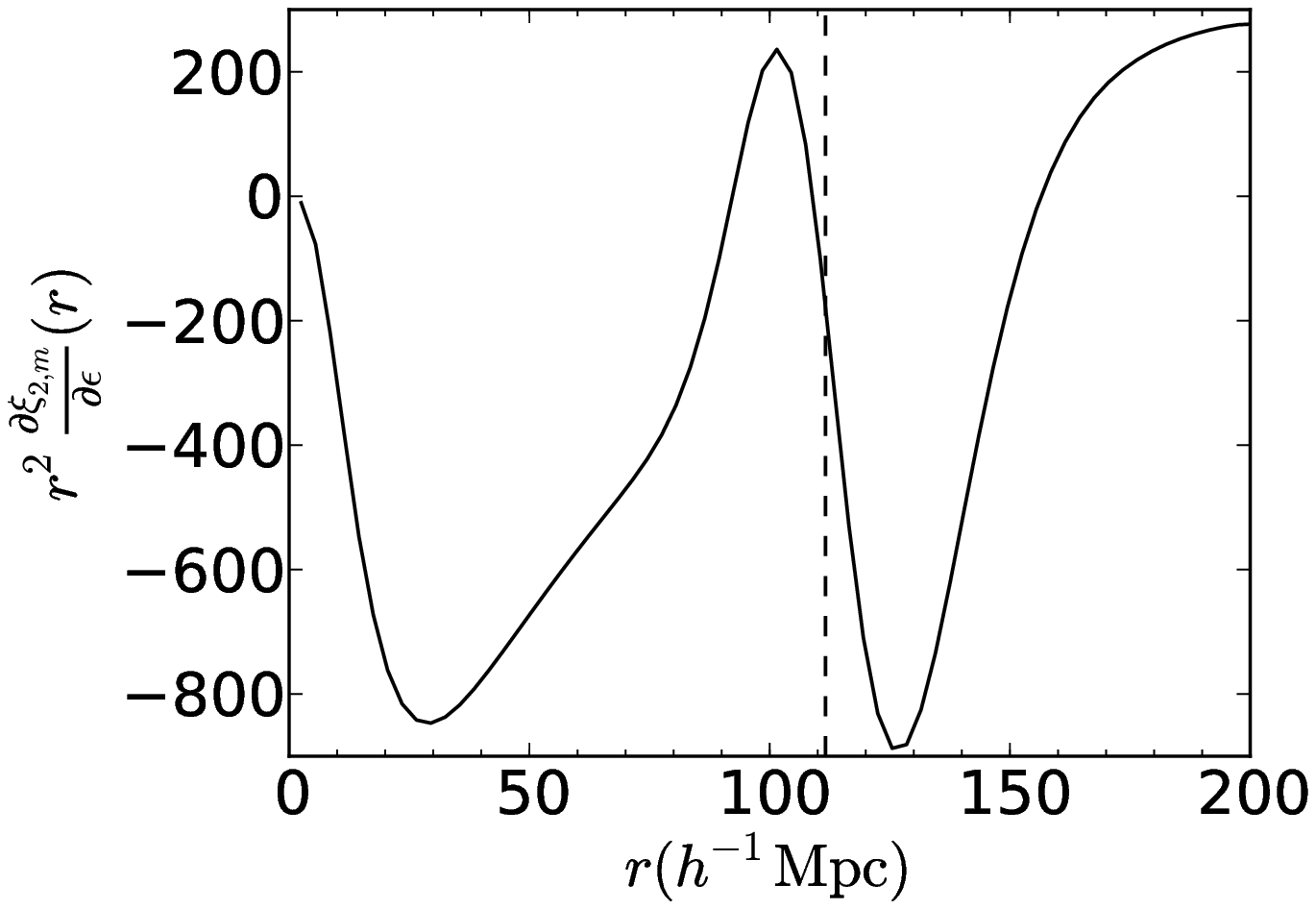, width=0.45\linewidth, clip=}
\end{tabular}
\caption{Derivatives of the model quadrupole with respect to
$\Sigma_\perp$ (top left), $\Sigma_\parallel$ (top right), $\Sigma_s$
(middle left), $\beta$ (middle right), $\alpha$ (bottom left) and
$\epsilon$ (bottom right). The plotted derivatives illustrate how
the model changes with these various parameters and is especially
interesting near the BAO scale marked by the dashed line. Note that near
the acoustic scale, the $\Sigma_\perp$, $\Sigma_\parallel$ and $\Sigma_s$
cases look like derivatives of a Gaussian with respect to its width. The
$\beta$ case looks like the derivative of a Gaussian with respect to its
height. The $\epsilon$ case looks like the derivative of a Gaussian with
respect to its center. These behaviours are all different. We see that the
$\Sigma_\perp$ and $\Sigma_\parallel$ derivatives are similar in nature
at the acoustic scale but opposite in sign. The $\Sigma_\parallel$ and
$\Sigma_s$ derivatives, however, are of the same sign and show the same
up-down-up structure near the BAO scale, but differ at small scales. The
$\beta$ derivative only shows a single peak near the acoustic scale. We
also see that the $\Sigma_\perp$, $\Sigma_\parallel$, $\Sigma_s$ and
$\beta$ derivatives are symmetric about the acoustic scale while the
$\alpha$ and $\epsilon$ derivatives are anti-symmetric. The $\alpha$
derivative has the most structure near the acoustic scale. The $\epsilon$
derivative shows a simple up-down structure. Despite their opposite
symmetries near the acoustic scale, the various crests and troughs of
the $\alpha$ and $\epsilon$ derivatives will be partially degenerate
with the other parameters. However, given reasonable priors on the other
parameters, the model will not be allowed to explore these degeneracies
and we will recover robust measurements of $\epsilon$.
\label{fig:derfig}}
\end{figure*}

The above observations of the model quadrupole behaviour can be
re-cast into Figure \ref{fig:derfig}. The panels of this figure show
the derivatives of the quadrupole model with respect to each parameter
($\Sigma_\perp$ - top left, $\Sigma_\parallel$ - top right, $\Sigma_s$ -
middle left, $\beta$ - middle right, $\alpha$ - bottom left and $\epsilon$
- bottom right). The dashed line marks the acoustic scale. The plotted
derivatives show the variation of the model with these parameters and
are especially interesting near the acoustic scale.

One can see that the behaviour of the $\Sigma_\perp$ derivative near
the BAO scale is exactly opposite to the $\Sigma_\parallel$ case. In
the former we see a down-up-down structure and in the latter we see an
up-down-up structure. The $\Sigma_s$ derivative shows similar behaviour
to the $\Sigma_\parallel$ derivative near the acoustic scale. However,
its small-scale behaviour is much different. The $\beta$ derivative
shows only a single peak near the acoustic scale. Note that all of these
derivatives are symmetric about the acoustic scale.

The $\alpha$ and $\epsilon$ derivatives are different from the others
in that they are both anti-symmetric about the acoustic scale. This
reflects their ability to shift the BAO feature. The difference between
$\alpha$ and $\epsilon$ lies mainly in the monopole where $\alpha$
can significantly shift the BAO but $\epsilon$ cannot; $\epsilon$ only
shifts the BAO in the quadrupole. Near the acoustic scale, the quadrupole
$\alpha$ derivative shows a large amount of structure while the $\epsilon$
derivative shows a simple up-down structure. The crests and troughs of
both of these will be partially degenerate with the other parameters
despite their opposite symmetries near the acoustic scale. However,
if we place well-informed priors on the other parameters, we limit the
model from exploring these degeneracies thereby obtaining reasonably
robust measurements of $\epsilon$.

A similar conclusion can be reached by noting that the first three cases
look like the derivative of a Gaussian with respect to its width. The
$\beta$ case looks like the derivative of a Gaussian with respect to its
height and the $\epsilon$ case looks like the derivative of a Gaussian
with respect to its center. All of these behaviours are different.

\subsection{Covariance Matrix Formalism} \label{sec:covform}

The simplest analytic form for the $\xi(r)$ covariance matrix assumes
that primordial overdensities are drawn from a Gaussian random field in
which all Fourier modes grow independently. In this limit, the covariance
between two multipole moments $\ell$ and $\ell'$ is
\begin{eqnarray}
C_{ij}(\xi_\ell(r_i),\xi_{\ell'}(r_j)) &=& \frac{2(2\ell+1)(2\ell'+1)}{V} 
\nonumber \\
&& \cdot
\int \frac{k^3 d\log(k)}{2\pi^2} j_\ell(kr_i)j_{\ell'}(kr_j) P^2_{\ell\ell'}(k)
\nonumber \\
\label{eqn:gaussc}
\end{eqnarray}
where $V$ is the survey volume, the $j_\ell(kr)$ are the spherical Bessel
functions of order $\ell$ and $P^2_{\ell\ell'}(k)$ is defined as
\begin{equation}
P^2_{\ell\ell'}(k) = \frac{1}{2} \int^{1}_{-1} 
\left[ P(k,\mu) + \frac{1}{\bar{n}} \right]^2 L_\ell(\mu) L_{\ell'}(\mu) d\mu.
\label{eqn:pell}
\end{equation}
Here, $P(k,\mu)$ is the 2D power spectrum and $\bar{n}$ is the mean
number density of galaxies. The $1/\bar{n}$ term corresponds to Poisson
shot-noise. However, it is well-known that non-linear structure growth
introduces coupling between the various modes \citep{MWP99, SE05, JK06,
ESW07, Hea07, GBS07, Ma07, Aea08, CS08, SBA08, Sea08, Smith08, PdW09,
Tea09}. These effects can be largely included by using non-linear
models for the 2D power spectrum and shot-noise terms (e.g. as shown
in \citealt{Xea12}). In the above, we have also ignored the redshift
dependence of $\bar{n}$ above, however, we will describe a method that
allows its inclusion in \S\ref{sec:fitting}.

In practice, the correlation functions we measure are binned and hence, we
must also account for this binning in the Gaussian covariance matrix. If
we calculate the correlation function in bins with lower bounds $r_1$
and upper bounds $r_2$, the binned covariance matrix is
\begin{eqnarray}
C_{ij}(\xi_\ell(r_i),\xi_{\ell'}(r_j)) &=& \frac{2(2\ell+1)(2\ell'+1)}{V}
\frac{3}{r_{i2}^3 - r_{i1}^3} \frac{3}{r_{j2}^3 - r_{j1}^3} \nonumber \\
&& \cdot \int_{r_{i1}}^{r_{i2}} r^2 dr \frac{d\Omega}{4\pi} 
\int_{r_{j1}}^{r_{j2}} r'^2 dr' \frac{d\Omega'}{4\pi} \nonumber \\
&& \cdot \int \frac{k^3 d\log(k)}{2\pi^2}j_\ell(kr)j_{\ell'}(kr')
P^2_{\ell\ell'}(k). \nonumber \\
\end{eqnarray}
In the case of the monopole and quadrupole relevant to this study, the
above expression can be split into the $\ell=\ell'=0$ case, the $\ell=0$
and $\ell'=2$ (or vice versa) case and the $\ell=\ell'=2$. The first
case was shown to have the form \citep{Xea12}
\begin{equation}
C_{ij}(\xi_0(r_i),\xi_0(r_j)) = \frac{2}{V} 
\int \frac{k^3 d\log(k)}{2\pi^2} \mathcal J_0(kr_i) \mathcal J_0(kr_j) 
P^2_{00}(k)
\label{eqn:mmcov}
\end{equation}
where 
\begin{equation}
\mathcal J_0(kr) = \frac{3}{r_2^3-r_1^3}
\bigg[ \frac{r^2j_1(kr)}{k} \bigg]^{r_2}_{r_1}
\end{equation}
and 
\begin{equation}
j_1(kr) = \frac{\sin(kr)}{(kr)^2} - \frac{\cos(kr)}{kr}.
\end{equation}
Here, $[f(x)]^a_b$ is standard notation for $f(a) - f(b)$ for any function
$f$. The second case has the form
\begin{equation}
C_{ij}(\xi_0(r_i),\xi_2(r_j)) =
\frac{10}{V} \int \frac{k^3 d\log(k)}{2\pi^2} \mathcal J_0(kr_i)
\mathcal J_2(kr_j) P^2_{02}(k)
\label{eqn:mqcov}
\end{equation}
where
\begin{equation}
\mathcal J_2(kr) = \frac{3}{r_2^3 - r_1^3} \bigg[
\frac{3\rm{si}(kr)}{k^3} - \frac{1}{k}
\bigg( \frac{3r}{k} j_0(kr) + r^2j_1(kr) \bigg) \bigg]^{r_2}_{r_1},
\end{equation}
\begin{equation}
\rm{si}(x) = \int_{0}^x \frac{\sin(x')}{x'}dx'
\end{equation}
and
\begin{equation}
j_0(kr) = \frac{\sin(kr)}{kr}.
\end{equation}
Note that due to symmetry, we have $C_{ij}(\xi_0(r_i),\xi_2(r_j)) =
C_{ij}(\xi_2(r_i),\xi_0(r_j))$. Finally, the last case has the form
\begin{equation}
C_{ij}(\xi_2(r_i),\xi_2(r_j)) =
\frac{50}{V} \int \frac{k^3 d\log(k)}{2\pi^2} \mathcal J_2(kr_i)
\mathcal J_2(kr_j) P^2_{22}(k).
\label{eqn:qqcov}
\end{equation}

Collectively, we can then write
\begin{eqnarray}
C_{ij}(\xi_\ell(r_i),\xi_{\ell'}(r_j)) &=&
\frac{2(2\ell+1)(2\ell'+1)}{V} \nonumber \\
&& \cdot \int \frac{k^3 d\log(k)}{2\pi^2} 
\mathcal J_\ell(kr_i) \mathcal J_{\ell'}(kr_j) P^2_{\ell\ell'}(k). \nonumber \\
\label{eqn:bincov}
\end{eqnarray}

These forms for the binned Gaussian covariance matrix form the basis
for deriving a covariance matrix for our data. We extend the method for
approximating the mock covariances using a modified form of the binned
Gaussian covariance matrix as described in \citet{Xea12}. This will be
outlined in \S\ref{sec:fitting}.

\section{Analysis} \label{sec:analysis}

\subsection{Reconstruction} \label{sec:recon}

Non-linear structure growth degrades and shifts the acoustic
peak. Reconstruction was initially proposed to partially remove these
effects \citep{Eea07}. In the linear theory description of structure
growth, overdensities are small and hence remain largely in place as they
accrete more matter and grow. However, at low redshifts, this description
becomes increasingly less suitable as some overdensities grow to masses
large enough that they begin exerting significant gravitational pulls
on each other. This gives rise to pairwise relative velocities between
particles separated by $\sim100\hMpc$. These coherent flows that form
large-scale structure are the dominant source of smearing of the BAO
signal. The peculiar motions of particles within a gravitationally bound
structure are subdominant. Reconstruction attempts to undo these coherent
motions in the matter density field and arrive back at something that
more closely resembles linear theory. This translates into a sharpening
up of the acoustic peak in the correlation function which allows us
to gain a better centroiding of its location and hence a more precise
measure of the acoustic scale. It also removes some of the shifting of
the BAO due to non-linear structure growth, which has been shown to be
a $\sim0.5\%$ effect at $z=0$ \citep{Sea10,Mea11}.

Our reconstruction algorithm is described in detail in \citet{Pea12}. A
more pedagogical discussion of reconstruction can also be found
there. This method is extended from the original reconstruction algorithm
proposed in \citet{Eea07} and its theoretical basis is established in
\citet{PWC09} and \citet{Nea09}. A basic outline is given below.

The ultimate goal of reconstruction is to infer the matter displacement
field that arises due to non-linear structure growth from the observed
galaxy density field. Then, we can shift the galaxies back along their
inferred displacement vectors to place them where they would have been
in linear theory. This is simple if we consider only the first order
displacements $\mathbf{\Psi}$. In this case
\begin{equation}
\nabla \cdot \mathbf{\Psi} + \beta \nabla \cdot (\Psi_s \mathbf{\hat{s}}) = 
-\frac{\delta_{\rm gal}}{b}
\label{eqn:rec}
\end{equation}
where $\Psi_s = \mathbf{\Psi} \cdot \mathbf{\hat{s}}$ is the displacement
in the line-of-sight direction \citep{ND94}, $\delta_{\rm gal}$ is
the galaxy density field, $b$ is the large-scale galaxy bias (which is
roughly constant) and hence $\delta_{\rm gal}/b$ is an approximation
of the matter density field. The second term in this equation arises
due to large-scale redshift-space distortions caused by the coherent
infall of galaxies towards overdense regions \citep{Kaiser87}. This
implies that as an additional bonus, reconstruction can also correct
for linear redshift-space distortions. The $\beta$ parameter governs the
amount of anisotropy introduced by the Kaiser effect. It is defined as
$\beta = f/b$ where $f = d\log D(a)/d\log(a) \sim \Omega_m(a)^{0.55}$
is the linear growth rate, $D(a)$ is the linear growth function and $a$
is the scale factor \citep{Cea92, L05}. If we assume that $\mathbf{\Psi}$
is irrotational (i.e. curl-free), we can write $\mathbf{\Psi} = \nabla
\phi$. After selecting appropriate values of $b$ and $\beta$, we can
solve for the scalar field $\phi$ and then the displacement field
$\mathbf{\Psi}$ using a finite difference approach.

Areas that do not fall within the survey or are masked out by the survey
need to be accounted for as the gravitational potential (and hence the
displacement field) is sensitive to these regions. We embed the survey
into a larger Gaussian realization \citep{HR91, Zea95} constrained
to match the density field where observed. The exact implementation
is described in detail in \citet{Pea12}, and we refer the reader to
the description there for more details. When doing this embedding,
one has a choice to either set unconstrained Fourier modes to zero
(Wiener filtering) or sample them from an assumed power spectrum. We
explicitly show that our distance constraints do not depend on this
choice in \S\ref{sec:mockres}.

\subsection{Computation} \label{sec:comp}

The computation of our correlation functions is tied to our reconstruction
algorithm. We bin our correlation functions in $3\hMpc$ bins starting at
$2.5\hMpc$ and going up to $197.5\hMpc$. A list of the steps involved
is given below.

\newcounter{lcounter}
\begin{list}{\hspace{0.5cm}\Roman{lcounter})~}{\usecounter{lcounter}}

\item Obtain a set of randomly distributed points that have the same
angular and radial selection function as the survey.

\item Compute the unreconstructed correlation function from the data
using the Landy-Szalay estimator \citep{LS93},
\begin{equation}
\xi(r,\mu) = \frac{DD(r,\mu)-2DR(r,\mu)+RR(r,\mu)}{RR(r,\mu)}
\end{equation}
where $DD$, $DR$ and $RR$ are the number of galaxy-galaxy, galaxy-random
and random-random pairs that are separated by $r$ and $\mu$. We apply
FKP weighting \citep{FKP94} for each object as
\begin{equation}
w_i = \frac{1}{1+\bar{n}(z_i)P(k_0)}
\end{equation}
where $\bar{n}(z_i)$ is the number density at the redshift of the object
$z_i$ and $P(k_0)=40000 h^{-3}\;\rm{Mpc}^3$ is the approximate value of
the power spectrum at the BAO scale. 

\item Estimate the galaxy bias $b$ and the anisotropy parameter $\beta$
from the unreconstructed correlation function. We use fiducial values
of $b=2.2$ and $\beta=0.3$.

\item Embed the survey in a larger volume and smooth the density field
using a Gaussian (again, we use a smoothing length of $15\hMpc$). Generate
a constrained Gaussian realization matching the observed density to fill
in the masked and unobserved regions.

\item Estimate the displacement field $\mathbf{\Psi}$ using Equation
(\ref{eqn:rec}) and shift the galaxies by $-\mathbf{\Psi}-f(\Psi_s
\mathbf{\hat{s}})$ to partially undo the effects of non-linear structure
growth and large-scale redshift-space distortions. This is the essence
of reconstruction.

\item Obtain another set of randomly generated particles with the same
radial and angular selection function as the survey. Shift these by
$-\mathbf{\Psi}$ and denote as $S$.

\item Compute the reconstructed correlation function using the
Landy-Szalay estimator
\begin{equation}
\xi(r, \mu) = \frac{DD(r,\mu) - 2DS(r,\mu) + SS(r,\mu)}{RR(r,\mu)}.
\end{equation}

\end{list}

\subsection{Fitting} \label{sec:fitting}

We construct models of the monopole and quadrupole in a fiducial
cosmology to measure the position of the BAO in the data relative to
the model (parameterized by $\alpha$) and the degree to which it is
anisotropic in the data (parameterized by $\epsilon$). We base our fitting
templates for the monopole and quadrupole on Equations (\ref{eqn:mono}) \&
(\ref{eqn:quad}). That is, we define fitting models of the form
\begin{eqnarray}
\xi_{0,m}(r) &=& B_0^2 \xi_{0,t}(\alpha r) + \frac{2}{5} \epsilon
\bigg[ 3\xi_{2,t}(\alpha r) + \frac{d\xi_{2,t}(\alpha r)}{d\log(r)} \bigg] 
+ A_0(r) \nonumber \\ 
\label{eqn:monot}\\
\xi_{2,m}(r) &=& 2 B_0^2 \epsilon \frac{d\xi_{0,t}(\alpha r)}{d\log(r)} 
+ \bigg( 1 + \frac{6}{7}\epsilon \bigg) \xi_{2,t}(\alpha r) 
+ \frac{4}{7} \epsilon \frac{d\xi_{2,t}(\alpha r)}{d\log(r)} \nonumber \\
&& + \frac{4}{7} \epsilon \bigg[5 \xi_{4,t}(\alpha r) + 
\frac{d\xi_{4,t}(\alpha r)}{d\log(r)} \bigg] + A_2(r)
\label{eqn:quadt}
\end{eqnarray}
where
\begin{equation}
A_\ell(r) = \frac{a_{\ell,1}}{r^2} + \frac{a_{\ell,2}}{r} + a_{\ell,3}.
\label{eqn:fida}
\end{equation}
The $A_\ell(r)$ are composed of linear nuisance terms used to marginalize
out broadband effects such as scale-dependent bias and redshift-space
distortions as in \citet{Xea12}. The $B_0^2$ term adjusts the amplitude
of the monopole template $\xi_{0,t}$. We infer the galaxy bias $b$
from the multiplicative offset, $b^2$, between this template and the
measured correlation function at $r=50\hMpc$. We then use this offset to
normalize the full models, $\xi_{0,m}$ and $\xi_{2,m}$ in Equations
(\ref{eqn:monot}) \& (\ref{eqn:quadt}), to the data. This ensures that
$B_0^2\sim1$ as it is primarily the monopole fit that sets this term
and $\epsilon$ is very small. In practice we perform our fits in the
non-linear parameter $\log(B_0^2)$ to prevent $B_0^2$ from going negative
which would be unphysical. We adopt a Gaussian prior on $\log(B_0^2)$
with standard deviation 0.4 and centered at 0 to prevent $B_0^2$ from
wandering too far from 1 as described in \citet{Xea12}. 

In addition, we apply a 10\% Gaussian prior on $1+\epsilon$ to limit noise
from dragging $\epsilon$ to unrealistically large values. In fitting the
mocks without the prior, only 2 have measured $\epsilon$ values $>0.1$, so
this prior does not have a significant impact. To verify that this prior
is not cosmologically important, we use the CMB+allBAO $ow_0w_a$CDM Markov
Chain Monte Carlo results from \citet{Mehta12} to estimate the allowed
distribution of $\epsilon$. The distribution is nearly Gaussian with a
standard deviation of $\sim0.026$ which is much less than our 0.1 prior.

The monopole and quadrupole correlation function templates ($\xi_{0,t}(r)$
and $\xi_{2,t}(r)$) are derived from the 2D power spectrum $P_t(k,\mu)$
template (Equation \ref{eqn:tdp}) as described in \S\ref{sec:nonlin}. We
set $\Sigma_s=4\hMpc$. We let $\beta$ vary in our fits as it affords
us leverage on the amplitude of the quadrupole with which it is
partially degenerate. We put a prior on $\beta$ centered at $f/b
\sim \Omega_m(z)^{0.55}/b = 0.35$ before reconstruction and 0 after
reconstruction with 0.2 standard deviation in both cases. The choice of
$\beta=0$ as the center of the prior after reconstruction is because we
expect the Kaiser effect to be mostly removed. We fix $\Sigma_\perp =
6\hMpc$ and $\Sigma_\parallel = 10\hMpc$ in our pre-reconstruction fits
and $\Sigma_\perp=\Sigma_\parallel=3\hMpc$ in our post-reconstruction
fits. These values are approximated from the fit results to the average
correlation function of the mocks where we set $\Sigma_\parallel =
(1+f)\Sigma_\perp$ in the pre-reconstruction case due to the Kaiser effect
and $\Sigma_\parallel=\Sigma_\perp$ in the post-reconstruction case
due to the expected removal of Kaiser squashing by reconstruction. 

We simultaneously fit the monopole and the quadrupole for 4 non-linear
parameters $\log(B_0^2)$, $\beta$, $\alpha$ and $\epsilon$, in addition to
the linear nuisance parameters in $A_\ell(r)$. The non-linear parameters
are handled using a simplex algorithm and the linear parameters are
obtained using a least-squares method nested within this simplex. That
is, for each set of non-linear parameters, the least-squares algorithm
returns the corresponding best-fit linear parameters. The simplex steps
through the non-linear parameter space until the best-fit values are
obtained. To determine the best-fit values, we minimize the $\chi^2$
goodness-of-fit indicator given by
\begin{equation}
\chi^2 = (\vec{m} - \vec{d})^T C^{-1} (\vec{m}-\vec{d})
\end{equation}
where $\vec{m}$ is a column vector of the model at each step in the
simplex and $\vec{d}$ is the data. Both of these must contain both the
monopole and quadrupole values in sequence. $C$ is the covariance matrix
described below. We use a fiducial fitting range of $50<r<200\hMpc$
which corresponds to fitting 50 points in both the monopole and the
quadrupole. This gives $2\times50 - \rm{\#\;of\;fit\; parameters} = 100-10
= 90$ degrees-of-freedom (dof) in the fit. Using this technique we obtain
best-fit values of our parameters of interest, the isotropic dilation
$\alpha$ and the anisotropic warping $\epsilon$ of the BAO signal.

In addition, we can calculate the probability distribution
$p(\alpha,\epsilon)$ by fitting for the other parameters at various grid
values of these two parameters and measuring the best-fit $\chi^2$. This
is feasible because $p(\vec{x}) \propto \exp(-\chi^2/2)$. The constant of
proportionality corresponds to the normalization that makes the integral
$\int p(\vec{x}) d\vec{x} = 1$. Then we can calculate $p(\alpha)$ and
$p(\epsilon)$ as
\begin{eqnarray}
p(\alpha) &=& \int p(\alpha,\epsilon) d\epsilon \label{eqn:padef}\\
p(\epsilon) &=& \int p(\alpha,\epsilon) d\alpha.
\end{eqnarray}
We can take the widths of these distributions ($\sigma_\alpha$ and
$\sigma_\epsilon$) as measurements of the errors on $\alpha$ and
$\epsilon$ if $\alpha$ and $\epsilon$ have Gaussian posteriors. In
\S\ref{sec:mockres} we demonstrate using fit results to our mock
catalogues that $\alpha$ and $\epsilon$ are consistent with having been
drawn from Gaussian distributions. The covariance between $\alpha$ and
$\epsilon$ ($C_{\alpha\epsilon}$) can also be calculated and converted
into a correlation coefficient $\rho_{\alpha\epsilon}$. These are defined
\begin{eqnarray} 
\sigma^2_x &=& \int p(x) (x-\langle x \rangle)^2 dx \label{eqn:sdeq}\\ 
C_{\alpha\epsilon} 
&=& \int \int p(\alpha,\epsilon) (\alpha - \langle \alpha \rangle) (\epsilon -
\langle \epsilon \rangle) d\alpha d\epsilon\\ \rho_{\alpha\epsilon}
&=& \frac{C_{\alpha\epsilon}}{\sigma_\alpha \sigma_\epsilon}
\label{eqn:cceq}
\end{eqnarray} 
where $\langle x \rangle$ is the mean of the distribution $p(x)$. For
our grids we pick the ranges $0.7<\alpha<1.3$ and $-0.3<\epsilon<0.3$ at
spacings of 0.0025 and 0.005 respectively. We also apply an additional
Gaussian prior on $\log(\alpha)$ with a width of 0.15 to suppress any
unphysical downturns in the $\chi^2$ distribution at small $\alpha$. These
$\alpha$ correspond to the acoustic peak in the model being pushed out
to larger scales where the fitter has an easier time hiding the peak
in the large errorbars. This procedure was scrutinized in detail in
\citet{Xea12}. Note that our grids are only used to compute $p(\alpha)$
and $p(\epsilon)$. We do not use them to infer the best-fit values of
these parameters, which are obtained through the non-linear simplex
optimization described above.

We obtain a smooth estimate of the covariance matrix using the method
described in \citet{Xea12} extended here to include the quadrupole
(see below). A detailed description of the method is given there. This
method relies on a form of the Gaussian covariance matrix that includes
some additional modification parameters. These allow us to adjust the
amount of shot-noise and sample variance to best match the covariances
calculated directly from the mock correlation functions. To derive values
for these modification parameters, we maximize the likelihood function
\begin{equation}
\mathcal L = \prod^{N}_{i=0} (2\pi)^{-q/2}(\rm{det}C)^{-1/2} exp{(-\chi^2_i/2)}.
\end{equation}
Here $N$ is the total number of mocks and $q$ is the number of points
to fit. We derive the parameters using the mock covariances between
$50<r<200\hMpc$ (50 monopole points and 50 quadrupole points) and
hence $q=50 \times 2=100$. $\chi^2_i = \vec{x}_i^TC^{-1}\vec{x}_i$ where
$\vec{x}_i$ is a column vector of dimension $q$ containing the difference
between the monopole and quadrupole of each mock and the average of
all mocks. $C$ is the modified form of the Gaussian covariance matrix
derived below and defined in Equation (\ref{eqn:modcov}). We use this
method since the mock covariances show evidence of noise and ideally
the covariance matrix should be smooth.

The method allows us to include the redshift dependence of the
galaxy number density $\bar{n}$ assuming that it has no angular
dependence. This is anchored in the observation that the covariance in
configuration space is just the transform of the variance in Fourier
space $P^2_{\ell\ell'}(k)/V$. Hence, we can imagine building up the
inverse variance as
\begin{equation}
I^2(k) = \int \frac{dV}{P^2_{\ell\ell'}(k)}
\end{equation}
where
\begin{equation}
dV = \frac{c}{H_0}\frac{R^2(z)}{\sqrt{\Omega_m(1+z)^3 + \Omega_\Lambda}}
dz d\Omega
\end{equation}
for a flat universe. Here, $R(z)$ is the comoving distance to redshift
$z$. We can then redefine the variance as $\mathfrak{P}^2_{\ell\ell'}(k) =
[I^2(k)]^{-1}$.

Swapping this new expression for the variance into Equation
(\ref{eqn:bincov}) gives
\begin{eqnarray}
C_{ij}(\xi_\ell(r_i),\xi_{\ell'}(r_j)) &=&
2(2\ell+1)(2\ell'+1) \nonumber \\
&& \cdot \int \frac{k^3d\log(k)}{2\pi^2} \mathcal J_\ell(kr_i)
\mathcal J_{\ell'}(kr_j) \mathfrak P^2_{\ell\ell'}(k). \nonumber \\
\end{eqnarray}
We can then insert the modification parameters $c_0$, $c_1$, $c_2$
and $c_3$ such that
\begin{eqnarray}
C_{ij}(\xi_\ell(r_i),\xi_{\ell'}(r_j)) &=&
2(2\ell+1)(2\ell'+1) \nonumber \\
&& \cdot \bigg[ \int \frac{k^3d\log(k)}{2\pi^2} \mathcal J_\ell(kr_i)
\mathcal J_{\ell'}(kr_j) \nonumber \\
&& \cdot \mathfrak P^2_{\ell\ell'}(k;c_0,c_1,c_2) \bigg] + c_3.
\label{eqn:modcov}
\end{eqnarray}
Here, where we have made the substitution
\begin{eqnarray}
P(k,\mu) + \frac{1}{\bar{n}} &\rightarrow&
\bigg[ c_0 P_{dw}(k,\mu) + \frac{c_1}{\bar{n}(z)} \bigg]
(1+\beta\mu^2)^2 F(k,\mu,\Sigma_s) \nonumber \\
&& + \frac{c_2}{\bar{n}(z)}
\end{eqnarray}
in Equation (\ref{eqn:pell}). The $c_0$ term adjusts the magnitude of
the sample variance. The $c_1$ term acts like a non-linear shot-noise
component and the $c_2$ term adjusts the magnitude of the standard
Poisson shot-noise contribution. The $c_3$ term can be associated with
the integral constraint, which appears as an additive offset in the
correlation function.

In calculating the covariance matrix, we set $\beta = f/b$ before
reconstruction and $\beta=0$ after reconstruction, again due to the
expected removal of large-scale redshift-space distortions. We test
several cases where we vary $\beta$ from these fiducial values and
find that changing $\beta$ affects the relative amplitudes of the
$C_{ij}(\xi_0(r_i),\xi_0(r_j))$ and $C_{ij}(\xi_2(r_i),\xi_2(r_j))$
terms. This can cause slight changes in the resulting $\sigma_\alpha$
and $\sigma_\epsilon$ at the 0.1\% level, which is not significant at
our current levels of statistical precision. The $\Sigma_s$ streaming
scale for the FoG is fixed at $4\hMpc$. We find very little difference
in the resulting modification parameters in cases where we allow
$\Sigma_s$ to vary. We fix $\Sigma_\perp$ and $\Sigma_\parallel$ to
their fiducial model values (recapped below) in our covariance matrix
calculations. The modification parameters we obtain are $c_0 = 1.06$,
$c_1=0.11$, $c_2=1.49$, $c_3=5.18 \times 10^{-8}$ before reconstruction,
and $c_0=1.12$, $c_1=0.05$, $c_2=1.58$, $c_3=8.82 \times 10^{-8}$
after reconstruction. With these modification parameters in hand, we
can construct a smooth approximation to the mock covariances from the
binned Gaussian covariance matrix using Equation (\ref{eqn:modcov}).

Our fiducial model parameters are summarized as follows: we define
our fiducial model before reconstruction to have $\Sigma_\perp =
6\hMpc$, $\Sigma_\parallel=10\hMpc$, $\Sigma_s = 4\hMpc$ and a prior
on $\beta$ centered on $\beta=0.35$. After reconstruction, we set
$\Sigma_\perp=\Sigma_\parallel=3\hMpc$ and center our $\beta$ prior on
0. Our fiducial fitting range is $50<r<200\hMpc$.

In \S\ref{sec:mockres} and \S\ref{sec:datares}, we present our measured
values of $\alpha$ and $\epsilon$ for the mocks (\S\ref{sec:sims}) and
actual survey data (\S\ref{sec:data}) using the fitting models defined in
Equations (\ref{eqn:monot}) \& (\ref{eqn:quadt}) and the modified Gaussian
covariance matrix described above. 

\section{Datasets} \label{sec:datasets}

\subsection{Simulations} \label{sec:sims}

We use the Large Suite of Dark Matter Simulations (LasDamas;
\citealt{Mea12}) to calibrate and test our reconstruction parameters,
fitting template and covariance matrix. The LasDamas collaboration
\footnote[1]{http://lss.phy.vanderbilt.edu/lasdamas} has provided
publicly available mock galaxy catalogues based on these simulations
for the Sloan Digital Sky Survey (SDSS) data release 7 (DR7) luminous
red galaxy (LRG) sample. In particular, we use the "gamma" release mock
catalogues referred to as \texttt{lrgFull}.

The LasDamas simulations were run assuming a flat $\Lambda$CDM cosmology
with $\Omega_b = 0.04$, $\Omega_m = 0.25$, $h=0.7$, $n_s=1.0$ and
$\sigma_8=0.8$. Although various box sizes were implemented, the 40
simulations used to construct the LRG mocks were 2.4$\hGpc$ on a side
with $1280^3$ particles in each. The initial particle positions were set
using second-order Lagrangian perturbation theory at $z=49$. To construct
mock galaxy catalogues from the simulations, the dark matter halos were
populated according to halo occupation parameters tuned to match the
observed clustering of the DR7 LRGs. In addition, the mock catalogues
include observational effects such as redshift-space distortions and
mimic the angular selection function of the LRG sample. The redshift
range covered by the mocks is $0.16<z<0.44$. We note that this is
slightly different to the flux-limited LRG sample described in the
following section that will be employed in this study. When fitting the
data, we account for this by extrapolating our covariance matrix using
the observed $\bar{n}(z)$ and the formalism described in the previous
section. In addition, we slightly downsample the $\bar{n}(z)$ distribution
in the mocks to better match the data. Our region of interest, the SDSS
Northern Galactic Cap, covers $\sim7200\;\rm{deg}^2$ on the sky. The
resulting geometry allows 4 mocks to be constructed from each simulation
and hence we have a total of 160 mocks for our analysis.

\subsection{SDSS DR7} \label{sec:data}

The observational dataset used in this study is the SDSS \citep{Yea00}
DR7 \citep{Aea09} LRG sample. The same dataset was also used in
the monopole-only BAO analysis of \citet{Pea12}, \citet{Xea12} \&
\citet{Mehta12}.

The SDSS has taken photometric observations of $\sim10,000\;\rm{deg}^2$ on
the sky and obtained spectroscopic followup of nearly a million of these
detected objects. It uses a dedicated 2.5m telescope \citep{Gea06} at
Apache Point Observatory which has a specially designed wide field camera
\citep{Gea98}. Photometric observations were taken in the $ugriz$ bands
\citep{Fea96,Sea02} by drift scanning the sky under favourable conditions
\citep{Hea01}. These images were then fed through an automated pipeline
that performed the necessary astrometric and photometric calibrations. The
pipeline also detected and measured the photometric properties of the
observed objects \citep{Pea03, Iea04, Tucker06, Pea08}. Select subsamples
\citep{Strauss02, Eea01} were then designated for spectroscopic followup
using a 640 fiber spectrograph.

The DR7 LRG sample is part of the last data release of SDSS-II, the
second phase of SDSS which was completed in 2009. The LRG sample was
selected according to the prescription in \citet{Eea01}. This selection
was optimized to identify the most luminous (and hence most massive and
highly biased) galaxies which can be observed out to high redshifts. Since
the volume encompassed by equal angles on the sky increases with redshift,
we can probe the large volumes necessary for cosmological studies using
these luminous galaxies. The LRGs tend to be old systems with uniform
spectral energy distributions that exhibit a strong 4000\r{A} break. This
gives them a distinct colour-flux-redshift relation which allows them
to be uniformly selected over a wide redshift range. Our sample matches
exactly that of \citet{Kea10} and we refer the interested reader there
for details of its construction. We use the flux-limited LRG sample in
the SDSS Northern Galactic Cap only. This sample spans a redshift range of
$0.16<z<0.47$ and has a number density of $\sim 10^{-4} h^3\rm{Mpc}^{-3}$.

\section{Mock Catalogue Results} \label{sec:mockres}

In this section we present the results of fits to the LasDamas mock
correlation functions before and after reconstruction. These were computed
and fit by taking the LasDamas cosmology as the fiducial cosmology. 

\begin{figure*}
\vspace{0.4cm}
\epsfig{file=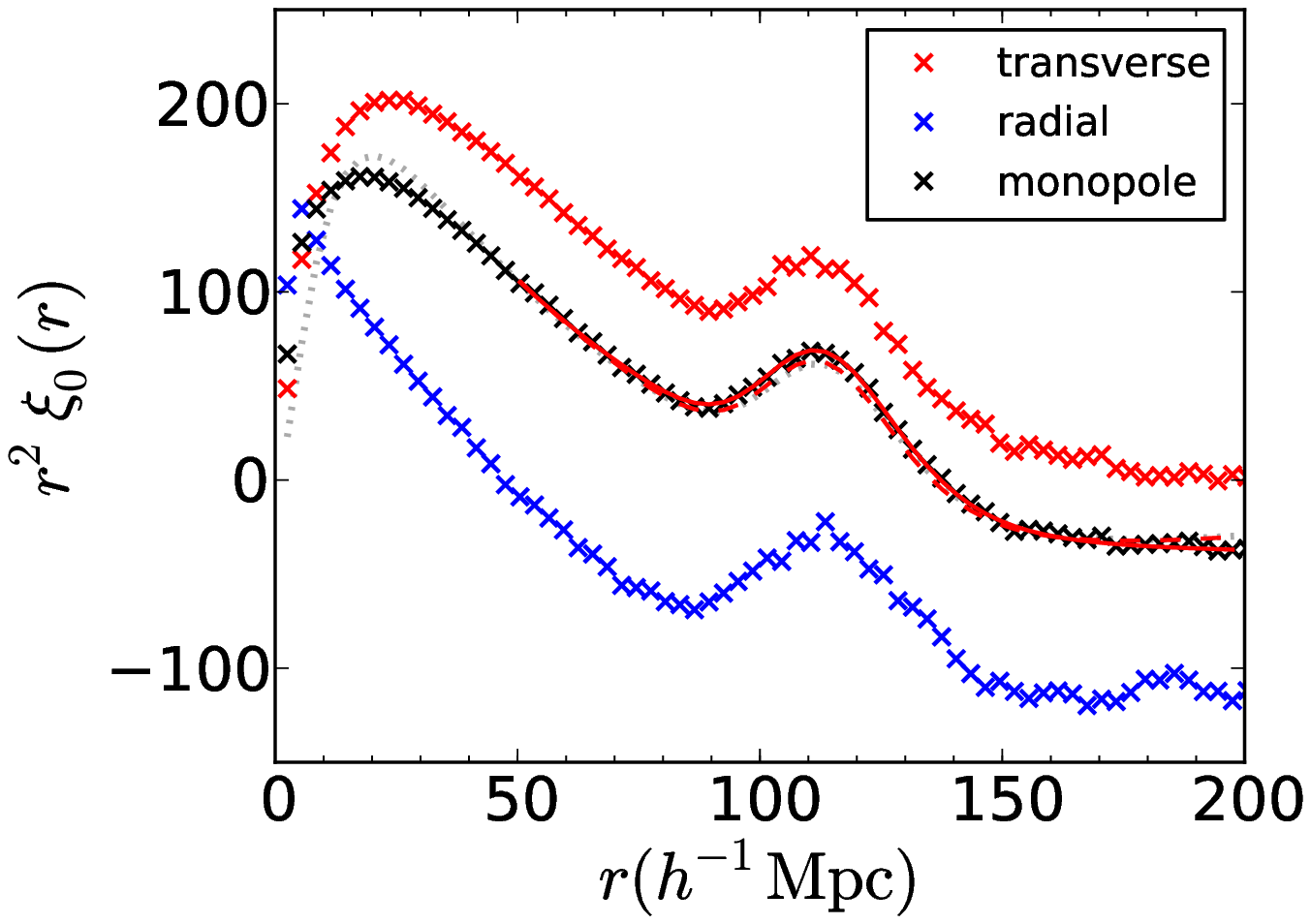, width=0.45\linewidth, clip=}
\hspace{0.5cm} 
\epsfig{file=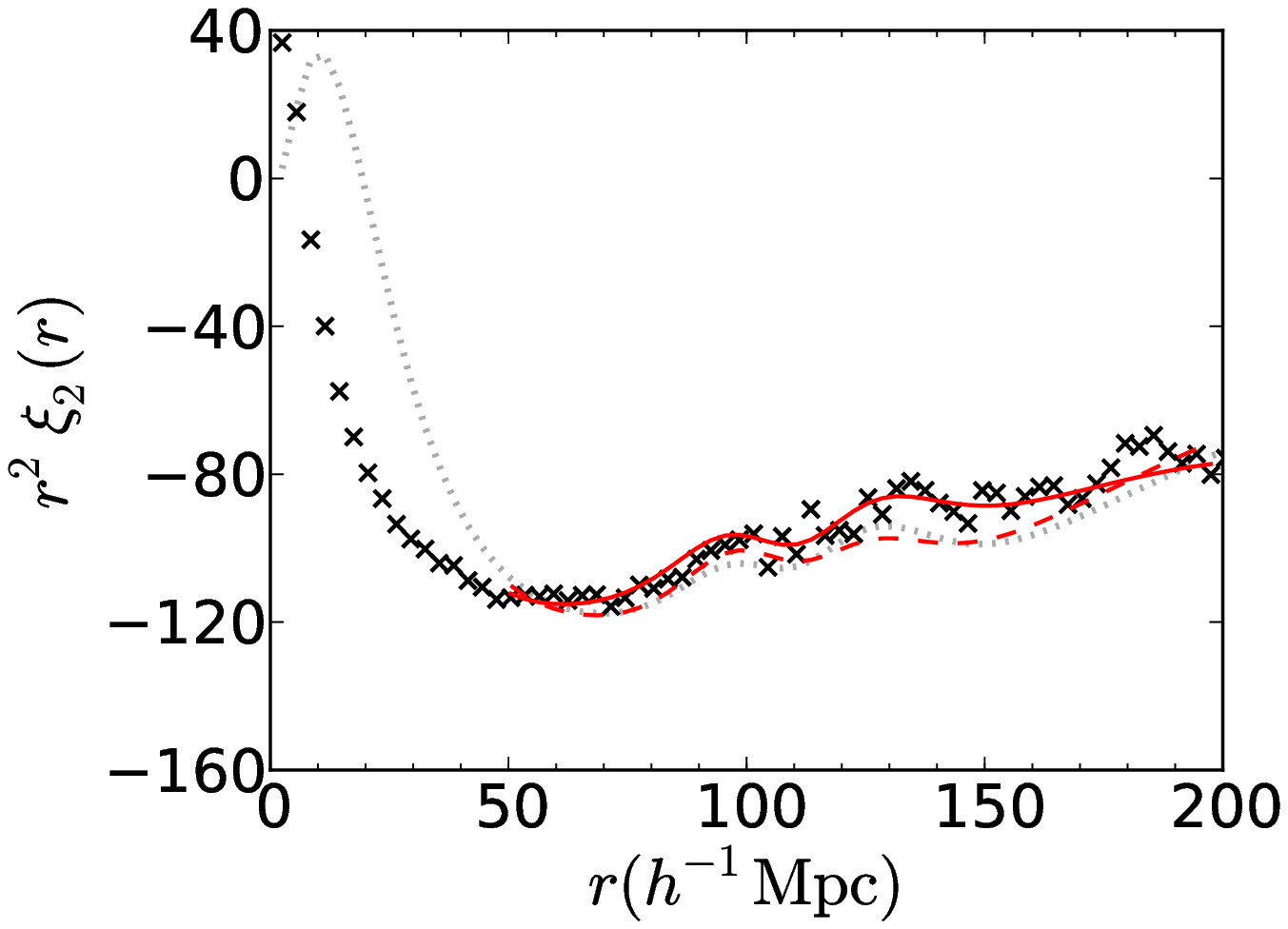, width=0.45\linewidth, clip=}
\caption{Average monopole (left) and quadrupole (right) of the 160 mock
catalogues before reconstruction. The monopole and the quadrupole at
large scales are similar to the fiducial templates (grey dotted lines,
identical to the solid lines plotted in Figures \ref{fig:epfig} \&
\ref{fig:varyfig}). The quadrupole on small scales ($r\lesssim50\hMpc$),
however, shows substantially different structure to the fiducial
template. The fit to the average of the monopole and quadrupole from
the mocks is overplotted in red. The solid line corresponds to a
fit using the fiducial $A_{0,2}(r)$ (Equation (\ref{eqn:fida})) and
the dashed line corresponds to a fit using $A_{0,2}(r)=0$. We allow
$\Sigma_\perp$ and $\Sigma_\parallel$ to vary in these fits and obtain
best-fit values of $6.3\hMpc$ and $10.4\hMpc$ respectively using the
fiducial $A_{0,2}(r)$. In the monopole case, the fit using the fiducial
$A_{0,2}(r)$ is very similar to the $A_{0,2}(r)=0$ fit. In the quadrupole,
the $A_{0,2}(r)=0$ fit is much worse around the acoustic scale. Overall,
$\chi^2$ decreased by $\sim33$ going from $A_{0,2}(r)=0$ to the fiducial
$A_{0,2}(r)$.
\label{fig:simfig}}
\end{figure*}

We plot the average monopole and quadrupole of the 160 mocks before
reconstruction in Figure \ref{fig:simfig}. The fiducial templates (solid
lines in Figures \ref{fig:epfig} \& \ref{fig:varyfig}) are shown as
the grey dotted lines. The monopole and the quadrupole at large scales
look similar to the fiducial templates. However, the small scales
($r \lesssim 50\hMpc$) in the quadrupole show substantially different
structure indicating that our model does not fit the data well at
these scales. This motivates our choice for the fiducial fitting range:
$50<r<200\hMpc$. The best-fit model to the monopole and quadrupole are
overplotted as the red lines where the solid line corresponds to using
the fiducial $A_{0,2}(r)$ and the dashed line corresponds to using
$A_{0,2}(r)=0$ (i.e. no broadband marginalization). The monopole fits
look very similar; however the fiducial $A_{0,2}(r)$ does much better
in the quadrupole near the acoustic scale with $\chi^2$ decreasing by
$\sim33$ relative to the $A_{0,2}(r)=0$ fit. We allow $\Sigma_\perp$
and $\Sigma_\parallel$ to vary in these fits and obtain best-fit values
of $6.3\hMpc$ and $10.4\hMpc$, motivating our choices in the fiducial
model. We keep $\Sigma_s$ fixed at $4\hMpc$.

\begin{figure*}
\vspace{0.4cm}
\centering
\begin{tabular}{cc}
\epsfig{file=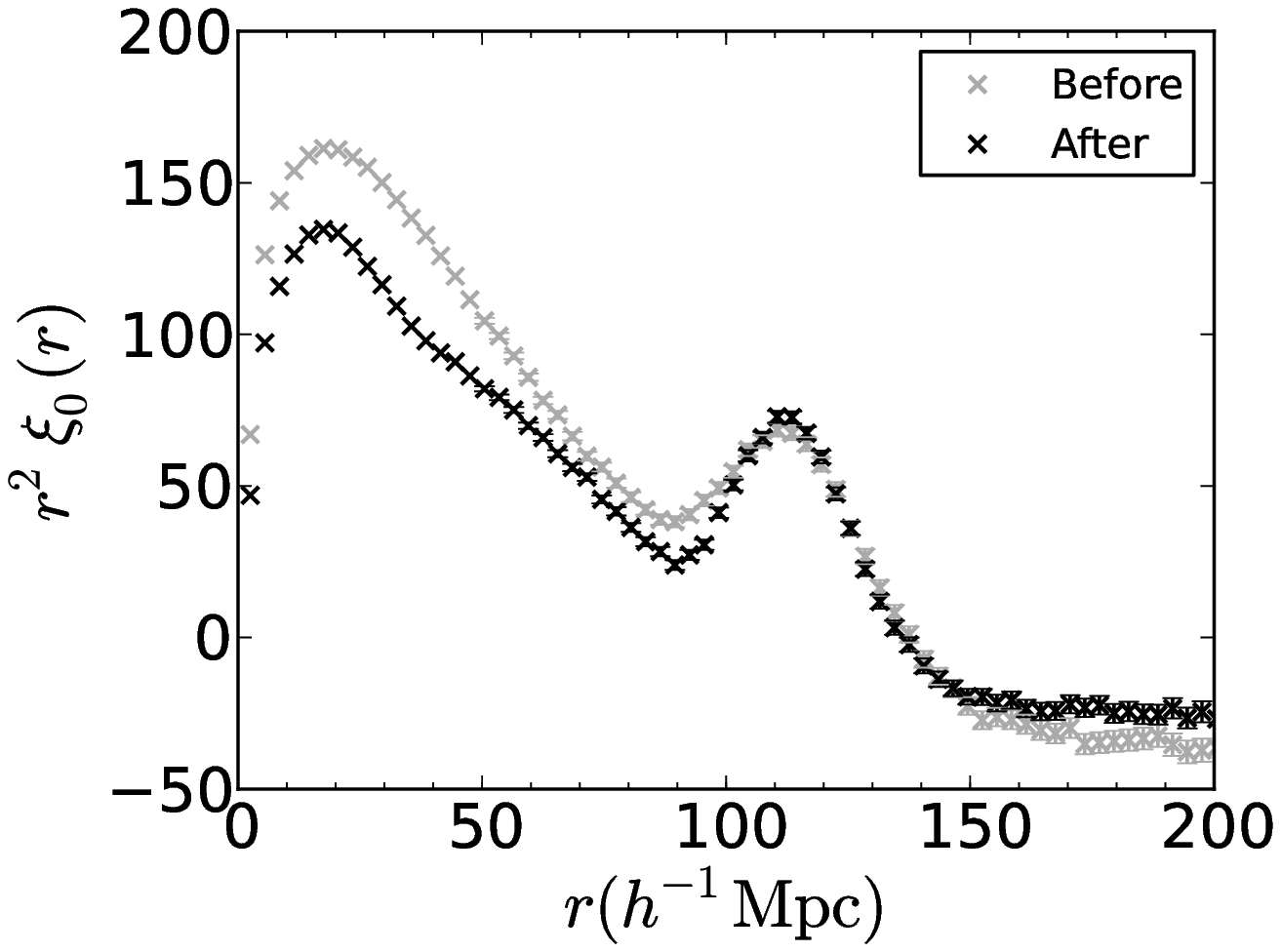, width=0.43\linewidth, clip=}&
\epsfig{file=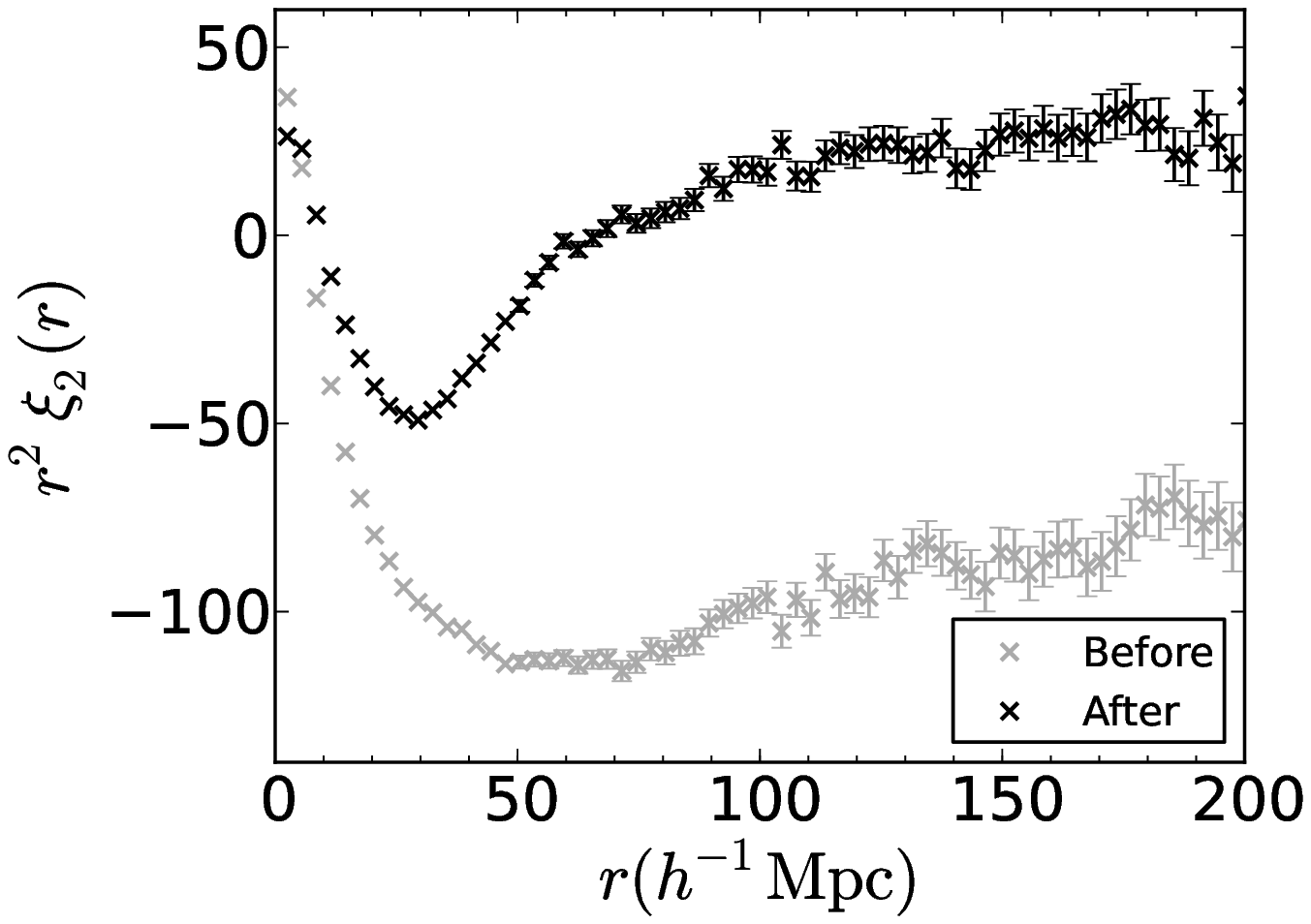, width=0.45\linewidth, clip=}
\end{tabular}
\caption{The average monopole (left) and quadrupole (right) of the 160
mocks before (gray crosses) and after (black crosses) reconstruction. One
can see that after reconstruction, the acoustic peak in the monopole has
sharpened up, indicating that reconstruction is effective at removing
the degradation of the BAO caused by non-linear structure growth. In
the quadrupole, the power at large-scales goes close to 0 which implies
that reconstruction was effective at removing the Kaiser effect. It
is not exactly zero due to some small anisotropy introduced by the
reconstruction technique itself (see Figure \ref{fig:realfig}). We note
that the quadrupole is multiplied by $r^2$ in this figure and hence the
magnitude of this anisotropy is exaggerated.
\label{fig:ursimfig}}
\end{figure*}

A comparison of the monopole and quadrupole before and after
reconstruction is shown in Figure \ref{fig:ursimfig}. As in \citet{Pea12},
we see the acoustic peak in the monopole appears less smeared after
reconstruction which indicates that our reconstruction technique was
effective at partially undoing non-linear evolution. This is also
reflected in the fact that after reconstruction (where we assume
the smearing is isotropic), a fit to the average of the mocks gave
a much smaller BAO smoothing scale of $\snl=2.9\hMpc$ as opposed
to the pre-reconstruction values of $\Sigma_\perp = 6.3\hMpc$ and
$\Sigma_\parallel=10.4\hMpc$. In addition, we see that the quadrupole
is nearly zero on large scales after reconstruction, which implies that
our reconstruction technique was also effective at partially removing the
Kaiser effect. The fact that the quadrupole is positive and not exactly 0
is likely due to some slight anisotropy introduced by reconstruction. This
is discussed in more detail below and shown not to significantly affect
our measurements of $\alpha$ and $\epsilon$.

By averaging the mocks we have effectively increased the survey volume
by a factor of 160 and hence the variance should decrease by an equal
amount. This means that the average of the mocks should have substantially
less noise. In addition, we know that we are computing the correlation
functions and fitting using the correct (LasDamas) cosmology. Hence,
we expect that the $\alpha$ and $\epsilon$ values measured from the
average of the mocks should be 1 and 0 respectively if our models are
unbiased (i.e. there should be no shift in the location of the peak
relative to the model and there should be no anisotropy). We find that
fitting the pre-reconstruction mock average gives $\alpha=1.005$ and
$\epsilon=0.003$ while post-reconstruction we measure $\alpha=1.002$
and $\epsilon=0.002$. The slight offset in $\alpha$ from 1 is not too
concerning as we expect non-linear structure growth to shift the peak
by $\lesssim0.5\%$ \citep{PdW09,Mea11}. The fact that $\alpha$ moves
closer to 1 after reconstruction is encouraging as reconstruction is
supposed to remove some effects of non-linear structure growth. The
small bias in $\epsilon$ is not significant at our current levels
of statistical precision and is likely the result of small mismatches
between the broadband model and the data. This will be discussed in more
detail shortly. We again emphasize that in these fits we have allowed
$\Sigma_\parallel$ and $\Sigma_\perp$ to vary. When we fit the individual
mocks, the signal-to-noise of the data is not sufficient for constraining
any of these parameters and hence we fix them in the fiducial model to
the values obtained in the averaged mock fits.

\begin{figure}
\vspace{0.4cm}
\centering
\epsfig{file=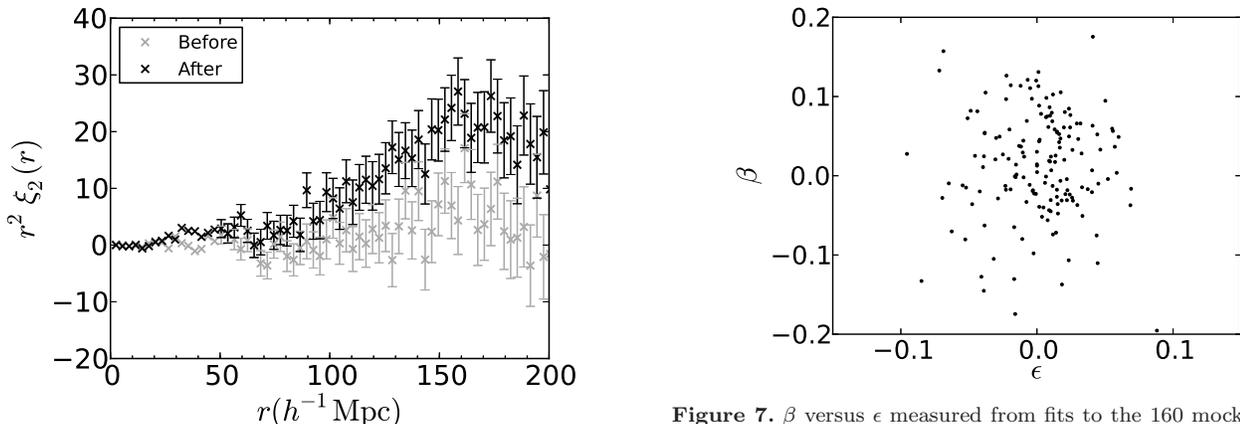, width=0.9\linewidth, clip=}
\caption{Average quadrupole from the mocks in real space before (grey)
and after (black) reconstruction. The quadrupole before reconstruction
is very close to 0 as we would expect in real space due to the lack of
redshift-space distortions. Our $\epsilon$ measurements are unbiased in
this case which suggests that the small biases we see in redshift space
are due to slight mismatches between our redshift-space distortion
models and the actual broadband in the data. After reconstruction,
the quadrupole at large scales acquires some additional power likely
due to the survey geometry and sample number density fluctuations as
a function of redshift. Our post-reconstruction real space $\epsilon$
measurements remain unbiased which suggests that this anisotropy can be
accounted for by our $A_2(r)$ nuisance terms.
\label{fig:realfig}}
\end{figure}

\begin{figure}
\vspace{0.4cm}
\centering
\epsfig{file=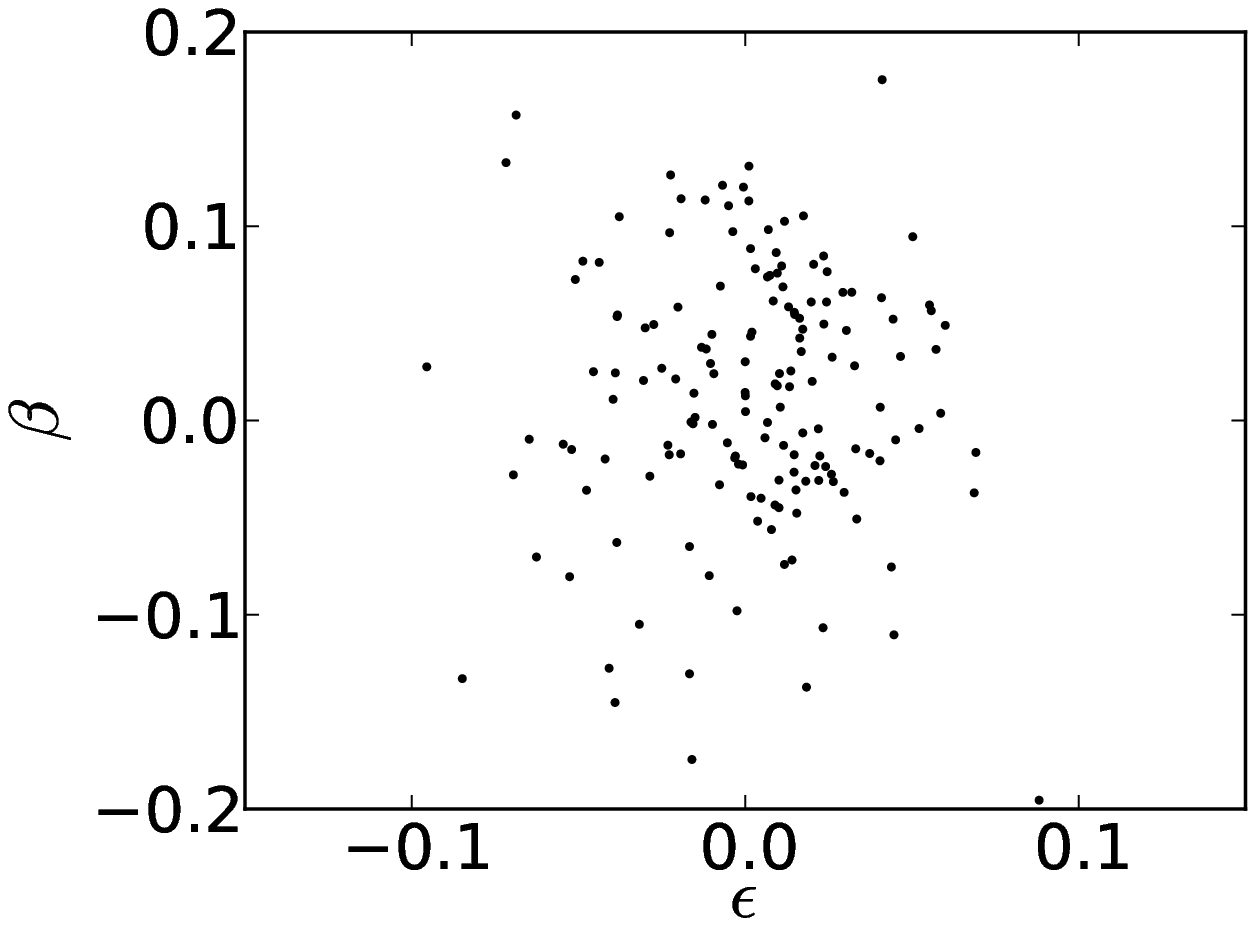,width=0.8\linewidth,clip=}
\caption{$\beta$ versus $\epsilon$ measured from fits to the 160
mocks after reconstruction. One can see that these two parameters are
uncorrelated. We see similar results before reconstruction as well.
\label{fig:epbet}}
\end{figure}

\begin{figure}
\vspace{0.4cm}
\centering
\begin{tabular}{c}
\epsfig{file=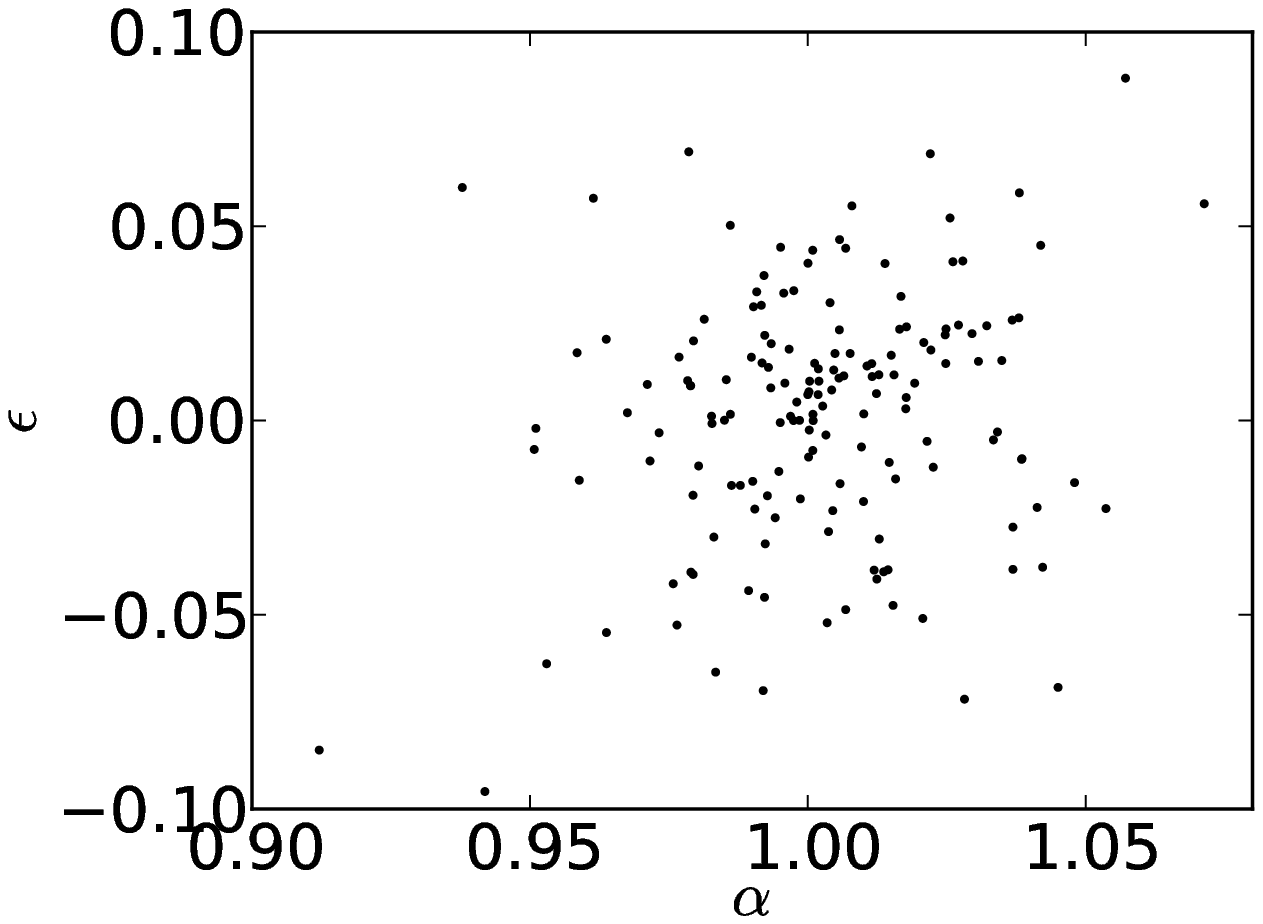, width=0.77\linewidth,clip=} \\
\hspace{0.6cm}\epsfig{file=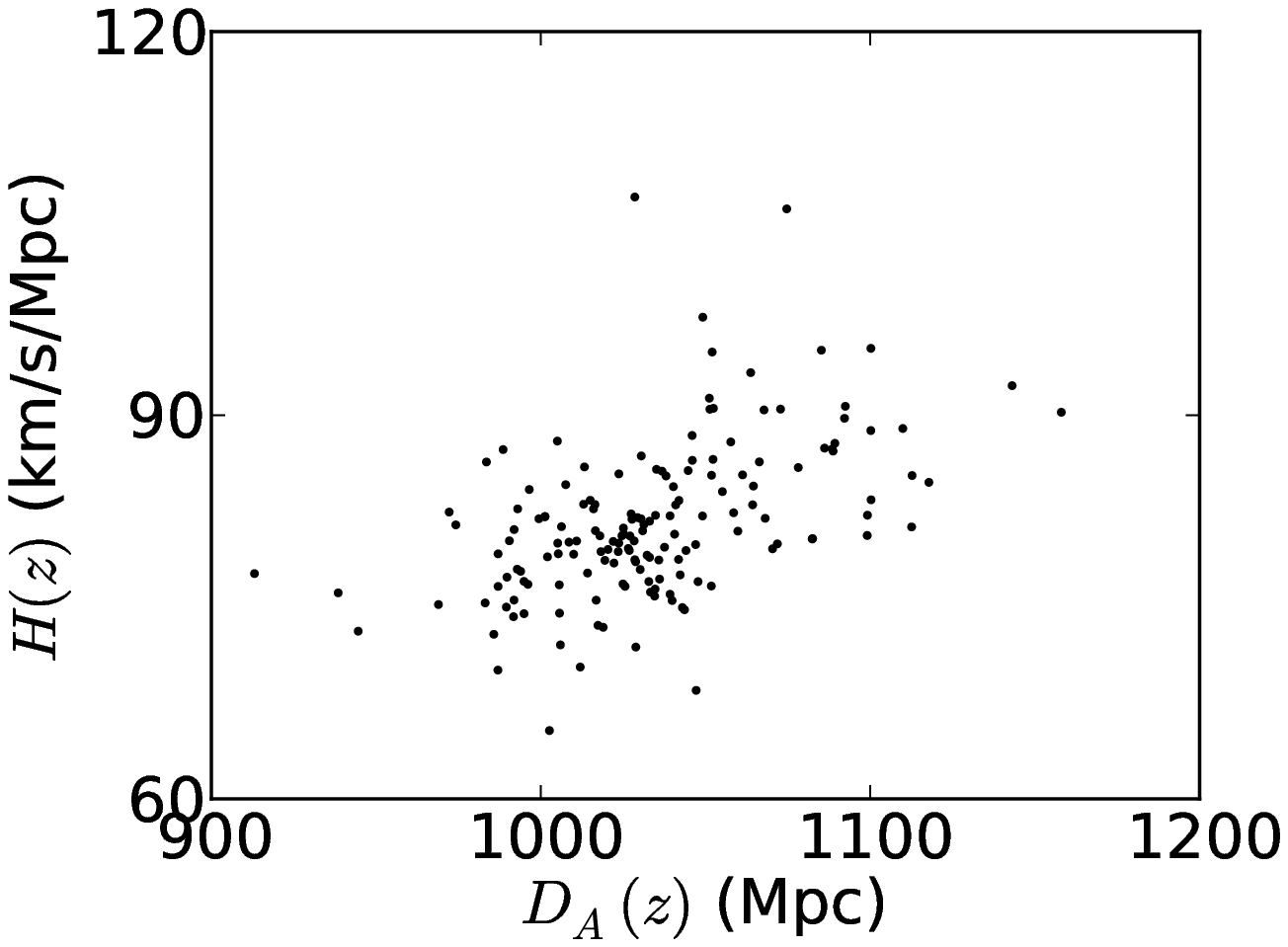, width=0.8\linewidth,clip=}
\end{tabular}
\caption{$\alpha$ versus $\epsilon$ (top) and $D_A(z)$ versus $H(z)$
(bottom) for the mocks after reconstruction. One can see that $\alpha$
and $\epsilon$ are not highly correlated. From the points plotted
we measure a correlation coefficient of 0.27 and analogously in the
pre-reconstruction case we measure 0.20. These are in excellent agreement
with Fisher matrix predictions ($\rho_{\alpha\epsilon}\sim0.21$). We
see a stronger correlation between $D_A$ and $H$ which we obtained by
combining $\alpha$ and $\epsilon$ as in Equations (\ref{eqn:daz}) \&
(\ref{eqn:hz}). We expect $\rho_{D_A H}\sim0.4$ and we find correlation
coefficients of 0.23 and 0.50 between our measured $D_A$ and $H$ values
before and after reconstruction respectively.
\label{fig:aedah}}
\end{figure}

\begin{figure}
\vspace{0.4cm}
\centering
\epsfig{file=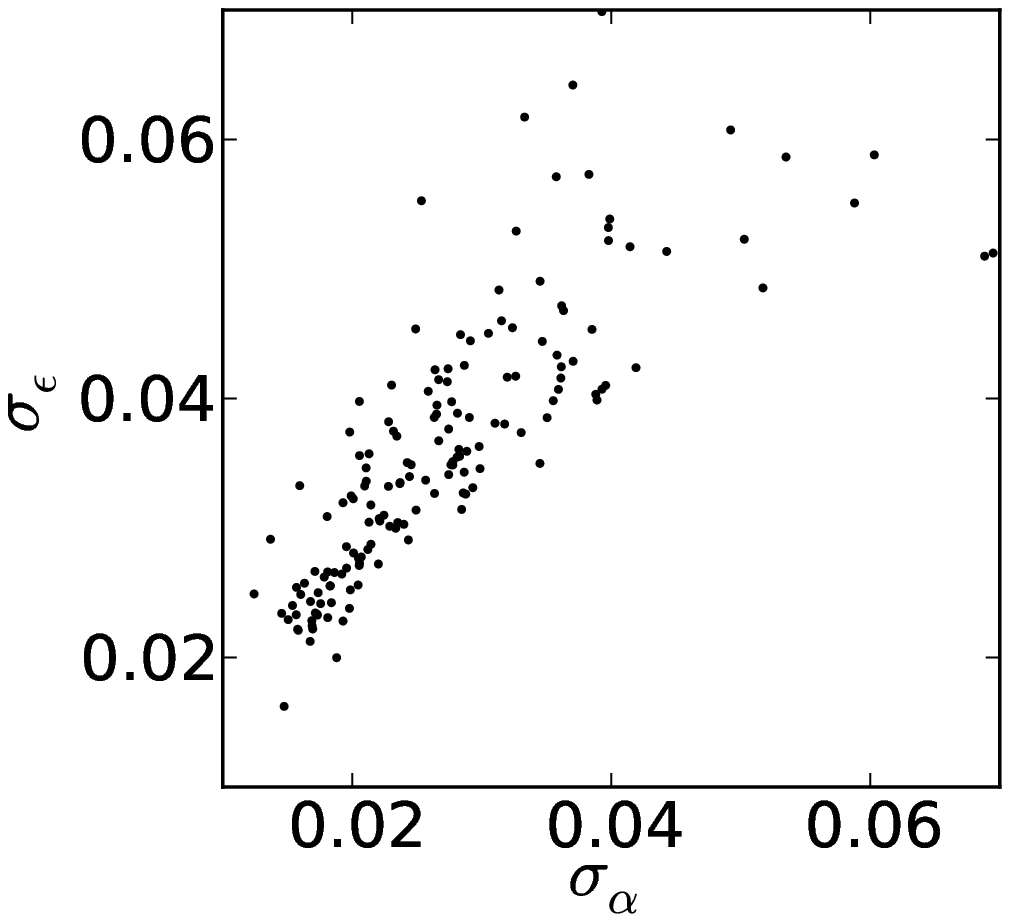, width=0.8\linewidth, clip=}
\caption{Post-reconstruction $\sigma_\alpha$ versus $\sigma_\epsilon$
for the mocks. We see that the errors on $\alpha$ and the errors on
$\epsilon$ are directly correlated with each other. This indicates that
mocks with poorer measurements of the acoustic scale $\alpha$ also
have poorer measurements of the BAO anisotropy $\epsilon$. A similar
correlation exists in the pre-reconstruction case. The median ratios of
$\sigma_\epsilon/(1+\epsilon)$-to-$\sigma_\alpha/\alpha$ are $\sim1.24$
and $\sim1.38$ before and after reconstruction respectively. Fisher
matrix arguments predict a ratio of $\sim1.2$.
\label{fig:sasefig}}
\end{figure}

We measure $\alpha$ and $\epsilon$ for each mock using the fitting
procedure and fiducial model outlined in \S\ref{sec:fitting}. We also
estimate the errors $\sigma_\alpha$, $\sigma_\epsilon$ and the correlation
coefficient $\rho_{\alpha\epsilon}$ for each mock using Equations
(\ref{eqn:sdeq}) \& (\ref{eqn:cceq}). Before reconstruction, we measure
a mean $\langle \alpha \rangle = 1.003 \pm 0.003$ with an rms scatter
between the mocks of $0.034$ and a median $\tilde{\alpha} = 1.008$ with
16th/84th percentiles of the mocks corresponding to $^{+0.030}_{-0.036}$
(these will henceforth be denoted the quantiles). For $\epsilon$ we
measure a mean $\langle \epsilon \rangle = 0.001 \pm 0.003$ with an rms
scatter between the mocks of $0.037$ and a median $\tilde{\epsilon}
= 0.004$ with quantiles $^{+0.032}_{-0.037}$. After reconstruction,
we measure a mean $\langle \alpha \rangle = 1.002 \pm 0.002$ with an
rms scatter between the mocks of $0.024$ and a median $\tilde{\alpha} =
1.002$ with quantiles $^{+0.023}_{-0.022}$. For $\epsilon$ we measure a
mean $\langle \epsilon \rangle = 0.002 \pm 0.003$ with an rms scatter
between the mocks of $0.032$ and a median $\tilde{\epsilon} = 0.007$
with quantiles $^{+0.023}_{-0.037}$. These values were calculated after
rejecting the mocks where the acoustic signal is too weak to obtain an
accurate centroiding of the BAO peak. This corresponds to making a cut
in $\sigma_\alpha$ at 0.07 and discarding the mocks that lie above this
cut as demonstrated in \citet{Xea12}. Before reconstruction, 9 mocks out
of 160 lie above this cut and after reconstruction, 0 lie above this cut.

\begin{table}
\centering
\caption{$\epsilon$ statistics for various mock combinations. The first
column indicates the number of mocks we have combined ($m$). The second
column quotes the mean $\epsilon$ we measure with the standard error on
the mean. The third column shows the rms of the mocks. The fourth column
quotes the median $\epsilon$ and the fifth column quotes the quantiles.
\label{tab:eptab}}
\begin{tabular}{lcccc}
\hline
$m$&
$\langle\epsilon\rangle$&
rms&
$\tilde{\epsilon}$&
Qtls \\
\hline
\multicolumn{5}{c}{Redshift Space without Reconstruction}\\
\hline

1&$0.001 \pm 0.003$&0.037&0.004&$^{+0.032}_{-0.037}$\\
2&$0.001 \pm 0.003$&0.029&0.006&$^{+0.023}_{-0.030}$\\
4&$0.001 \pm 0.003$&0.019&-0.001&$^{+0.017}_{-0.010}$\\
8&$0.002 \pm 0.003$&0.012&0.002&$^{+0.009}_{-0.013}$\\

\hline
\multicolumn{5}{c}{Redshift Space with Reconstruction}\\
\hline

1&$0.002 \pm 0.003$&0.032&0.007&$^{+0.023}_{-0.037}$\\
2&$0.003 \pm 0.002$&0.018&0.006&$^{+0.012}_{-0.020}$\\
4&$0.003 \pm 0.002$&0.012&0.003&$^{+0.012}_{-0.011}$\\
8&$0.003 \pm 0.002$&0.008&0.006&$^{+0.004}_{-0.007}$\\

\hline
\end{tabular}
\end{table}

The median $\epsilon$ in the pre- and post-reconstruction cases are
different from 0 and from the mean at $\gtrsim1$-2 times the error on the
mean. In addition, the post-reconstruction quantiles are asymmetric,
implying that the posterior $\epsilon$ distribution deviates from
Gaussian. These appear to be in part due to the intrinsic noise in
the data and in part due to a slight mismatch between the model and
the data. To reduce noise, we effectively increase the spatial volume
of the data by combining our 160 mocks into groups of 2, 4 and 8, and
re-perform our fits. In general, we see a better agreement between the
mean and median $\epsilon$. The quantiles remain mildly asymmetric in some
cases but overall we see improved agreement. The rms scatter decreases by
roughly the expected amount ($\sim\sqrt{m}$, where $m=2,4\;\rm{or}\;8$)
if we consider $\epsilon$ to be Gaussian. These results are summarized
in Table \ref{tab:eptab}.

\begin{figure*}
\vspace{0.4cm}
\centering
\subfigure[Comparison of results obtained using a larger fitting range
($30<r<200\hMpc$) versus the fiducial fitting range. Here, we see an
average $\Delta\alpha$ that is $\sim-0.003$ before reconstruction. The
error on the mean is a factor of $\sim\sqrt{160}$ smaller than the scatter
indicated by the quantiles, which makes this $0.003$ shift statistically
significant. This is caused by the templates being poor matches to the
data at low $r$. While the $A_{0,2}(r)$ terms attempt to compensate for
this, accurate fitting of the BAO scale is compromised. The significant
$\Delta\epsilon$ in this case is rooted in the same cause. The fact that
$\Delta\alpha$ and $\Delta\epsilon$ are both 0 after reconstruction
suggests that the post-reconstruction model is better matched to the
data.]
{
\begin{tabular}{cc}
\epsfig{file=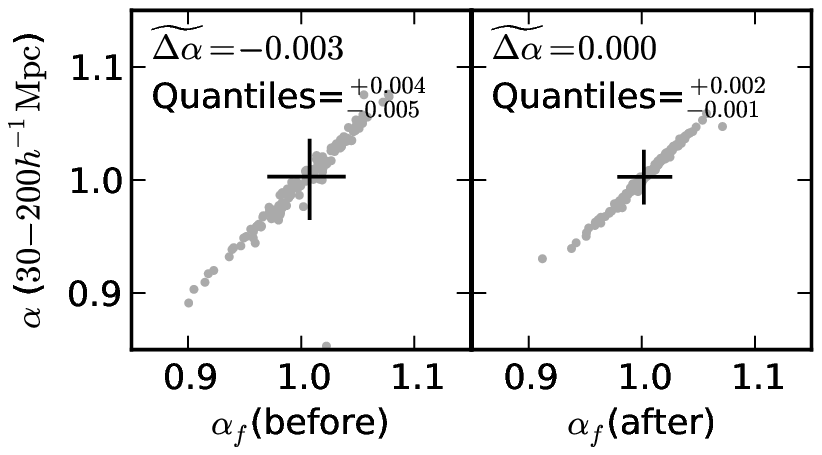, width=0.45\linewidth, clip=}&
\epsfig{file=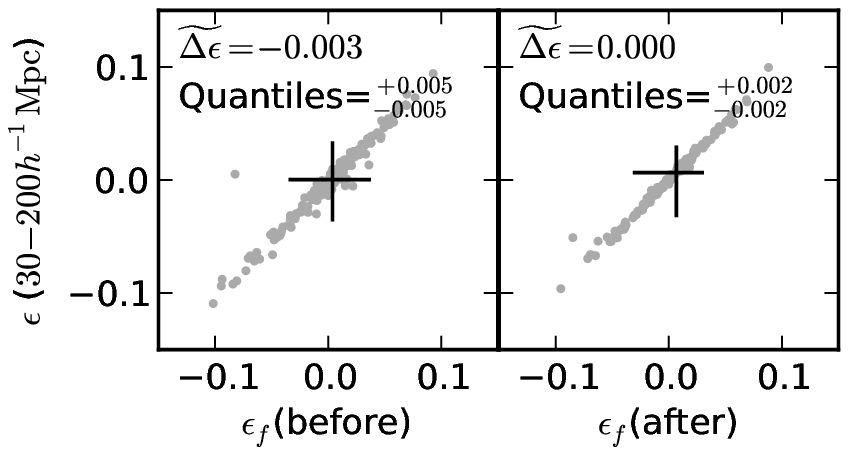, width=0.465\linewidth, clip=}
\end{tabular}
\label{fig:compsfig_fr}
}
\vspace{0.5cm}
\subfigure[Comparison of results obtained using $\Sigma_s=0\hMpc$ versus
the fiducial value of $\Sigma_s=4\hMpc$. Here, we see that the average
$\Delta\epsilon$ is different from 0 when considering the error on the
mean which is a factor of $\sim\sqrt{160}$ smaller than the scatter
indicated by the quantiles. This is caused by the quadrupole model being
a less optimal match to the data. We note that these small shifts in
$\alpha$ and $\epsilon$ are certainly not detectable in each mock, which
have much larger errors on $\alpha$ and $\epsilon$ than $\sim0.003$.]
{
\begin{tabular}{cc}
\epsfig{file=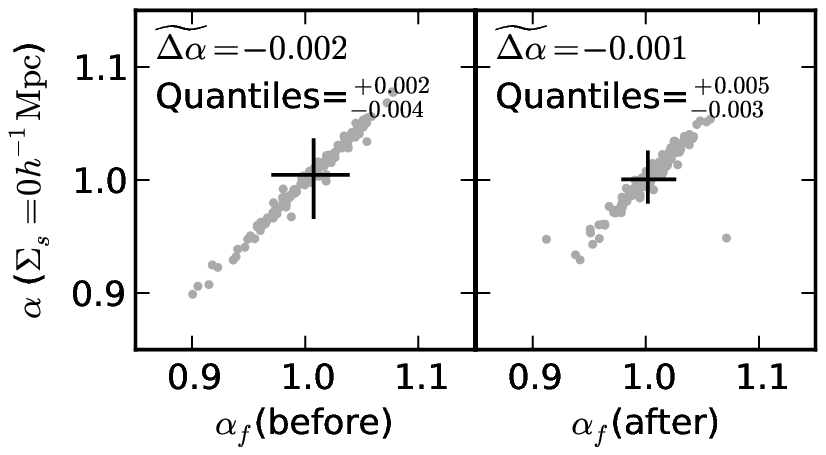, width=0.45\linewidth, clip=}&
\epsfig{file=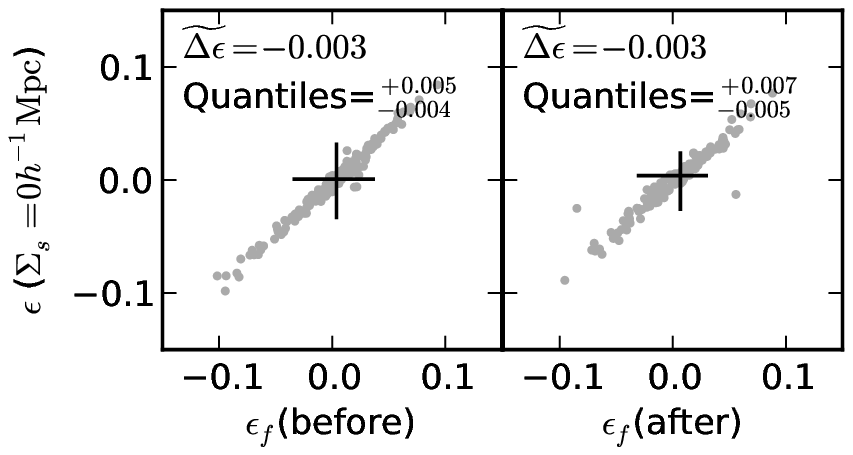, width=0.465\linewidth, clip=}
\label{fig:compsfig_ss}
\end{tabular}
}
\caption{The robustness of our fitting model as demonstrated by $\alpha$
and $\epsilon$ scatter plots for some sample cases. Results for other
changes to the fiducial model are given in Table \ref{tab:alphas}. The
plotted $\alpha$ and $\epsilon$ values were obtained through fitting
the mock correlation functions before and after reconstruction. The
plots are presented in pairs: $\alpha$ on the left and $\epsilon$ on the
right. The first plot in each pair shows the pre-reconstruction results
and the second plot shows the post-reconstruction results. The black
crosses indicate the medians and quantiles of the mock measurements. If
we obtain consistent measurements of $\alpha$ and $\epsilon$ with a
model that has parameters slightly different to the fiducial model,
then we should see $\Delta\alpha$ and $\Delta\epsilon$ values that are
$\sim0$. We see that this is true at the level of our current statistical
precision. Therefore, our fitting model is reasonably robust against
small changes to the fiducial model parameters.
\label{fig:compsfig}}
\end{figure*}

We see that there is a persistent bias in $\epsilon$ towards
non-zero values that is currently below our detection threshold. This
bias is $\lesssim1\sigma$ significant before reconstruction and only at
the $1$-$1.5\sigma$ level after reconstruction. To further test this,
we split the mocks into 2 groups of 80 which reduces the noise in the
data. After re-performing the fits, we find $\langle\epsilon\rangle \sim
0.002$ both before and after reconstruction. These values agree with the
fit results to the average of the 160 mocks described above. This suggests
that the persistent bias in $\epsilon$ is not due to noise but is rather a
result of some mismatch between the model and the data. In our fits we fix
$\Sigma_\perp$, $\Sigma_\parallel$ and $\Sigma_s$, and use the nuisance
terms in $A_2(r)$ to account for any other mismatch in the broadband shape
between the model and the data. However $\Sigma_\perp$, $\Sigma_\parallel$
and $\Sigma_s$ are partially degenerate with $\epsilon$, so if they
are fixed at non-optimal values that cannot be fully compensated by
$A_2(r)$ (see Figure \ref{fig:compsfig_ss}), the fitter can resort to
adjusting $\epsilon$. We stress however, that such small biases in our
redshift-space measurements of $\epsilon$ are below the current detection
limit in a single DR7 realization as indicated by the rms of the mocks.

We can gain additional insights by analyzing the real-space mocks which
do not have redshift-space distortions and therefore do not require
$\Sigma_s$ or anisotropic $\snl$ in the model. We find that these give
non-biased measures of $\epsilon$ in both the pre- and post-reconstruction
cases. In the pre-reconstruction case we measure the mean $\epsilon$
to be $\langle \epsilon \rangle = 0.003\pm0.003$ with an rms between the
mocks of 0.037. The median is $\tilde{\epsilon} = 0.002$ with quantiles
$^{+0.040}_{-0.040}$. After reconstruction we measure $\langle \epsilon
\rangle = 0.001\pm0.002$ with a mock rms of 0.027 and $\tilde{\epsilon}
= -0.002$ with quantiles $^{+0.030}_{-0.023}$. One can see that the mean
and median $\epsilon$ are consistent with each other and with 0. Fitting
the average of the 160 mocks gives $\epsilon=0.001$ and $0.000$ before
and after reconstruction respectively; again implying a largely unbiased
measurement of $\epsilon$ in real space. 

\begin{figure*}
\vspace{0.4cm}
\centering
\begin{tabular}{cc}
\epsfig{file=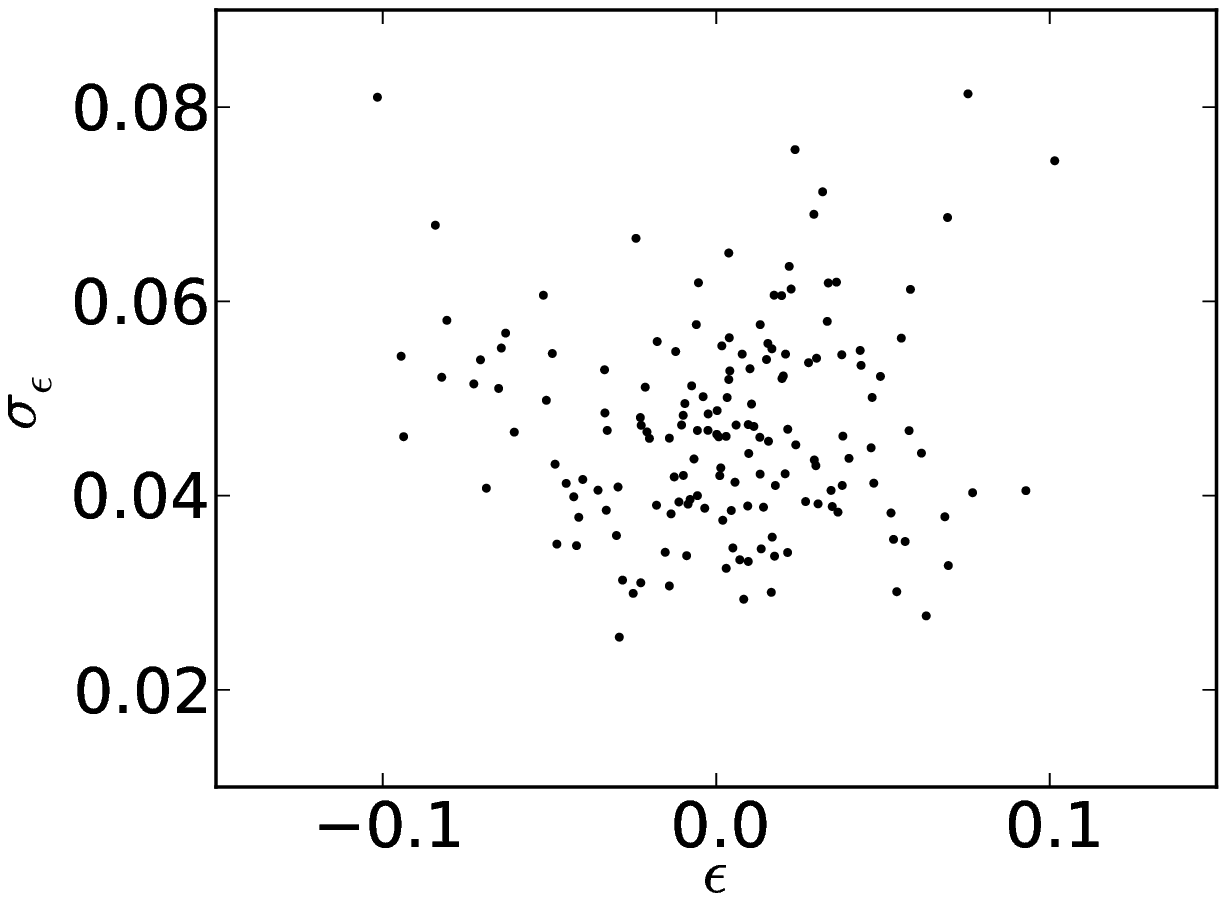, width=0.45\linewidth, clip=} &
\epsfig{file=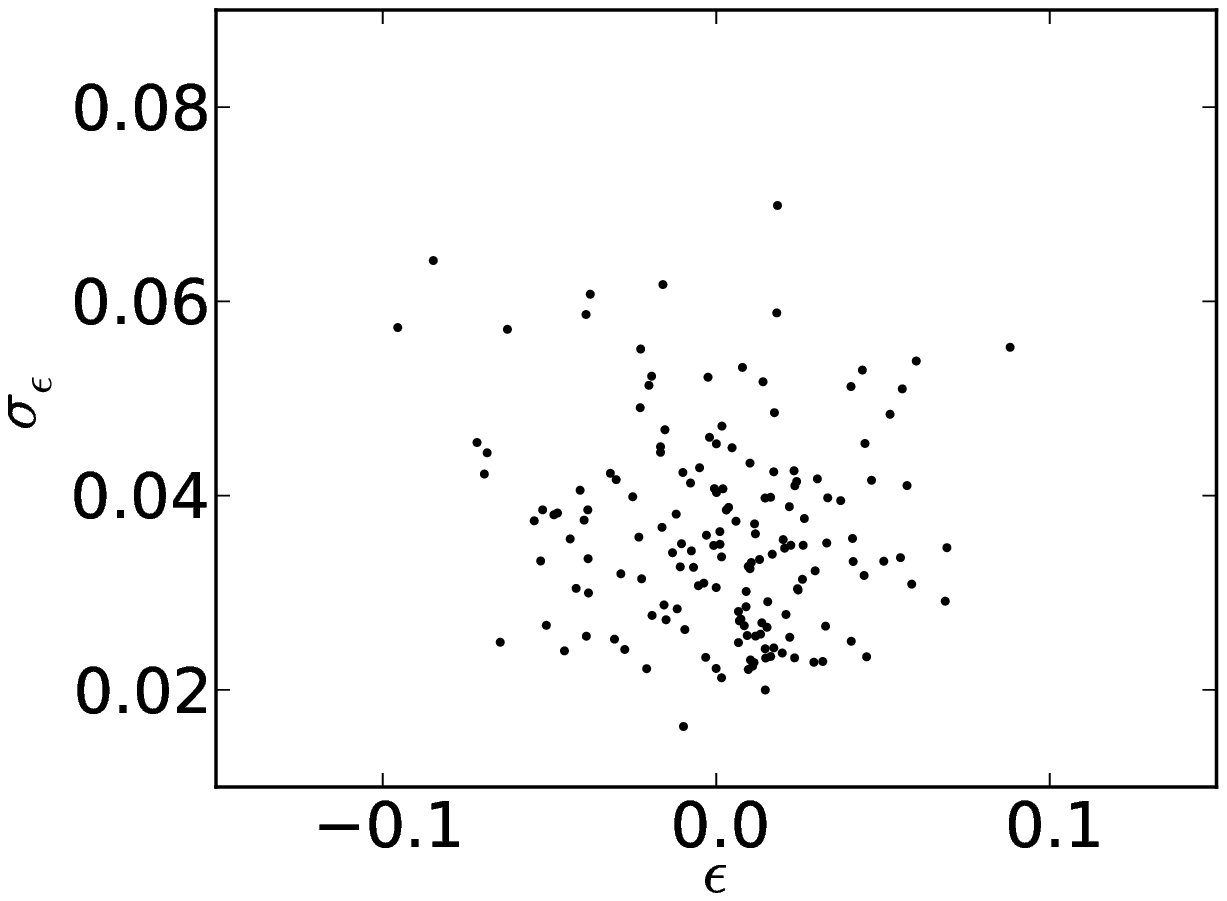, width=0.45\linewidth, clip=} 
\end{tabular}
\caption{$\epsilon$ versus $\sigma_\epsilon$ before (left) and after
(right) reconstruction for the 160 mocks. One can see that reconstruction
decreses the scatter in the measured $\epsilon$ values (this is further
highlighted in Figure \ref{fig:epcomp}). While reconstruction does
decrease the average error on $\epsilon$, we see that the errors we
measure are still fairly large compared to the errors on $\alpha$
(see Figure \ref{fig:sasefig}). 
\label{fig:epsig}}
\end{figure*}

An interesting artifact we do find is that reconstruction
appears to introduce some broadband anisotropy as shown in Figure
\ref{fig:realfig}. Here we have plotted the mean of the real-space
quadrupoles before (grey) and after (black) reconstruction. We see that
the quadrupole is nearly 0 before reconstruction as expected since
there should not be any anisotropies in real space. However, after
reconstruction, the quadrupole acquires some additional large-scale
power. The reconstruction displacement vectors may take on a subtle
anisotropy due to the variation of number density with redshift or the
survey geometry (i.e. if it is wider than it is deep). Fortunately,
this broadband anisotropy introduced by reconstruction is fairly smooth
and can be removed by the $A_2(r)$ nuisance parameters as evidenced by
our unbiased measures of $\epsilon$ in post-reconstruction real space.

\begin{figure}
\vspace{0.4cm}
\centering
\epsfig{file=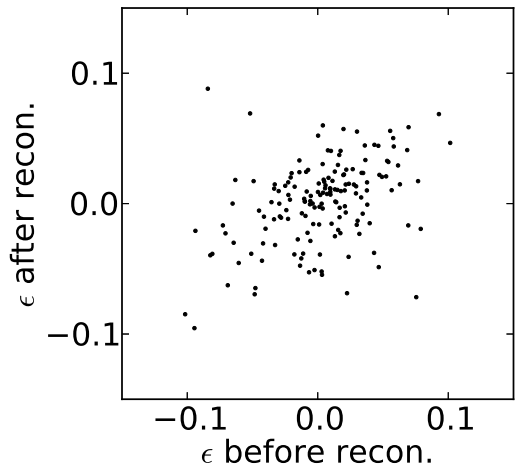, width=0.8\linewidth, clip=}
\caption{$\epsilon$ before reconstruction versus $\epsilon$ after
reconstruction fit from the 160 mocks. The slope of a linear fit to these
points is less than 1 implying that the post-reconstruction $\epsilon$
values have smaller scatter.
\label{fig:epcomp}}
\end{figure}

To build more intuition for the parameters we fit in redshift space
and to demonstrate their inter-dependencies, we show various scatter
plots of these quantities in Figures \ref{fig:epbet}, \ref{fig:aedah},
\ref{fig:sasefig}, \ref{fig:epsig} \& \ref{fig:epcomp}. Figure
\ref{fig:epbet} shows the values of $\beta$ and $\epsilon$ we obtain from
our fits to the 160 mocks after reconstruction. Our pre-reconstruction
results are similar. One can see that these two parameters are not
correlated with each other.

The top panel of Figure \ref{fig:aedah} shows the $\epsilon$ versus
$\alpha$ values we measure from the mocks after reconstruction. Again
we see that these two parameters are not highly correlated. The
correlation coefficient between $D_A$ and $H$ is predicted to be
$\rho_{D_A H}\sim0.4$ \citep{SE07}. This subsequently predicts a
$\sigma_H/H$-to-$\sigma_{D_A}/D_A$ ratio $\sim2$ (i.e. the fractional
error of the Hubble parameter is twice that of the angular diameter
distance). Using these values, a Fisher matrix argument shows that
we should expect $\rho_{\alpha\epsilon}\sim0.21$ (see Appendix
\ref{app:fisher}). The correlation coefficient between the $\alpha$
and $\epsilon$ values plotted in Figure \ref{fig:aedah} is 0.20 and the
corresponding pre-reconstruction value is 0.27. Both are in excellent
agreement with the Fisher matrix prediction. The bottom panel of Figure
\ref{fig:aedah} shows our $\alpha$ and $\epsilon$ measurements translated
into measurements of $D_A$ and $H$ using Equations (\ref{eqn:daz})
\& (\ref{eqn:hz}). In the plotted post-reconstruction case, the
correlation coefficient between $D_A$ and $H$ is $\sim0.50$ and in the
pre-reconstruction case it is $\sim0.23$, which are not too different
from our assumed $\rho_{D_A H}=0.4$.

Figure \ref{fig:sasefig} shows $\sigma_\epsilon$ versus $\sigma_\alpha$
for the mocks after reconstruction. We see that the errors on
$\alpha$ and $\epsilon$ are correlated which implies that mocks
with poorer measurements of the acoustic scale (i.e. larger
$\sigma_\alpha$ values) also have poorer measurements of the BAO
anisotropy (i.e. larger $\sigma_\epsilon$ values). We see a similar
correlation in the pre-reconstruction results. Taking the ratio
of $\sigma_\epsilon/(1+\epsilon)$-to-$\sigma_\alpha/\alpha$, we
find a median $\sim1.24$ before reconstruction and $\sim1.38$ after
reconstruction. Fisher matrix arguments predict a ratio of $\sim1.2$
(Equation \ref{eqn:sigae}), which is similar to what we see.

Figure \ref{fig:epsig} shows the values of $\epsilon$ we
measure versus $\sigma_\epsilon$ before (left) and after (right)
reconstruction. Reconstruction clearly decreases the scatter in
$\epsilon$ which is again highlighted in Figure \ref{fig:epcomp},
showing the values of $\epsilon$ before and after reconstruction. While
the two are correlated, the post-reconstruction values have smaller
scatter as evidenced by the locus of points having a slope shallower
than 1:1. From Figure \ref{fig:epsig} we also see that reconstruction
decreases $\sigma_\epsilon$, our estimated error on $\epsilon$. However,
these $\sigma_\epsilon$ values are still large compared to $\sigma_\alpha$
(see Figure \ref{fig:sasefig}). 

\begin{table*}
\caption{Fitting results from the mocks for various models. The model is given in column 1. The median $\alpha$ is given in column 2 with the 16th/84th percentiles from the mocks given in column 3 (these are denoted as the quantiles in the text, hence the label Qtls in the table). The median $\epsilon$ is given in column 6 with corresponding quantiles in column 7. The median difference in $\alpha$ on a mock-by-mock basis between the model listed in column 1 and the fiducial model is given in column 4 with corresponding quantiles in column 5. The analogues for $\epsilon$ are given in columns 8 and 9. The mean $\chi^2$/dof is given in column 10.}
\label{tab:alphas}

\begin{tabular}{@{}lccccccccc}

\hline
Model&
$\tilde{\alpha}$&
Qtls&
$\widetilde{\Delta\alpha}$&
Qtls&
$\tilde{\epsilon}$&
Qtls&
$\widetilde{\Delta\epsilon}$&
Qtls&
$\langle\chi^2\rangle/dof$\\

\hline
\multicolumn{10}{c}{Redshift Space without Reconstruction}\\
\hline

Fiducial $[f]$ &
$1.008$&
$^{+0.030}_{-0.036}$&
--&
--&
$0.004$&
$^{+0.032}_{-0.037}$&
--&
--&
92.01/90\\
\\[-1.5ex]
Fit w/ $(\Sigma_\perp,\Sigma_\parallel) \rightarrow (8,8) h^{-1}\rm{Mpc}$. &
$1.007$&
$^{+0.029}_{-0.039}$&
$0.001$&
$^{+0.003}_{-0.003}$&
$0.001$&
$^{+0.032}_{-0.037}$&
$-0.002$&
$^{+0.003}_{-0.003}$&
92.06/90\\
\\[-1.5ex]
Fit w/ $\Sigma_s \rightarrow 0 h^{-1}\rm{Mpc}$. &
$1.005$&
$^{+0.031}_{-0.037}$&
$-0.002$&
$^{+0.002}_{-0.004}$&
$0.001$&
$^{+0.031}_{-0.034}$&
$-0.003$&
$^{+0.005}_{-0.004}$&
91.85/90\\
\\[-1.5ex]
Fit w/ $A_2(r)=poly2$. &
$1.007$&
$^{+0.031}_{-0.037}$&
$0.000$&
$^{+0.002}_{-0.001}$&
$0.005$&
$^{+0.032}_{-0.036}$&
$0.001$&
$^{+0.005}_{-0.005}$&
93.16/91\\
\\[-1.5ex]
Fit w/ $A_2(r)=poly4$. &
$1.006$&
$^{+0.030}_{-0.035}$&
$0.000$&
$^{+0.001}_{-0.001}$&
$0.002$&
$^{+0.036}_{-0.034}$&
$-0.000$&
$^{+0.007}_{-0.006}$&
91.06/89\\
\\[-1.5ex]
Fit w/ $30<r<200\hMpc$ range. &
$1.003$&
$^{+0.032}_{-0.037}$&
$-0.003$&
$^{+0.004}_{-0.005}$&
$0.000$&
$^{+0.032}_{-0.035}$&
$-0.003$&
$^{+0.005}_{-0.005}$&
106.04/104\\
\\[-1.5ex]
Fit w/ $70<r<200\hMpc$ range. &
$1.007$&
$^{+0.029}_{-0.036}$&
$0.000$&
$^{+0.002}_{-0.002}$&
$0.004$&
$^{+0.032}_{-0.040}$&
$-0.000$&
$^{+0.003}_{-0.003}$&
79.39/76\\
\\[-1.5ex]
Fit w/ $50<r<150\hMpc$ range. &
$1.005$&
$^{+0.030}_{-0.041}$&
$-0.000$&
$^{+0.004}_{-0.006}$&
$0.004$&
$^{+0.036}_{-0.043}$&
$-0.001$&
$^{+0.008}_{-0.007}$&
54.37/58\\
\hline
\multicolumn{10}{c}{Redshift Space with Reconstruction}\\
\hline
Fiducial $[f]$ &
$1.002$&
$^{+0.023}_{-0.022}$&
--&
--&
$0.007$&
$^{+0.023}_{-0.037}$&
--&
--&
92.68/90\\
\\[-1.5ex]
Fit w/ $(\Sigma_\perp,\Sigma_\parallel) \rightarrow (2,4) h^{-1}\rm{Mpc}$. &
$1.002$&
$^{+0.023}_{-0.021}$&
$-0.000$&
$^{+0.001}_{-0.001}$&
$0.008$&
$^{+0.022}_{-0.037}$&
$0.000$&
$^{+0.001}_{-0.001}$&
92.73/90\\
\\[-1.5ex]
Fit w/ $\Sigma_s \rightarrow 0 h^{-1}\rm{Mpc}$. &
$1.001$&
$^{+0.024}_{-0.020}$&
$-0.001$&
$^{+0.005}_{-0.003}$&
$0.004$&
$^{+0.020}_{-0.030}$&
$-0.003$&
$^{+0.007}_{-0.005}$&
92.27/90\\
\\[-1.5ex]
Fit w/ $A_2(r)=poly2$. &
$1.002$&
$^{+0.024}_{-0.022}$&
$0.000$&
$^{+0.001}_{-0.000}$&
$0.008$&
$^{+0.023}_{-0.037}$&
$0.001$&
$^{+0.002}_{-0.002}$&
94.23/91\\
\\[-1.5ex]
Fit w/ $A_2(r)=poly4$. &
$1.002$&
$^{+0.023}_{-0.022}$&
$0.000$&
$^{+0.001}_{-0.001}$&
$0.005$&
$^{+0.025}_{-0.037}$&
$-0.001$&
$^{+0.004}_{-0.004}$&
91.68/89\\
\\[-1.5ex]
Fit w/ $30<r<200\hMpc$ range. &
$1.003$&
$^{+0.022}_{-0.023}$&
$0.000$&
$^{+0.002}_{-0.001}$&
$0.006$&
$^{+0.022}_{-0.038}$&
$0.000$&
$^{+0.002}_{-0.002}$&
106.12/104\\
\\[-1.5ex]
Fit w/ $70<r<200\hMpc$ range. &
$1.002$&
$^{+0.023}_{-0.021}$&
$0.000$&
$^{+0.001}_{-0.001}$&
$0.006$&
$^{+0.022}_{-0.036}$&
$-0.000$&
$^{+0.002}_{-0.002}$&
79.99/76\\
\\[-1.5ex]
Fit w/ $50<r<150\hMpc$ range. &
$1.002$&
$^{+0.024}_{-0.023}$&
$-0.001$&
$^{+0.004}_{-0.004}$&
$0.008$&
$^{+0.022}_{-0.037}$&
$-0.001$&
$^{+0.005}_{-0.005}$&
54.65/58\\
\\[-1.5ex]
Recon. w/ $\beta \rightarrow 0.24$. &
$1.002$&
$^{+0.023}_{-0.022}$&
$-0.000$&
$^{+0.001}_{-0.001}$&
$0.005$&
$^{+0.023}_{-0.033}$&
$0.000$&
$^{+0.002}_{-0.003}$&
92.49/90\\
\\[-1.5ex]
Recon. w/ $\beta \rightarrow 0.36$. &
$1.002$&
$^{+0.022}_{-0.020}$&
$0.000$&
$^{+0.001}_{-0.002}$&
$0.005$&
$^{+0.024}_{-0.038}$&
$-0.000$&
$^{+0.002}_{-0.002}$&
92.89/90\\
\\[-1.5ex]
Recon. w/ $b \rightarrow 1.8$. &
$1.001$&
$^{+0.022}_{-0.021}$&
$-0.000$&
$^{+0.006}_{-0.005}$&
$0.006$&
$^{+0.025}_{-0.041}$&
$-0.000$&
$^{+0.006}_{-0.006}$&
92.61/90\\
\\[-1.5ex]
Recon. w/ $b \rightarrow 2.6$. &
$1.003$&
$^{+0.023}_{-0.023}$&
$0.001$&
$^{+0.004}_{-0.004}$&
$0.004$&
$^{+0.025}_{-0.036}$&
$-0.001$&
$^{+0.006}_{-0.005}$&
92.65/90\\
\\[-1.5ex]
Recon. w/ Wiener Filter. &
$1.004$&
$^{+0.020}_{-0.022}$&
$-0.000$&
$^{+0.004}_{-0.003}$&
$0.005$&
$^{+0.024}_{-0.035}$&
$-0.000$&
$^{+0.004}_{-0.004}$&
92.69/90\\
\\[-1.5ex]
Recon. on $\Omega_m=0.4$ case.$^{1}$ &
$0.832$&
$^{+0.021}_{-0.019}$&
$-0.171$&
$^{+0.010}_{-0.010}$&
$0.020$&
$^{+0.028}_{-0.037}$&
$0.017$&
$^{+0.014}_{-0.014}$&
92.61/90\\
\hline
\end{tabular}

\medskip
\flushleft
$^{1}$ $\alpha=1$ and $\epsilon=0$ in the LasDamas cosmology correspond to $\alpha=0.832$ and $\epsilon=0.013$ in this $\Omega_m=0.4$ cosmology according to Equations (\ref{eqn:daz} \& \ref{eqn:hz}).

\end{table*}

Next we test the robustness of our fitting model to changes in various
model parameters. A full list of these results are found in Table
\ref{tab:alphas} for changes of $\Sigma_\perp$, $\Sigma_\parallel$,
$\Sigma_s$, fitting range and form of $A_2(r)$ both before and after
reconstruction. In the table, $poly2$ corresponds to an $A_2(r) = a_1/r^2
+ a_2/r$ and $poly4$ corresponds to an $A_2(r) = a_1/r^2 + a_2/r + a_3 +
a_4r$. We see that the scatter in $\epsilon$ between the mocks can show
$\sim10\%$ variations quite often. This again indicates the noisiness
of our $\epsilon$ measurements.

Figure \ref{fig:compsfig} shows scatter plots in $\alpha$ and $\epsilon$
for a few sample cases. Here, $\Delta\alpha$ and $\Delta\epsilon$ are
the differences between the $\alpha$ and $\epsilon$ values measured
using the slightly altered model and the fiducial model. We expect
the average $\Delta\alpha$ and $\Delta\epsilon$ to be 0 within the
errors if our measurements of $\alpha$ and $\epsilon$ are consistent
between the two models. We see that in all cases, $\Delta\alpha=0$ and
$\Delta\epsilon=0$ fall within the scatter predicted by the quantiles on
a mock-by-mock basis. However, the errors on the average $\Delta\alpha$
and $\Delta\epsilon$ are on the order of $\sqrt{160}$ times smaller
than the scatter implied by the mocks. This indicates that in a few
cases, we detect a significant shift in the average value of $\alpha$
and $\epsilon$ measured.

In particular, this occurs in the pre-reconstruction cases where we have
changed the fitting range. The $\alpha$ and $\epsilon$ scatter plots
for the $30<r<200\hMpc$ fitting range versus the fiducial fitting range
cases are shown in Figure \ref{fig:compsfig_fr}. The plots are shown in
pairs with $\alpha$ on the left and $\epsilon$ on the right. The first
plot in each pair corresponds to the pre-reconstruction case and the
second plot to the post-reconstruction case. We see that on average,
the larger fitting range gives slightly smaller values of $\alpha$ and
$\epsilon$. If we begin fitting at $r=30\hMpc$ where the errorbars are
smaller, the fitter forces the model to match the data at these small
scales where we know the templates (especially the quadrupole) are not
faithful representations of the data. The $A_{0,2}(r)$ marginalization
terms compensate for this at the expense of accurately fitting the BAO
scale. After reconstruction, $\Delta\alpha$ and $\Delta\epsilon$ are
both 0 which suggests that the model is better matched to the data in
this case.

We also see average $\Delta\epsilon$ values that are
significantly different from zero in the $\Sigma_s=0\hMpc$ case
both before and after reconstruction. This is illustrated in Figure
\ref{fig:compsfig_ss}. $\Sigma_s=0\hMpc$ implies that we exclude FoG from
the model, which is unrealistic as it is implemented in the mocks. The
likely culprit here is again the mismatch between the data and the model
at small $r$ especially in the quadrupole. In addition, if we compare
the dotted line in Figure \ref{fig:ssfig} (the $\Sigma_s=0\hMpc$ case)
and the average quadrupole in Figure \ref{fig:simfig}, we see that
the quadrupole BAO feature in this model is a poorer fit to the data
overall. The mocks show more of a crest-trough-crest structure near the
BAO scale as in the fiducial parameter template (solid lines in Figure
\ref{fig:varyfig}) whereas the trough in the $\Sigma_s=0\hMpc$ case is
much weaker. This is further affirmed by the fact that the dotted line
in Figure \ref{fig:snlfig} (isotropic $\snl$) has a similar looking
BAO feature and shows a similar discrepancy in $\epsilon$ relative
to the fiducial model before reconstruction. The fitter can partially
compensate for these differences through adjusting the value of $\epsilon$
which also gives rise to crests and troughs near the BAO scale, although
with different structure than those introduced through $\Sigma_\perp$,
$\Sigma_\parallel$ and $\Sigma_s$. Hence one must pick a quadrupole
model that has a BAO feature fairly well matched to the data to avoid
biasing the $\epsilon$ values measured.

Despite these offsets in the median $\alpha$ and $\epsilon$ for different
fitting models, we note that at the statistical precision of current
datasets, we would not be able to detect any of these changes. For the DR7
mocks, the average $\sigma_\alpha$ is $\sim0.04$ before reconstruction
and $\sim0.03$ after reconstruction. The average $\sigma_\epsilon$
are even larger at $\sim0.05$ before reconstruction and $\sim0.04$ after
reconstruction. Hence, assuming that $\sigma_\alpha$ and $\sigma_\epsilon$
characterize the error on $\alpha$ and $\epsilon$, a $0.003$ shift will
fall entirely within the expected errors. Therefore, our fitting model
is reasonably robust against small changes to model parameters and our
measured $\alpha$ and $\epsilon$ values are largely unbiased.

\begin{figure}
\vspace{0.4cm}
\centering
\epsfig{file=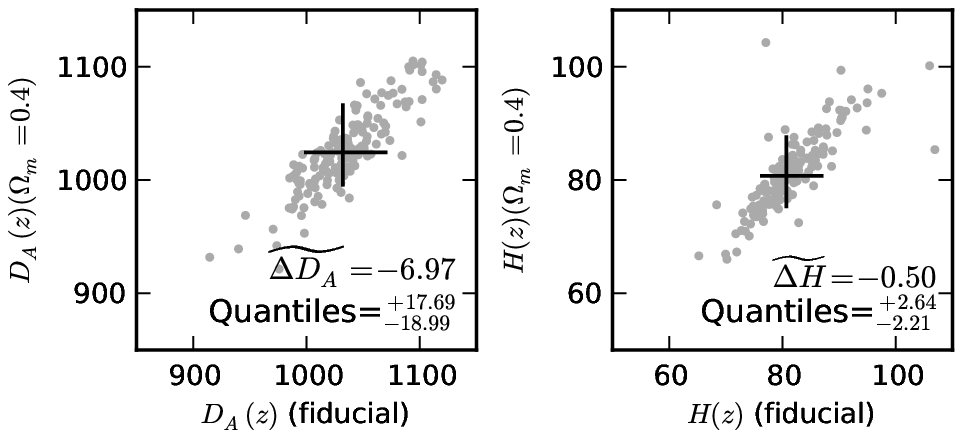, width=0.98\linewidth, clip=}
\caption{$D_A(z)$ and $H(z)$ scatter plots obtained by plotting those
measured using an $\Omega_m=0.4$ cosmology versus the true LasDamas
cosmology. $D_A(z)$ is in units of Mpc and $H(z)$ is in units of
km/s/Mpc. These values were calculated using Equations (\ref{eqn:daz}) \&
(\ref{eqn:hz}) and assuming $r_s=159.71$ Mpc. The median $\Delta D_A$ and
$\Delta H$ values are significantly different from 0 when approximating
the error on the median as the scatter predicted by the quantiles
divided by $\sqrt{160}$. Such a discrepancy may be due to our median
redshift not being exactly $z=0.35$ as assumed. It could also be due to
the breakdown of Taylor assumptions made in deriving our fitting model;
in the $\Omega_m=0.4$ cosmology, our measured $\alpha$ and $\epsilon$
values deviate substantially from 1 and 0 as shown in the last row of
Table \ref{tab:alphas}. We again emphasize that while this difference
is detectable in the median $\alpha$ and $\epsilon$ of the mocks, it is
not significant in each individual mock.
\label{fig:om40fig}}
\end{figure}

\begin{figure}
\vspace{0.4cm}
\centering
\epsfig{file=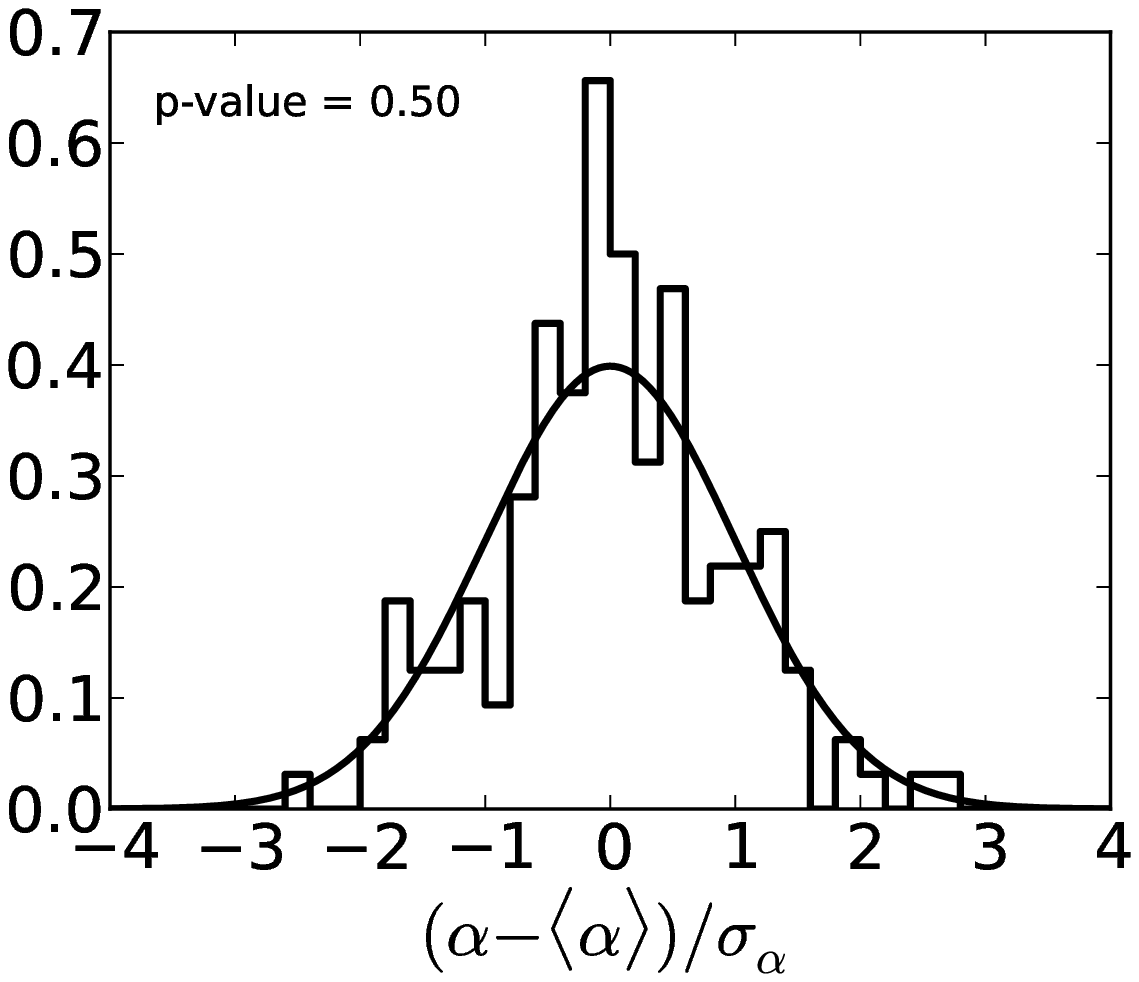, width=0.8\linewidth, clip=}
\epsfig{file=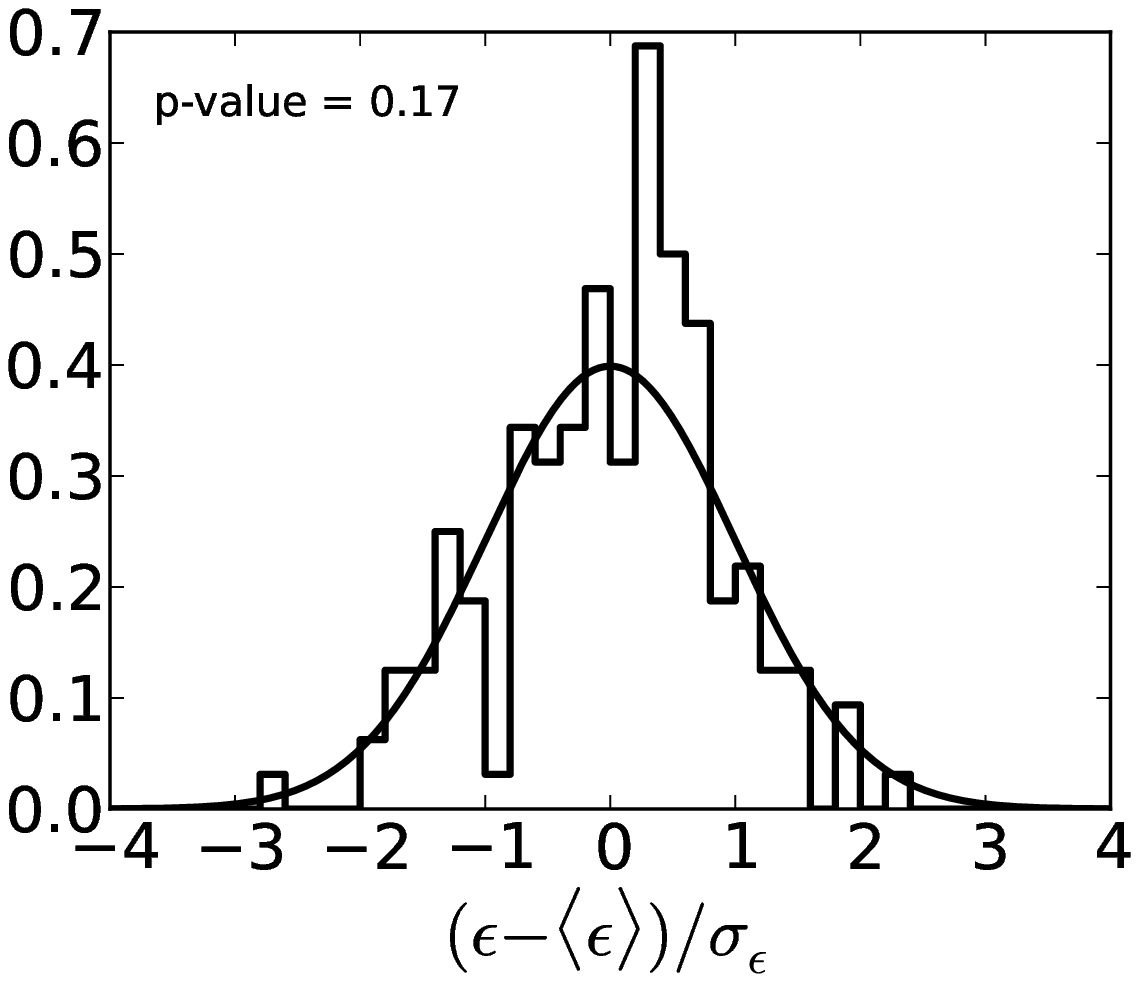, width=0.8\linewidth, clip=}
\caption{Histograms of $(\alpha - \langle\alpha\rangle)/\sigma_\alpha$
(top) and $(\epsilon - \langle\epsilon\rangle)/\sigma_\epsilon$ (bottom)
after reconstruction. These are a measure of the signal-to-noise
of our measured $\alpha$ and $\epsilon$ values. The overplotted
black lines correspond to the unit normal. We perform a K-S test
to see how likely these distributions are drawn from a unit normal
distribution. The p-values or probabilities are indicated on the
plots and imply that $\alpha$ and $\epsilon$ both have finite chances
of being drawn from Gaussian distributions. This verifies that the
standard deviations $\sigma_\alpha$ and $\sigma_\epsilon$ we calculate
from $\chi^2(\alpha,\epsilon)$ characterize the errors on $\alpha$ and
$\epsilon$ reasonably well. A similar conclusion holds for our $\alpha$
and $\epsilon$ values before reconstruction.
\label{fig:snrfig}}
\end{figure}

We perform similar exercises for various different reconstruction
parameters such as the bias and $\beta$ values we input to the
algorithm. The fiducial reconstruction parameters we use are $b=2.2$
and $\beta=0.3$. We also tested using simple Wiener filtering to
interpolate between masked regions. This differs from the constrained
Gaussian realizations in that unobserved modes are set to 0 instead of
being drawn from a fiducial power spectrum. Table \ref{tab:alphas} shows
the results of these tests. It indicates that the unobserved modes do not
affect our measurements of $\alpha$ and $\epsilon$ given our statistical
precision, and that our fitting model effectively marginalizes away
any broadband signal that reconstruction introduces when incorrect
values of the fiducial parameters are used. Higher precision studies of
possible systematics from reconstruction due to survey boundaries will
be necessary for future surveys. However, this goes beyond the scope
and goals of this current paper.

Finally we calculate and perform our fits using a fiducial cosmology
that is significantly different to the LasDamas cosmology. This forces
a stronger anisotropic BAO signal to appear in the quadrupole. We pick a
cosmology with $\Omega_m=0.4$ that preserves the matter-to-baryon ratio
of LasDamas. We also fix $\Omega_m h^2$ which implies $h=0.553$. We
convert the measured $\alpha$ and $\epsilon$ values to $D_A(z)$ and
$H(z)$ using Equations (\ref{eqn:daz}) \& (\ref{eqn:hz}) and compare
these to the values measured using the fiducial cosmology. This is
illustrated in Figure \ref{fig:om40fig}. The equations listed above only
allow us to infer $D_A(z)/r_s$ and $H(z)r_s$. We have assumed $r_s =
159.71$ Mpc, which is the sound horizon in the LasDamas cosmology,
to obtain the $D_A(z)$ and $H(z)$ values plotted in the figure. In
the LasDamas cosmology (which is the true cosmology in our mocks),
$D_A(z) = 1032$ Mpc and $H(z) = 81.8$ km/s/Mpc at $z=0.35$. Taking the
ratio $\widetilde{\Delta D_A}/ D_A(z)$ and $\widetilde{\Delta H}/H(z)$
implies that on average our measurements of $D_A(z)$ and $H(z)$ using
the $\Omega_m=0.4$ cosmology and the true LasDamas cosmology differ
by $\sim0.7\%$ and $\sim0.6\%$ respectively. Dividing the scatter
indicated by the quantiles by $\sqrt{160}$ suggests that these average
offsets are significant, although again, in a single mock, we would not
be able to detect these offsets. The fiducial $D_A(z)$ and $H(z)$ are
calculated assuming a median redshift of $z=0.35$; however, if the true
median redshift were slightly different, such discrepancies would not
be unexpected. In addition, our models for $\alpha$ and $\epsilon$ are
based on Taylor expansions around 1 and 0 respectively. When the fitting
model is constructed using a fiducial cosmology that is extremely wrong,
the $\alpha$ and $\epsilon$ values we measure will deviate significantly
from 1 and 0 as shown in the last row of Table \ref{tab:alphas}. Our
first-order Taylor assumption may be breaking down at this point,
further affecting our measurements. In this case, one can iteratively
change the fiducial cosmology and re-fit for $\alpha$ and $\epsilon$
until values closer to 1 and 0 are obtained.

We verify our assumption that the second moments of $p(\alpha)$
and $p(\epsilon)$ are good indicators of the errors on $\alpha$
and $\epsilon$ in Figure \ref{fig:snrfig}. The top panel of this
figure shows a normalized histogram of $(\alpha - \langle \alpha
\rangle)/\sigma_\alpha$ after reconstruction and the bottom panel shows
the analogue for $\epsilon$. The unit normal is overplotted. We perform
K-S tests on these distributions and list the p-values in the plots. These
give the probability that the plotted distribution is drawn from a
unit normal. One can see that the post-reconstruction p-values are 0.50
and 0.17 for $\alpha$ and $\epsilon$ respectively. For comparison, the
corresponding pre-reconstruction p-values are 0.40 and 0.19. These values
indicate that there are finite probabilities that $\alpha$ and $\epsilon$
have Gaussian posteriors. Hence the standard deviations $\sigma_\alpha$
and $\sigma_\epsilon$ we calculate from $\chi^2(\alpha,\epsilon)$
characterize the errors on $\alpha$ and $\epsilon$ reasonably well.

\section{DR7 Results} \label{sec:datares}

\subsection{Anisotropic Results}

Now that we have verified the robustness of our techniques and obtained
a better understanding of the anisotropic signal from our mocks, we
can proceed to the actual SDSS DR7 LRG data. To calculate our fitting
model for the data, we use the flat $\Lambda$CDM cosmology predicted by
WMAP7: $H_0 = 70.2 \pm 1.4$ km/s/Mpc, $\Omega_b h^2 = 0.02255 \pm 0.054$,
$\Omega_c h^2=0.1126 \pm 0.0036$, $n_s = 0.968 \pm 0.012$ and $\sigma_8 =
0.816 \pm 0.024$ \citep{Kea11}. For the covariance matrix, we again use
the modified Gaussian covariance matrix discussed in \S\ref{sec:fitting}
with the modification parameters derived from the mocks and the WMAP7
cosmology.

\begin{figure*}
\vspace{0.4cm}
\centering
\begin{tabular}{c}
\epsfig{file=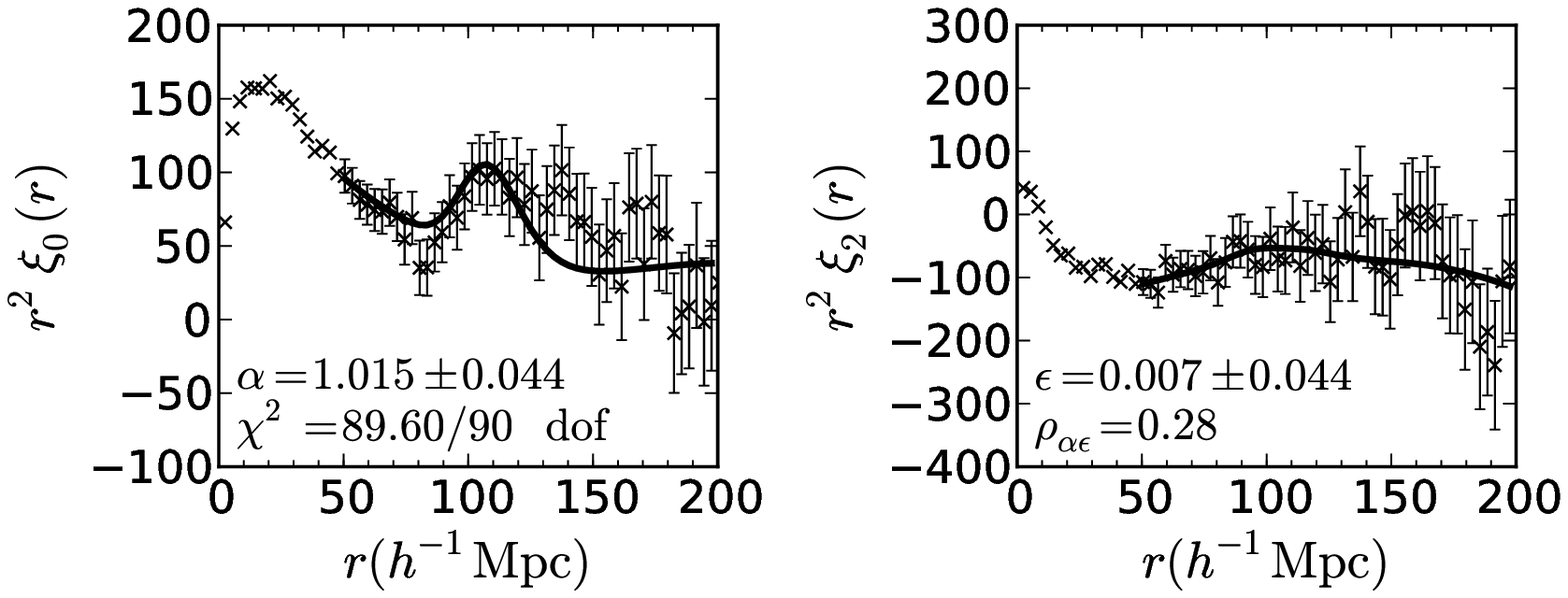,width=0.9\linewidth,clip=}\\
\epsfig{file=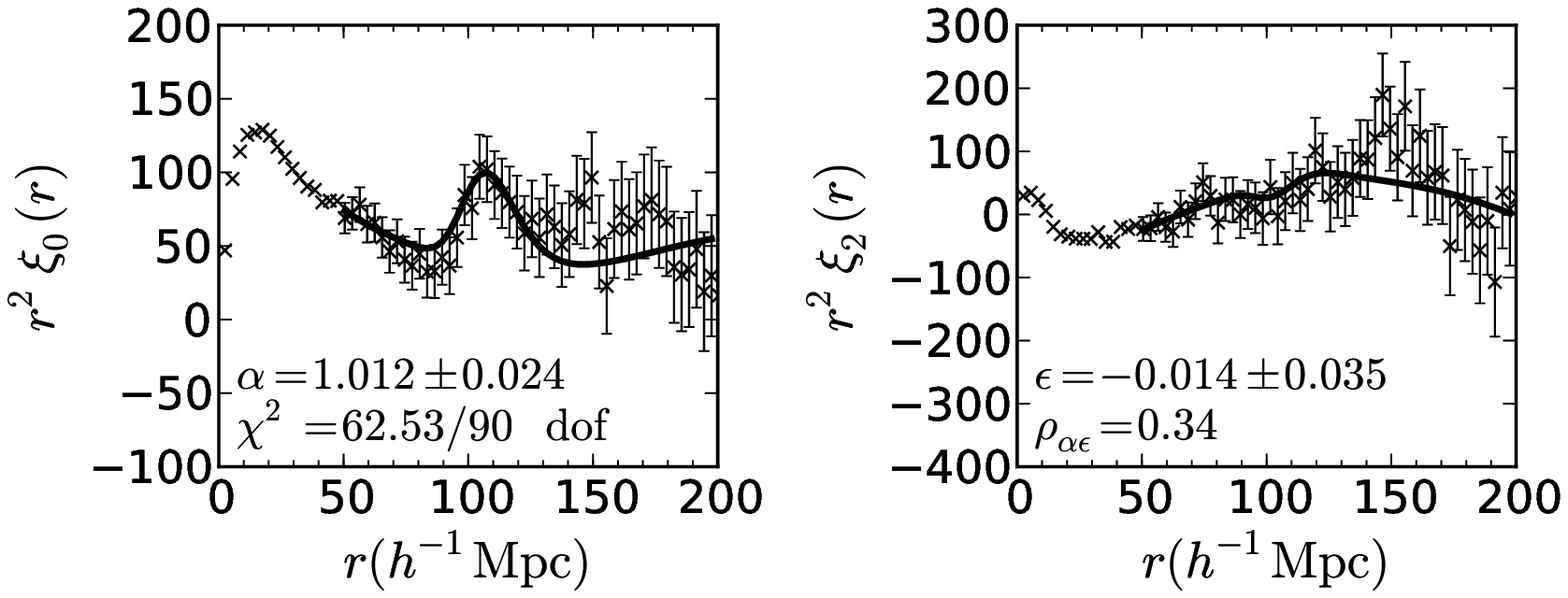,width=0.9\linewidth,clip=}
\end{tabular}
\caption{DR7 fit results before (top row) and after (bottom row)
reconstruction. These values imply a 3.6\% measurement of $D_A(z)$
and an 8.3\% measurement of $H(z)$ after reconstruction. We see that
the acoustic peak has sharpened up significantly after reconstruction
as expected. The error on $\alpha$ decreases by a factor of 1.8 and
the error on $\epsilon$ decreases by a factor of 1.3 as a result. The
quadrupole is nearly 0 at $\sim100\hMpc$ after reconstruction, indicating
the effectiveness of our Kaiser correction. The deviation from 0 at
larger $r$ is likely some anisotropy introduced by reconstruction.
\label{fig:dr7fig}}
\end{figure*}

The results of our fits are shown in Figure \ref{fig:dr7fig}. The
pre-reconstruction results are in the top row and the post-reconstruction
results are in the bottom row. The acoustic peak appears much sharper
after reconstruction, again indicating the effectiveness of our
technique in undoing non-linear evolution. This is reflected in the
decrease in error on $\alpha$ and $\epsilon$ by factors of 1.8 and 1.3
respectively after reconstruction. The quadrupole near $100\hMpc$ scales
is much closer to 0 after reconstruction which implies that our partial
removal of the Kaiser effect was also successful. The deviation from 0
at larger scales again indicates that reconstruction is introducing some
additional anisotropy. This is as expected based on our analysis of the
mock catalogues in \S\ref{sec:mockres} (see Figure \ref{fig:realfig}).

\begin{figure*}
\vspace{0.4cm}
\centering
\begin{tabular}{cc}
\epsfig{file=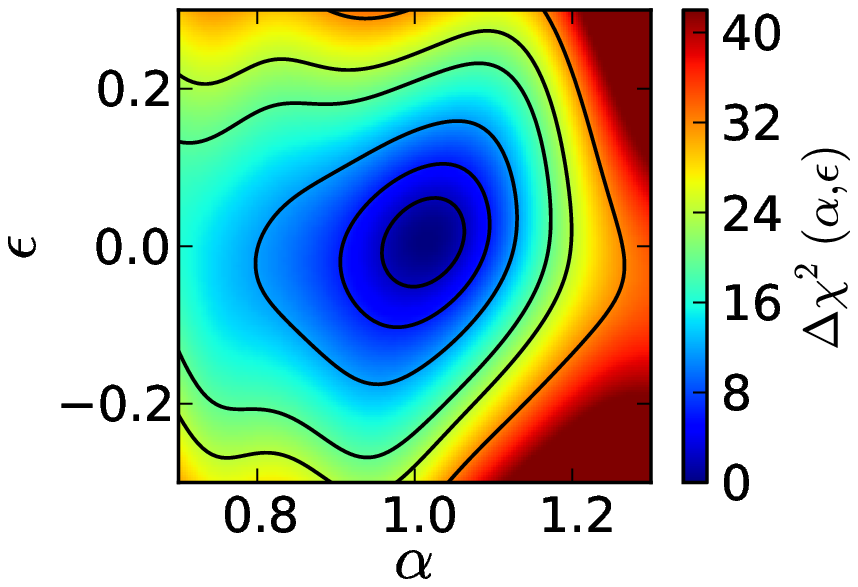,width=0.355\linewidth,clip=}&
\epsfig{file=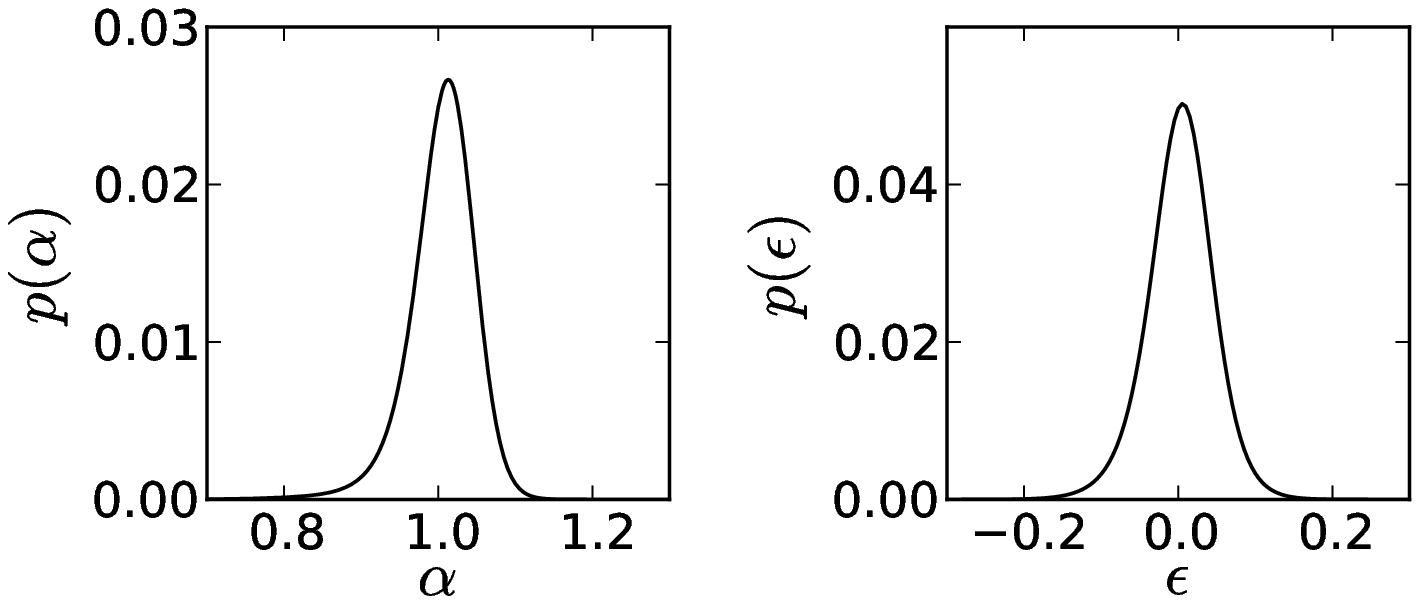,width=0.6\linewidth,clip=}\\
\epsfig{file=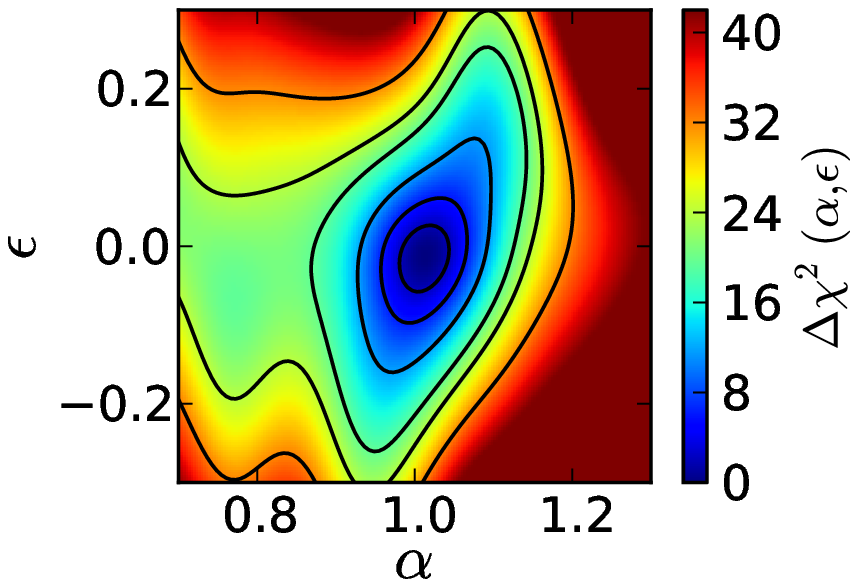,width=0.355\linewidth,clip=}&
\epsfig{file=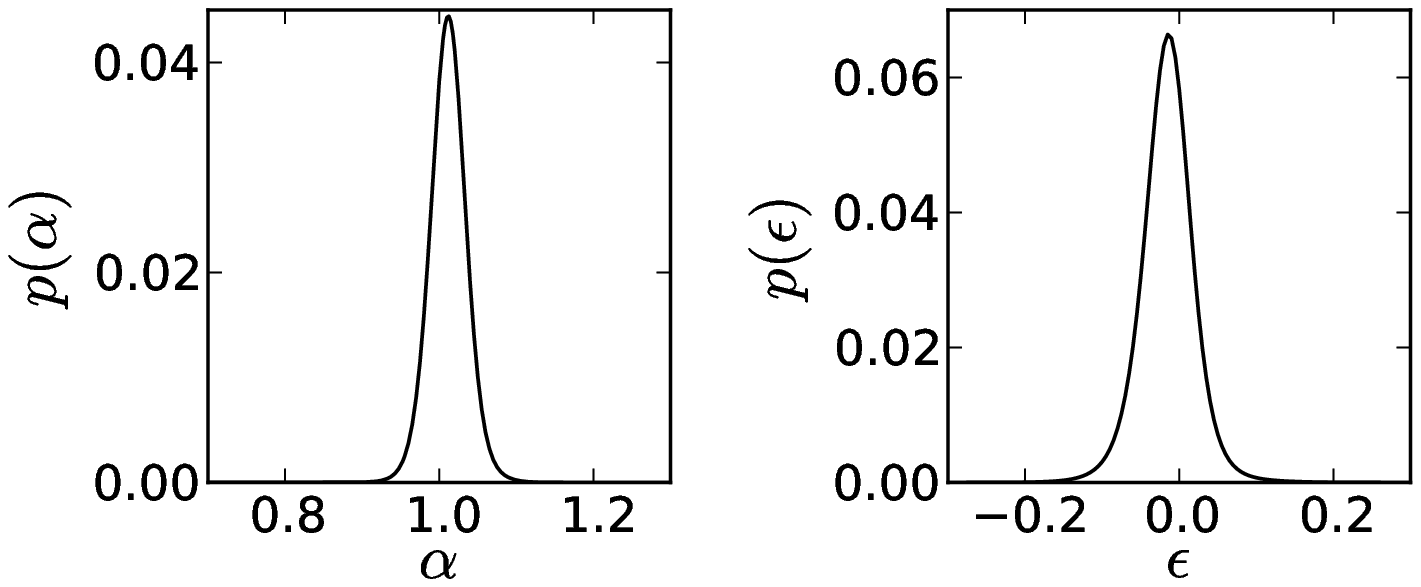,width=0.6\linewidth,clip=}\\
\end{tabular}
\caption{The $\Delta\chi^2({\alpha,\epsilon})$ distribution (column 1)
and the derived $p(\alpha)$ and $p(\epsilon)$ distributions (columns
2 \& 3) for DR7. The pre-reconstruction results are in the top row and
the post-reconstruction results are in the bottom row. Contour levels
corresponding to 1-6$\sigma$ for a 2D distribution are overplotted. We
apply a 0.15 prior in $\log(\alpha)$ to suppress the unphysical downturn
at low $\alpha$ which corresponds to the acoustic peak being pushed out
to large $r$. The errorbars are much larger here and the fitter has an
easier time hiding the peak inside the errors. The plateauing of the
distribution at small $\alpha$ is a result of this. We see that after
reconstruction, the contours become much tighter. This corresponds to the
tightening of $p(\alpha)$ and $p(\epsilon)$ after reconstruction seen in
columns 2 \& 3 due to the sharpening up of the acoustic peak. We also
see that $p(\alpha)$ and $p(\epsilon)$ are nearly Gaussian and hence
the second moments $\sigma_\alpha$ and $\sigma_\epsilon$ characterize
the errors on $\alpha$ and $\epsilon$ well.
\label{fig:gridfig}}
\end{figure*}

The first column of Figure \ref{fig:gridfig} shows the
$\Delta\chi^2(\alpha,\epsilon)=\chi^2({\alpha,\epsilon})-\chi^2_{min}$
distribution measured at various grid points in $\alpha$ and
$\epsilon$. As described in \S\ref{sec:fitting}, our $\alpha$ grid points
are separated by 0.0025 in the range $0.7<\alpha<1.3$ and our $\epsilon$
grid points are separated by 0.005 in the range $-0.3<\epsilon<0.3$. The
1 through 6$\sigma$ confidence levels for a 2D distribution are
overplotted. The pre-reconstruction results are shown in the top row
and the post-reconstruction results are shown in the bottom row. As we
go to smaller $\alpha$, the acoustic peak in the model is being pushed
out to larger scales where the errorbars are larger. Hence it is much
easier for the fitter to hide the peak within the errors. Although we
have applied a 0.15 prior in $\log(\alpha)$ to suppress this unphysical
downturn in $\chi^2$, the distribution still plateaus in this region. One
can also see that after reconstruction, the $\chi^2({\alpha,\epsilon})$
distribution is much tighter at the center, indicating that the best-fit
values are much better measured. This corresponds to the smaller errorbars
we see in $\alpha$ and $\epsilon$ after reconstruction.

The second and third columns of Figure \ref{fig:gridfig} show the $\alpha$
and $\epsilon$ probability distributions derived from the $\chi^2$
grid. One can see that both of these are fairly Gaussian so we can
quantify the errors on $\alpha$ and $\epsilon$ as the second moments of
these distributions, $\sigma_\alpha$ and $\sigma_\epsilon$. These values
are summarized in Table \ref{tab:dr7fin} for both before (top row) and
after (bottom row) reconstruction. The smaller standard deviations after
reconstruction accompany the sharpening up of the acoustic peak. This
corresponds to the tightening of the contours in the $\chi^2$ distribution
shown in column 1.

Our measured $C_{\alpha\epsilon}$, $\sigma_\alpha$ and $\sigma_\epsilon$
imply correlation coefficients of $\rho_{\alpha\epsilon}=0.28$ and
$0.34$ before and after reconstruction. These values are slightly larger
than the expected $\rho_{\alpha\epsilon}\sim0.21$ from Fisher matrix
arguments. However, given the large rms of $\rho_{\alpha\epsilon}$ from
the mocks of $\sim0.35$ both before and after reconstruction, our DR7
results are not significantly different from the Fisher matrix prediction.

\begin{figure}
\vspace{0.4cm}
\centering
\epsfig{file=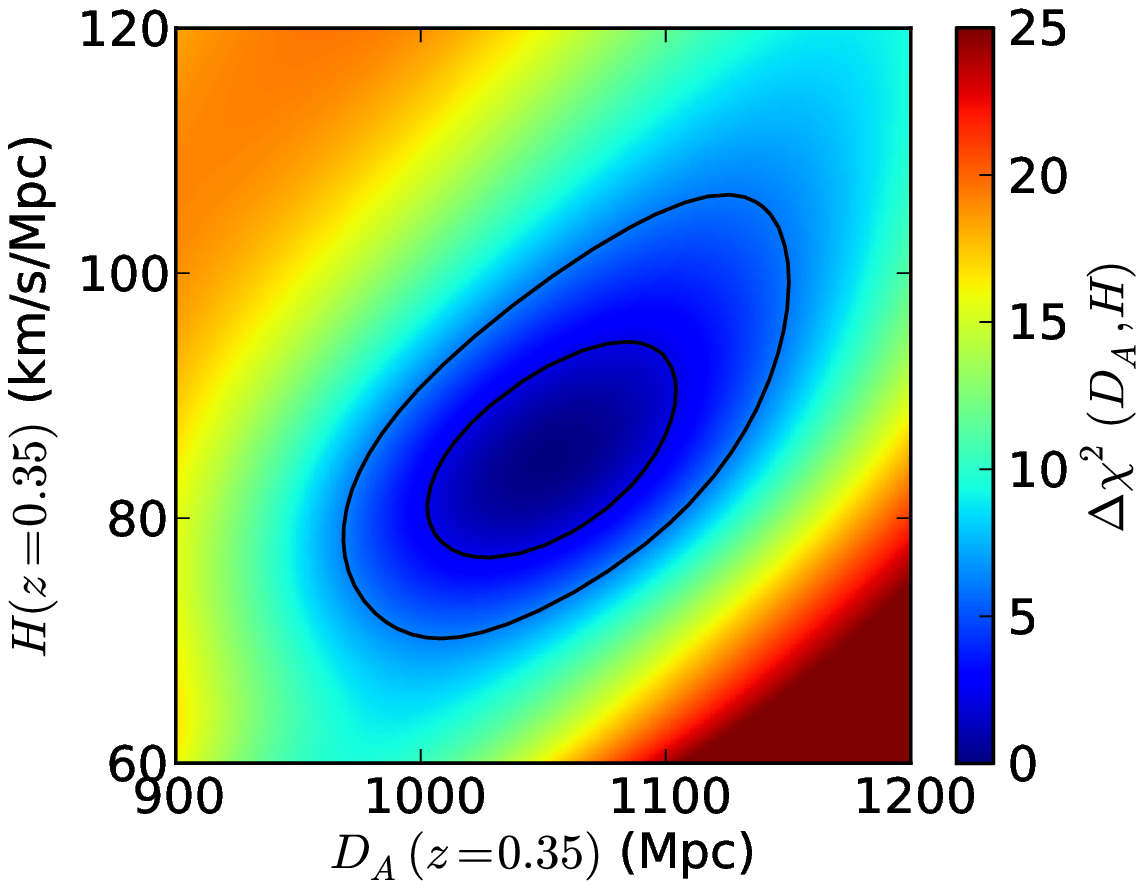,width=0.9\linewidth,clip=}
\caption{The post-reconstruction SDSS DR7
$\Delta\chi^2(D_A,H)=\chi^2(D_A,H)-\chi^2_{min}$ distribution with
1 and 2$\sigma$ contours overplotted. We measure $D_A(z)=1050\pm38$
Mpc and $H(z)=84.4\pm7.0$ km/s/Mpc at $z=0.35$. The tilted elliptical
contours clearly indicate a correlation between our $D_A$ and $H$
measurements. The correlation coefficient is $\rho_{D_AH}=0.57$.
\label{fig:dahgrid}}
\end{figure}

\begin{table*}
\centering
\caption{Summary of key measurements from DR7 data. Columns 2 and 3
list the $\alpha$ and $\epsilon$ values we measure. Column 4 lists the
covariance between $\alpha$ and $\epsilon$ while column 5 lists their
correlation coefficient. Columns 6 and 7 list the distance constraints
we obtain to $z=0.35$ from our measured $\alpha$ and $\epsilon$
values. Columns 8 and 9 translate these relative distance measures into
more tangible quantities assuming $r_s=152.76$ Mpc as in the WMAP7
cosmology. The pre-reconstruction results are listed in the top row
(Before) and the post-reconstruction results are listed in the bottom row
(After).
\label{tab:dr7fin}}
\begin{tabular}{lccccccccc}
\hline
&$\alpha$&
$\epsilon$&
$C_{\alpha\epsilon}$&
$\rho_{\alpha\epsilon}$&
$D_A(z)/r_s$&
$H(z)r_s$&
$D_A(z)$&
$H(z)$&
$\rho_{D_AH}$\\
&&&&&&(km/s)&(Mpc)&(km/s/Mpc)\\
\hline
Before&
$1.015\pm0.044$&
$0.007\pm0.044$&
0.00054&
0.28&
$6.751\pm0.352$&
$12339\pm1330$&
$1031\pm54$&
$80.8\pm8.7$&
0.26\\
After&
$1.012\pm0.024$&
$-0.014\pm0.035$&
0.00029&
0.34&
$6.875\pm0.246$&
$12895\pm1070$&
$1050\pm38$&
$84.4\pm7.0$&
0.57\\
\hline
\end{tabular}
\end{table*}

Using our measured $\alpha$, $\sigma_\alpha$, $\epsilon$,
$\sigma_\epsilon$ and $C_{\alpha\epsilon}$ values from the DR7 data,
we can use Equations (\ref{eqn:daz}), (\ref{eqn:hz}), (\ref{eqn:edaz})
\& (\ref{eqn:ehz}) to determine $D_A(z=0.35)$ and $H(z=0.35)$. These
results are again summarized in Table \ref{tab:dr7fin}. In
addition, Figure \ref{fig:dahgrid} shows the post-reconstruction
$\Delta\chi^2(D_A,H)=\chi^2(D_A,H)-\chi^2_{min}$ contour plot obtained
from the $\alpha$-$\epsilon$ grid in Figure \ref{fig:gridfig}. The
values listed in the table and plotted in the figure were obtained
assuming $r_s=152.76$ Mpc and using the fiducial WMAP7 values:
$D_{A,f}(z=0.35)/r_{s,f}=6.69$ and $H_f(z=0.35)r_{s,f}=12689$ km/s. We
see that post-reconstruction we have a $\sim3.6\%$ measurement of
$D_A(z)$ and $\sim8.3\%$ measurement of $H(z)$ from SDSS DR7. Note
that our measures of $D_A$ and $H$ are correlated; this is also clearly
evident from the contour plot. Before reconstruction, $\rho_{D_AH}=0.26$
and after reconstruction $\rho_{D_AH}=0.57$. These are similar to the
$\rho_{D_AH}\sim0.4$ predicted by \citet{SE07}.

\begin{figure}
\vspace{0.4cm}
\centering
\epsfig{file=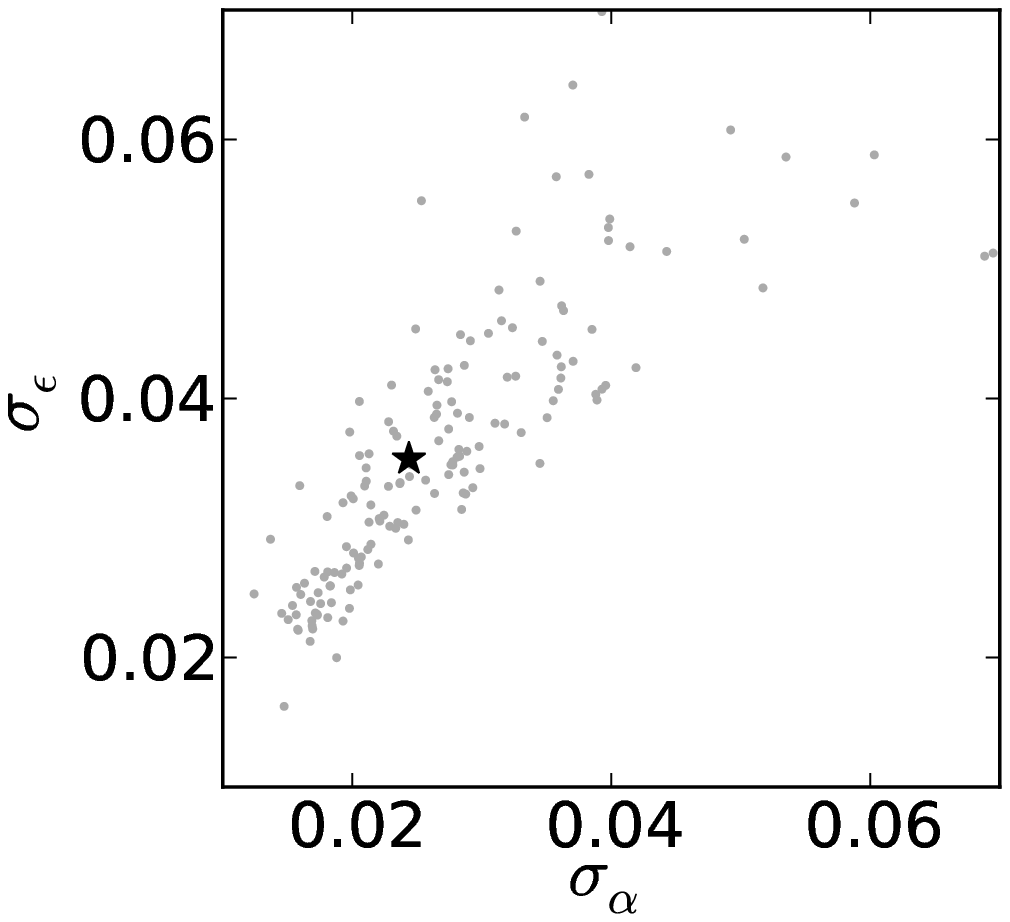, width=0.8\linewidth, clip=}
\caption{Post-reconstruction $\sigma_\alpha$ versus $\sigma_\epsilon$
for the mocks with the DR7 point overplotted as the black star. Note that
the mock points are identical to Figure \ref{fig:sasefig}. We see that
our DR7 measurement falls nicely within the locus of mock points. The
DR7 $\sigma_\epsilon/(1+\epsilon)$-to-$\sigma_\alpha/\alpha$ ratio is
$\sim1.0$ before reconstruction and $\sim1.5$ after reconstruction,
roughly consistent with the Fisher matrix prediction of $\sim1.2$.
\label{fig:sase_dr7}}
\end{figure}

We see that our DR7 $\sigma_\alpha$ and $\sigma_\epsilon$
measurements fall nicely within the locus of mock
points as shown in Figure \ref{fig:sase_dr7} for the
post-reconstruction case. Note that the mock results shown
in this figure are identical to Figure \ref{fig:sasefig}. The
$\sigma_\epsilon/(1+\epsilon)$-to-$\sigma_\alpha/\alpha$ ratio we obtain
is $\sim1.0$ before reconstruction and $\sim1.5$ after reconstruction
which is roughly consistent with the Fisher matrix prediction of
$\sim1.2$. Lastly, we note that our $\sigma_H/H$-to-$\sigma_{D_A}/D_A$
ratio is $\sim2$, consistent with the predictions of \citet{SE07}
and the assumption that went into our Fisher matrix predictions (see
Appendix \ref{app:fisher}).

\begin{table*}
\caption{DR7 fitting results for various models. The model is given in column 1. The measured $\alpha$ values are given in column 2 and the measured $\epsilon$ values are given in column 3. The $\chi^2$/dof is given in column 4.}
\label{tab:dr7_alphas}

\begin{tabular}{@{}lccc}

\hline
Model&
$\alpha$&
$\epsilon$&
$\chi^2/dof$\\

\hline
\multicolumn{4}{c}{Redshift Space without Reconstruction}\\
\hline

Fiducial $[f]$ &
$1.015 \pm 0.044$&
$0.007 \pm 0.044$&
89.60/90\\
\\[-1.5ex]
Fit w/ $(\Sigma_\perp,\Sigma_\parallel) \rightarrow (8,8) h^{-1}\rm{Mpc}$. &
$1.012 \pm 0.045$&
$0.009 \pm 0.042$&
89.77/90\\
\\[-1.5ex]
Fit w/ $\Sigma_s \rightarrow 0 h^{-1}\rm{Mpc}$. &
$1.018 \pm 0.040$&
$0.007 \pm 0.037$&
89.60/90\\
\\[-1.5ex]
Fit w/ $A_2(r)=poly2$. &
$1.018 \pm 0.043$&
$0.013 \pm 0.044$&
91.42/91\\
\\[-1.5ex]
Fit w/ $A_2(r)=poly4$. &
$1.015 \pm 0.044$&
$0.006 \pm 0.045$&
89.58/89\\
\\[-1.5ex]
Fit w/ $30<r<200\hMpc$ range. &
$1.018 \pm 0.039$&
$0.004 \pm 0.042$&
105.03/104\\
\\[-1.5ex]
Fit w/ $70<r<200\hMpc$ range. &
$1.016 \pm 0.050$&
$0.008 \pm 0.049$&
82.43/76\\
\\[-1.5ex]
Fit w/ $50<r<150\hMpc$ range. &
$1.019 \pm 0.042$&
$0.001 \pm 0.046$&
47.10/58\\
\hline
\multicolumn{4}{c}{Redshift Space with Reconstruction}\\
\hline
Fiducial $[f]$ &
$1.012 \pm 0.024$&
$-0.014 \pm 0.035$&
62.53/90\\
\\[-1.5ex]
Fit w/ $(\Sigma_\perp,\Sigma_\parallel) \rightarrow (2,4) h^{-1}\rm{Mpc}$. &
$1.012 \pm 0.025$&
$-0.014 \pm 0.036$&
62.48/90\\
\\[-1.5ex]
Fit w/ $\Sigma_s \rightarrow 0 h^{-1}\rm{Mpc}$. &
$1.013 \pm 0.021$&
$-0.013 \pm 0.027$&
61.83/90\\
\\[-1.5ex]
Fit w/ $A_2(r)=poly2$. &
$1.013 \pm 0.025$&
$-0.011 \pm 0.036$&
65.61/91\\
\\[-1.5ex]
Fit w/ $A_2(r)=poly4$. &
$1.013 \pm 0.025$&
$-0.011 \pm 0.036$&
61.92/89\\
\\[-1.5ex]
Fit w/ $30<r<200\hMpc$ range. &
$1.014 \pm 0.023$&
$-0.013 \pm 0.033$&
68.39/104\\
\\[-1.5ex]
Fit w/ $70<r<200\hMpc$ range. &
$1.012 \pm 0.027$&
$-0.016 \pm 0.040$&
54.50/76\\
\\[-1.5ex]
Fit w/ $50<r<150\hMpc$ range. &
$1.017 \pm 0.023$&
$-0.009 \pm 0.033$&
31.95/58\\
\\[-1.5ex]
Recon. w/ $\beta \rightarrow 0.24$. &
$1.014 \pm 0.024$&
$-0.016 \pm 0.034$&
68.77/90\\
\\[-1.5ex]
Recon. w/ $\beta \rightarrow 0.36$. &
$1.013 \pm 0.024$&
$-0.013 \pm 0.035$&
67.05/90\\
\\[-1.5ex]
Recon. w/ $b \rightarrow 1.8$. &
$1.014 \pm 0.025$&
$-0.017 \pm 0.035$&
66.75/90\\
\\[-1.5ex]
Recon. w/ $b \rightarrow 2.6$. &
$1.015 \pm 0.024$&
$-0.012 \pm 0.034$&
77.09/90\\
\\[-1.5ex]
Recon. w/ Wiener Filter. &
$1.012 \pm 0.025$&
$-0.014 \pm 0.035$&
61.23/90\\
\hline
\end{tabular}

\end{table*}

We again test the robustness of our $\alpha$ and $\epsilon$
measurements to our fitting model. The results are listed in
Table \ref{tab:dr7_alphas}. We see that our $\alpha$ and $\epsilon$
measurements are always consistent with the results obtained using the
fiducial fitting parameters. Our $\sigma_\alpha$ and $\sigma_\epsilon$
measurements show $\sim10\%$ variations which are consistent with the
differences in scatter between the various cases seen in the mocks.

We also test the robustness of our reconstruction technique by varying
the input parameters and then re-performing our fits. The $\alpha$
and $\epsilon$ values we measure from these tests are also listed in
Table \ref{tab:dr7_alphas}. Again we see very consistent $\alpha$,
$\epsilon$, $\sigma_\alpha$ and $\sigma_\epsilon$ values between the
various cases. This indicates that our measurements of the acoustic
scale and anisotropy are robust against reconstruction parameters.

\subsection{Comparison with Past Works} \label{sec:past}

Three past papers, \citet{GCH11}, \citet{CW11} and \citet{CW12}, have
performed anisotropic BAO analyses using the DR7 LRG sample. Of these,
\citet{GCH11} found a peak in the clustering along the line-of-sight that
they interpret as a detection of the acoustic peak, but \citet{Kea10b}
show that the peak is consistent with the expected noise.

\citet{CW11}, hereafter CW12a, measure $D_A(z=0.35)$ and $H(z=0.35)$ by
fitting the 2D correlation function of the LRGs before reconstruction,
whereas \citet{CW12}, hereafter CW12b, fit the monopole and
quadrupole. The former measures $D_A(z=0.35)=1048^{+60}_{-58}$
Mpc and $H(z=0.35)=82.1^{+4.8}_{-4.9}$ km/s/Mpc at $z=0.35$
while the latter measures $D_A(z=0.35)=1057^{+88}_{-87}$ Mpc and
$H(z=0.35)=79.6^{+8.3}_{-8.7}$ km/s/Mpc. These values are all consistent
with our pre-reconstruction measurements, however, the magnitudes of the
errors are slightly different. Since our treatments of the covariance
matrix differ, this is not surprising.

A significant difference between our analysis and that of CW12a and CW12b
is that they use a Markov Chain Monte Carlo approach over the parameter
space $\{D_A(z), H(z), \beta, \Omega_mh^2, \Omega_bh^2, n_s, \Sigma_s,
\Sigma_{nl}\}$ to derive their $D_A$ and $H$ measurements. Both $\beta$
and $\Sigma_s$ measure anisotropy due to redshift-space distortions
in the broadband correlation function which we do not attempt in our
analysis. The method we employ uses the $A(r)$ term to marginalize
out broadband information and focuses only on using the anisotropic
information in the BAO. In addition, the FoG model employed by CW12a and
CW12b is roughly the square-root of our model. Recall that FoG arises
from a perceived change in a galaxy's cosmological redshift due to its
peculiar motion along the line-of-sight. We assume that the peculiar
velocity distribution within a halo is exponential and convolve this
with the density field directly, thus yielding the power of two in our
FoG model for the power spectrum. In the model used by CW12a and CW12b,
the galaxy pair-wise velocity is assumed to be exponentially distributed
which effectively results in the correlation function being convolved
with an exponential. This model is also well motivated as discussed in
detail in \citet{H98} and \citet{Aea08}.

\citet{RW11} has shown that accurately modeling broadband redshift-space
distortions in dark matter-only simulations is a challenging theory
problem and requires going beyond simple $\Sigma_s$ and $\beta$
parameterizations. They demonstrate that neglecting bispectrum and higher
order terms from the real to redshift space transformation results in
inaccurate models. However, \citet{Aea08} have shown that the model
used in CW12a and CW12b faithfully describes the clustering of particular
populations of galaxies. Further work will be necessary to determine
the robustness of such models across galaxy types, especially as the
statistical precision of the measurements increase. By comparison, this
work attempts to avoid these uncertainties by focusing only on the BAO
feature, at the cost of not using all the available information. It is
conceivable that any of these differences may give rise to the error
discrepancy seen between our results and those of CW12a and CW12b.

\begin{figure}
\vspace{0.4cm}
\centering
\epsfig{file=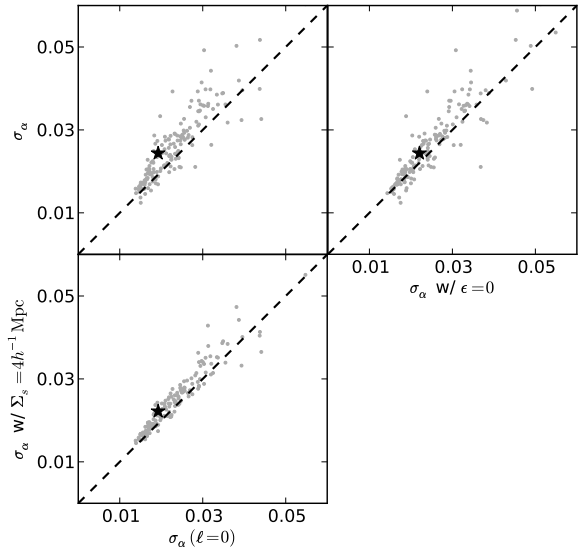,width=0.9\linewidth,clip=}
\caption{The $\sigma_\alpha$ values measured from the post-reconstruction
mocks for various fitting models. The DR7 results are overplotted as
the black stars and fall within the locus of mock points. (top left) The
$\sigma_\alpha$ values measured through the full monopole+quadrupole fits
versus the monopole-only results of \citet{Xea12}. One can see that the
full fits have $\sigma_\alpha$ that are larger on average. There is also
considerable scatter at large $\sigma_\alpha$ where the acoustic scale
is not as well measured. We emphasize however, that these variations in
$\sigma_\alpha$ are incredibly small and do not significantly affect
our results. (bottom left) $\sigma_\alpha$ from monopole+quadrupole
fits with $\Sigma_s=4\hMpc$ and $\snl=3\hMpc$ versus those from the
monopole-only fits with $\Sigma_s=0\hMpc$ and $\snl=4\hMpc$. The
degradation in $\sigma_\alpha$ is obvious and accounts for half the
increase in $\sigma_\alpha$ relative to the monopole-only case. This
again suggests that there is some mismatch between our fitting model and
the data. (top right) $\sigma_\alpha$ from full monopole+quadrupole fits
including $\epsilon$ versus those from the monopole+quadrupole fits with
$\epsilon=0$. The introduction of $\epsilon$ appears to cause most of the
scatter in $\sigma_\alpha$ and is responsible for the other half of the
$\sigma_\alpha$ increase relative to the monopole-only case. $\epsilon$
is known to have some correlation with $\alpha$ so this is not surprising.
\label{fig:scfig}}
\end{figure}

Our DR7 $\alpha$ measurements are consistent with the monopole-only
measurements of \citet{Xea12}, however, our errorbars on $\alpha$ are
a factor of 1.25 larger both before and after reconstruction. Although
this is a very small change in absolute terms, it is still worthy of
some investigation. The top left panel of Figure \ref{fig:scfig} shows
the $\sigma_\alpha$ values we measure for our full monopole+quadrupole
fits as described in \S\ref{sec:fitting} from the post-reconstruction
mocks versus the monopole-only results of \citet{Xea12}. The DR7 results
are overplotted as the black star. We see that, in general, the mocks
have larger $\sigma_\alpha$ values in our full monopole+quadrupole
fits. The scatter at large $\sigma_\alpha$ where the acoustic scale is
not well measured is also significantly bigger. Our DR7 $\sigma_\alpha$
measurement lies within the locus of mock points and hence the increase
we see is consistent with the mocks.

\begin{table*}
\centering
\caption{Changes in $\sigma_\alpha$ relative to the monopole-only fits
of \citet{Xea12} that arise when we introduce new fitting elements. In
the first row we introduce the quadrupole fitting with the new combined
monopole+quadrupole covariance matrix. The second row introduces our
$\beta$ fitting in addition to the new covariance matrix while the third
row introduces changes in $\Sigma_s$ and $\snl$ instead. The fourth row
combines the previous two and corresponds to fitting with the fiducial
model while forcing $\epsilon=0$. The last row introduces $\epsilon$
fitting and corresponds to our fiducial model results.
\label{tab:testres}}
\begin{tabular}{lcccc}
\hline
Parameters&\multicolumn{2}{c}{Before Recon.}&
\multicolumn{2}{c}{After Recon.}\\
&$\widetilde{\Delta \sigma_\alpha}$&Qtls&
$\widetilde{\Delta \sigma_\alpha}$&Qtls\\
\hline
New $C_{ij}$&0.0007&$^{+0.0014}_{-0.0015}$&0.0005&$^{+0.0020}_{-0.0016}$\\
New $C_{ij}$, adding $\beta$ fit&0.0007&$^{+0.0030}_{-0.0016}$&0.0006&$^{+0.0018}_{-0.0019}$\\
New $C_{ij}$, $\Sigma_s=4\hMpc$, new $\snl$&0.0015&$^{+0.0017}_{-0.0016}$&0.0013&$^{+0.0021}_{-0.0016}$\\
New $C_{ij}$, adding $\beta$ fit, $\Sigma_s=4\hMpc$, new $\snl$ (fiducial model w/ $\epsilon=0$)&0.0018&$^{+0.0024}_{-0.0017}$&0.0014&$^{+0.0014}_{-0.0016}$\\
New $C_{ij}$, adding $\beta$ \& $\epsilon$ fits, $\Sigma_s=4\hMpc$, new $\snl$ (fiducial model)&0.0036&$^{+0.0040}_{-0.0044}$&0.0022&$^{+0.0039}_{-0.0027}$\\
\hline
\end{tabular}
\end{table*}

Our monopole+quadrupole full fits have several differences
relative to the monopole-only fits of \citet{Xea12}. First, the
covariance matrix is expanded to include the quadrupole-quadrupole
and monopole-quadrupole covariances. Second, we introduce $\beta$ as
a fitting parameter. Third, we include FoG (i.e. $\Sigma_s=4\hMpc$)
in our fitting model; \citet{Xea12} have $\Sigma_s$ implicitly set
to $0\hMpc$. Since $\Sigma_s$ can induce some smearing of the BAO,
the $\snl$ value we use in the full fits is correspondingly smaller
($3\hMpc$ versus $4\hMpc$ in the monopole-only case). In addition,
in our pre-reconstruction fitting model, we introduce a non-isotropic
$\snl$. To understand which of these steps induces the greatest change in
$\sigma_\alpha$, we start by fitting the monopole+quadrupole using the
new covariance matrix and gradually add in the other changes. Before we
describe our results, we again stress that the changes in $\sigma_\alpha$
we see are very small and require probing some subtleties in our models
to understand. Our measurements of $\alpha$ and $\sigma_\alpha$ are still
reasonably robust against fitting parameters in a single DR7 realization
as shown in Tables \ref{tab:alphas} \& \ref{tab:dr7_alphas}.

The median changes in $\sigma_\alpha$ (relative to the monopole-only
case) as we add in more elements of the fitting are listed in Table
\ref{tab:testres} for the pre- and post-reconstruction mocks. The total
change between the monopole-only fits and the full monopole+quadrupole
fits is listed in the last row of this table. We find that changing
the covariance matrix (first row) increases $\sigma_\alpha$ by a small
amount and adding $\beta$ (second row) does not further degrade the
errors. Introducing $\Sigma_s$ and the accompanying change in $\snl$
(third row) appears to be a major contributor to the degradation of
$\sigma_\alpha$. This increases the median $\sigma_\alpha$ by about half
the total. Combining the $\beta$ fitting and the changes in $\Sigma_s$
and $\snl$ (fourth row) shows little additional degradation above the
previous case. Note that this corresponds to using the fiducial model with
$\epsilon$ fixed at 0. Finally, as mentioned above, the last row adds in
$\epsilon$ fitting and corresponds to the fiducial model. We see that this
step causes the other half of the total increase in $\sigma_\alpha$. It
also introduces a significant amount of scatter in $\sigma_\alpha$.

The steps that contribute the most to the $\sigma_\alpha$ increase are
shown in the bottom left and top right panels of Figure \ref{fig:scfig}
for the post-reconstruction case. In the bottom left we have plotted
the $\sigma_\alpha$ values measured from monopole+quadrupole fits with
$\Sigma_s=4\hMpc$ and $\snl=3\hMpc$ versus those measured from the
monopole-only fits with $\Sigma_s=0\hMpc$ and $\snl=4\hMpc$. The offset
between the two is obvious and again suggests that our FoG model is not
perfectly matched to the data.

The top right panel shows the $\sigma_\alpha$ values measured from
the mocks through the full monopole+quadrupole fits versus the
monopole+quadrupole fits with $\epsilon$ fixed at 0. In addition
to the obvious offset, we also see the appearance of significant
scatter. $\epsilon$ has small but non-zero correlation with $\alpha$,
implying a slight degeneracy between these 2 parameters. It is not
surprising that this extra covariance may increase $\sigma_\alpha$ and
its scatter. Again, this small degradation is not of great concern at
our current levels of statistical precision.

\section{Cosmological Implications} \label{sec:cosmo}

\begin{figure*}
\vspace{0.4cm}
\centering
\begin{tabular}{cc}
\epsfig{file=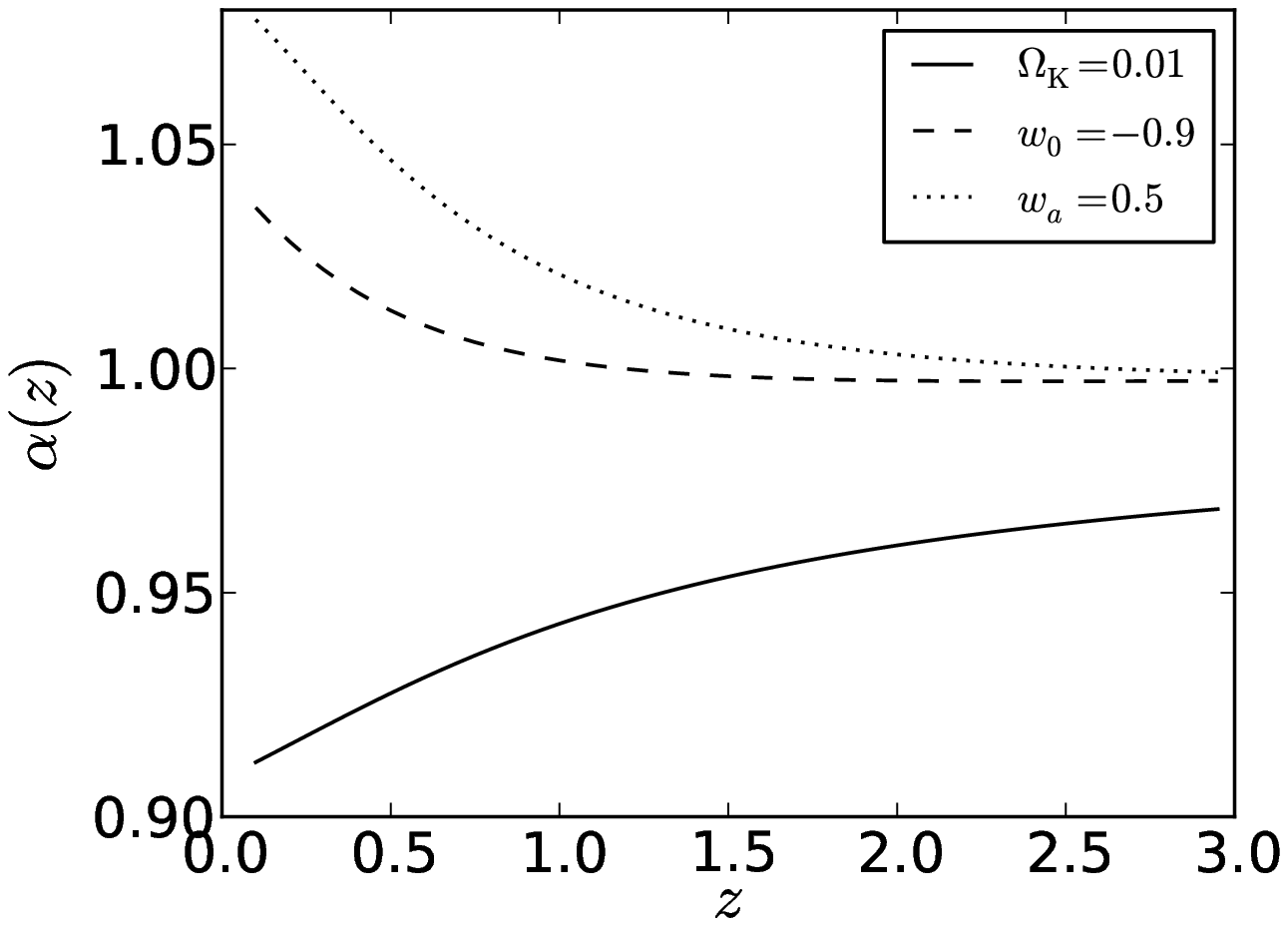,width=0.4\linewidth,clip=}&
\epsfig{file=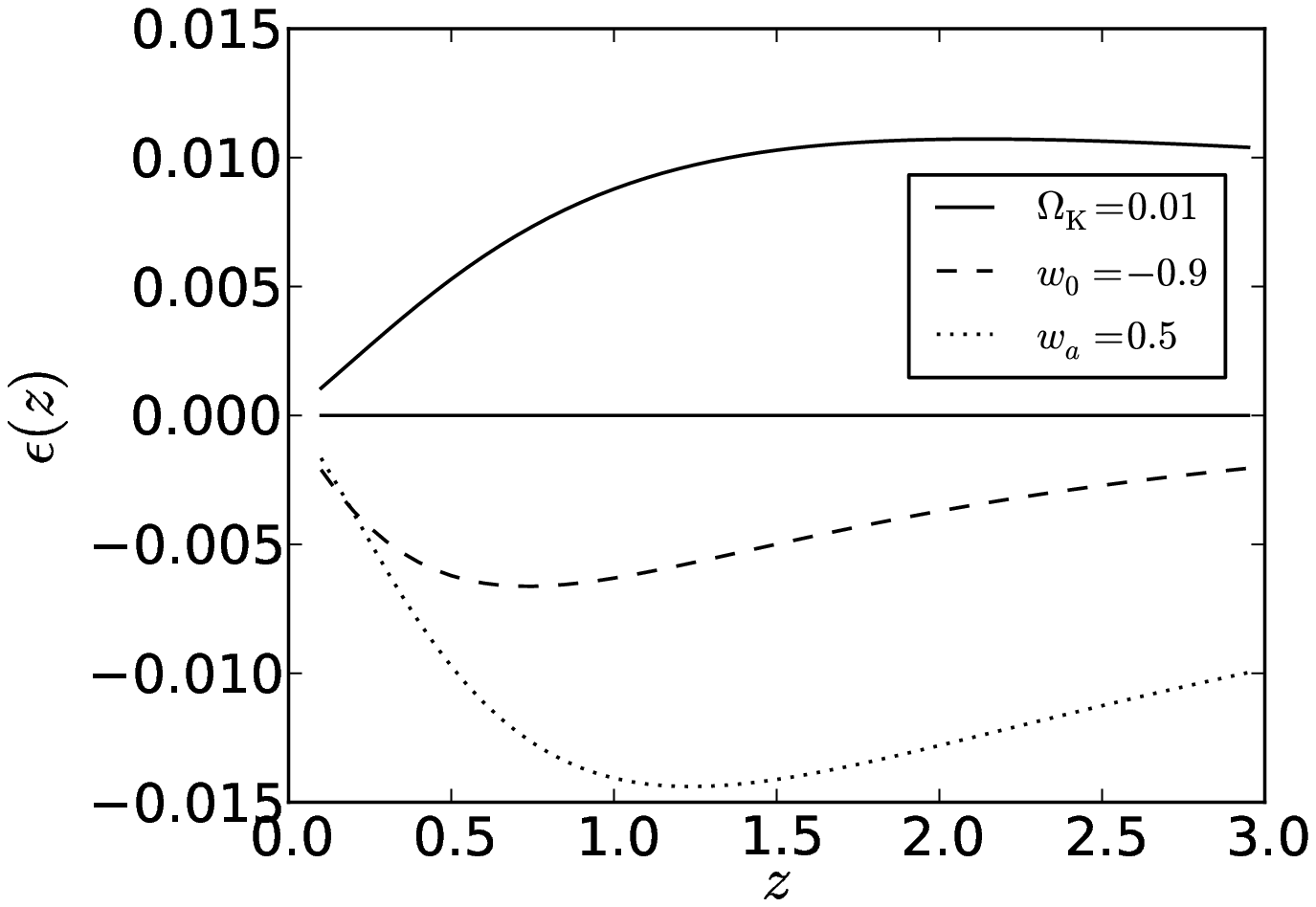,width=0.43\linewidth,clip=}
\end{tabular}
\caption{The expected variation in $\alpha(z)$ (left) and $\epsilon(z)$
(right) as we open up curvature or allow non-cosmological constant or
time-varying dark energy. We have taken the fiducial cosmology to be
the flat, $\Lambda$CDM cosmology predicted by WMAP7 as usual. One can
see that curvature and dark energy properties affect $\alpha$ more at
low $z$. At high $z$ it becomes increasingly difficult to distinguish
between non-cosmological constant and time-varying dark energy models
using measurements of $\alpha$. However, we see that $\epsilon$ is
affected by curvature and dark energy the most at higher $z$, peaking
at $z\sim1$. This suggests that the anisotropic BAO signal is stronger
and therefore offers more constraining power at higher redshifts.
\label{fig:cosmofig}}
\end{figure*}

In this section we will place our measurement of $\epsilon$
within the context of current cosmological constraints. To build
more intuition for how $\alpha(z)$ and $\epsilon(z)$ vary as we
change the amount of curvature or the nature of dark energy, we
look to Figure \ref{fig:cosmofig}. The left panel shows $\alpha$
as a function of redshift for a cosmology that has positive curvature
($\Omega_{\rm{K}}=0.1$), dark energy that is not a cosmological constant
($w_0=-0.9$) and time-varying dark energy ($w_a=0.5$). The analogous plot
for $\epsilon$ is shown in the right panel. Here we have again taken the
fiducial cosmology to be the flat $\Lambda$CDM cosmology predicted by
WMAP7. As we vary the cosmology, we fix $\Omega_m h^2$ and the distance to
the last scattering surface (i.e. the distance to $z=1089$, the redshift
of recombination). This guarantees that the sound horizon and the CMB
remain approximately unchanged in all the plotted cosmologies.

Recall that if the fiducial cosmology matches the true cosmology of
the universe, then we would expect $\alpha=1$ and $\epsilon=0$. We
see that introducing curvature and altering the nature of dark energy
both perturb $\alpha$ away from 1 the most at low redshift. However,
at higher redshift, it becomes increasingly difficult to distinguish
between non-cosmological constant models and time-varying dark energy
models using measurements of $\alpha$. The opposite is true for
$\epsilon$. We see that the effects of adding curvature or changing
the properties of dark energy are most prominent at larger redshifts,
peaking at $z\sim1$. This suggests that to exploit the anisotropic BAO
signal, we gain more leverage by going to higher $z$. However, we also
see that even the maximum difference in $\epsilon$ between the $w_a=0.5$
and $\Omega_{\rm{K}}=0.1$ cosmologies is smaller than our current error
on $\epsilon$ which indicates that we are not able to distinguish between
these cosmologies using our DR7 measurement. 

\begin{figure}
\vspace{0.4cm}
\centering
\epsfig{file=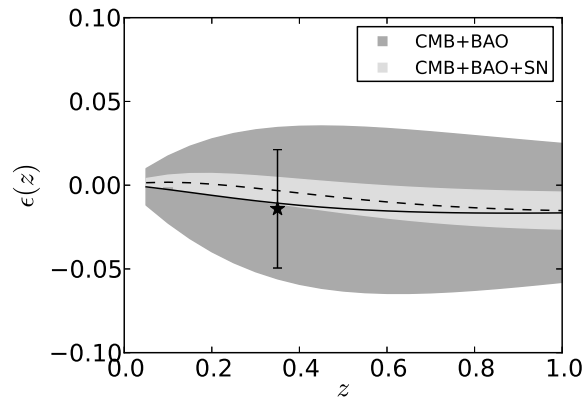,width=0.9\linewidth,clip=}
\caption{Our DR7 $\epsilon$ measurement at $z=0.35$ overplotted on
the 1$\sigma$ regions predicted by MCMC chains computed via combining
CMB+BAO (dark grey) and CMB+BAO+SN (light grey) datasets. The solid
black line corresponds to the mean $\epsilon(z)$ from the CMB+BAO
chain and the dashed black line is the analogue for the CMB+BAO+SN
chain. Here we have assumed a WMAP7 fiducial cosmology. The plotted
chains assumed a DETF cosmology \citep{Aea06} in which the universe is
allowed to be curved and the dark energy equation of state takes on the
form $w(a) = w_0 + (1-a)w_a$. The CMB data used is taken from WMAP7
\citep{Jarosik11,Larson11} and the SN distance constraints used are
taken from the SNLS3 \citep{Cea11}. The BAO distance constraints comes
from applying Equation (\ref{eqn:padef}) to the DR7 $p(\alpha,\epsilon)$
distribution in order to obtain $p(\alpha)$, collapsing our 2D constraint
into a single constraint on the spherically averaged distance $D_V$ to
$z=0.35$. This is equivalent to marginalizing over $\epsilon$. One can
see that our DR7 $\epsilon$ measurement overlaps the regions predicted
by our chains very well. In addition, we see that our errorbars on
$\epsilon$ are smaller than the region predicted by using the spherically
averaged constraint $p(\alpha)$ in the CMB+BAO case. Hence the anisotropic
information encoded in our measurement of $\epsilon$ provides additional
constraints on the parameter space of allowed cosmologies.
\label{fig:epcon}}
\end{figure}

Figure \ref{fig:epcon} shows our DR7 measurement of $\epsilon$ overplotted
on constraints derived from cosmological Markov Chain Monte Carlo (MCMC)
sampling. The MCMC method computes the likelihood that a set of input
cosmological parameters fits distance measures from Cosmic Microwave
Background (CMB) observations at high redshift, and Type Ia supernova
(SN) and BAO observations at low redshift. The number of steps in the
chain spent exploring a certain region in the cosmological parameter
space is proportional to the likelihood of that region representing the
true cosmology. Hence, we can infer $D_A(z)$ and $H(z)$ at each step in
the chain to compute $\epsilon(z)$ relative to some fiducial cosmology
(WMAP7 in our case). At each $z$, we can measure the mean and rms of the
$\epsilon$ distribution which can then be compared to our DR7 measurement
at $z=0.35$.

The grey regions in Figure \ref{fig:epcon} are derived from MCMC
chains exploring a cosmology which is allowed to have curvature and
varying dark energy with equation of state $w(a) = w_0 + (1-a)w_a$
where $a$ is the scale factor. Note that this is the most generalized
and least-constraining cosmology that is typically tested and is the
reference cosmology defined by the Dark Energy Task Force (the DETF
cosmology; \citealt{Aea06}). These chains were computed using CosmoMC, a
standard MCMC sampler \citep{LB02}. The dark grey and light grey regions
correspond to the 1$\sigma$ limits calculated from chains using CMB+BAO
data and CMB+BAO+SN data respectively. The CMB data is taken from WMAP7
\citep{Jarosik11,Larson11} and the SN distance contraints are taken from
the Supernova Legacy Survey 3 (SNLS3; \citealt{Cea11}).

The BAO distance measure used in the chains is based on our DR7
anisotropic measurements, however, we marginalize over $\epsilon$ and
only use the remaining isotropic or spherically-averaged information
encoded in $\alpha$. This is equivalent to using $p(\alpha)$, the
probability distribution for the spherically-averaged distance $D_V$
to $z=0.35$ (plotted in Figure \ref{fig:gridfig}), obtained by applying
Equation (\ref{eqn:padef}) to our measured $p(\alpha,\epsilon)$. Figure
\ref{fig:epcon} overplots our DR7 $\epsilon$ measurement on the
$\epsilon$ constraints obtained through our chains. One can see that
our DR7 measurement of $\epsilon$ is consistent with the CMB+BAO
and CMB+BAO+SN constraints at $z=0.35$. In the CMB+BAO case, our DR7
$\epsilon$ errorbar falls within the $\epsilon$ constraints obtained by
using the spherically-averaged distance measure. This suggests that the
DR7 anisotropic information we measure offers additional constraints on
the allowed cosmological parameter space.

\begin{table*}
\centering
\caption{Cosmological parameters measured from CosmoMC chains. The first
line listed for each cosmology is the DR7 monopole-only result from
\citet{Mehta12}. The second (bold) line is derived from our DR7 $\alpha$
and $\epsilon$ measurements.
\label{tab:cosmores}}
\begin{tabular}{llllllll}
\hline
Cosmology&
Datasets&
$\Omega_mh^2$&
$\Omega_m$&
$H_0$&
$\Omega_{\rm{K}}$&
$w_0$&
$w_a$\\
\hline
$o$CDM&CMB+BAO&0.1333(53)&0.278(15)&69.3(16)&-0.004(5)&--&--\\
&&\textbf{0.1342(50)}&\textbf{0.285(16)}&\textbf{68.7(18)}&\textbf{-0.003(5)}&
--&--\\
\\[-1.5ex]
$w$CDM&CMB+BAO&0.1349(57)&0.285(25)&69.0(39)&--&-0.97(17)&--\\
&&\textbf{0.1347(57)}&\textbf{0.289(26)}&\textbf{68.4(38)}&--&
\textbf{-0.95(17)}&--\\
\\[-1.5ex]
$ow_0w_a$CDM&CMB+BAO+SN&0.1346(53)&0.276(15)&69.9(19)&-0.010(7)&-0.90(16)&
-1.30(99)\\
&&\textbf{0.1352(52)}&\textbf{0.280(17)}&\textbf{69.6(21)}&\textbf{-0.010(7)}&
\textbf{-0.90(16)}&\textbf{-1.32(100)}\\
\hline
\end{tabular}
\end{table*}

The bold text in Table \ref{tab:cosmores} lists results from CosmoMC
chains run using our DR7 $\alpha$ and $\epsilon$ measurements. In
the table, $o$CDM refers to a cosmology in which $\Omega_{\rm{K}}$
is allowed to vary (i.e. the universe is allowed to be curved), $w$CDM
refers to a cosmology in which dark energy is allowed to vary from a
cosmological constant and $ow_0w_a$CDM is the DETF cosmology. Here we
have combined our full anisotropic BAO measurements at $z=0.35$ with
observations of the CMB from WMAP7 \citep{Kea11} and, in the case of the
DETF cosmology, supernovae from the SNLS3 \citep{Cea11}. We compare these
with the results from \citet{Mehta12} (non-bold text in the table) and
find consistent measurements and errors for the cosmologies tested. In
addition, using the $\epsilon$-marginalized $p(\alpha)$ distribution
obtained by applying Equation (\ref{eqn:padef}) to $p(\alpha,\epsilon)$
gives similarly consistent values of the cosmological parameters. However,
as expected, the errors can be $\sim10-20\%$ higher in this case since
we are not including the DR7 anisotropic constraints.

Recall that the errors on $\alpha$ we measure from DR7 are larger than
those measured by \citet{Xea12} and used in \citet{Mehta12}. The fact
that our chains produce similar errors on the cosmological parameters
implies that we are obtaining some constraint from $\epsilon$. Since
the fully anisotropic analysis presented here is a more careful
treatment of BAO measurements, the similarity in results confirms the
validity of past monopole-only BAO analyses. We also note that the
redshift of DR7 ($z=0.35$) is relatively low and as shown in Figure
\ref{fig:cosmofig}, the main constraining power of $\epsilon$ comes in
at larger redshifts. This implies that anisotropic analyses should be
more powerful than monopole-only analyses as we go to higher $z$.

The future is bright with the SDSS DR9 CMASS sample \citep{Aea12}
now in hand. The galaxies in this dataset are denser and more abundant
than DR7, and also at higher redshift ($z=0.57$). Therefore, the DR9
CMASS anisotropic BAO signal should be less noisy and more prominent,
implying a much tigher constraint on $\epsilon$.

\section{Conclusions} \label{sec:theend}

The differential clustering along the line-of-sight and transverse
directions that arise from assuming the wrong fiducial cosmology can be
used to directly constrain the angular diameter distance $D_A(z)$ and
the Hubble parameter $H(z)$. This anisotropy can be measured from the
BAO signal in the monopole and quadrupole moments of 2-point statistics
such as the correlation function studied in this work.

We have presented measurements of the anisotropic BAO signal ($\epsilon$)
from the SDSS DR7 LRG sample including density-field reconstruction. We
measured $\alpha=1.012\pm0.024$ and $\epsilon=-0.014\pm0.035$ which
translate into $D_A(z=0.35)=1050\pm38$ Mpc and $H(z=0.35)=84.4\pm7.0$
km/s/Mpc assuming $r_s=152.76$ Mpc. Note that these measurements of $D_A$
and $H$ are correlated with $\rho_{D_AH}=0.57$. We have demonstrated
that the methods for extracting $\epsilon$ outlined in this paper are
robust and applicable to future anisotropic BAO studies.

We have given a detailed account of the theoretical background motivating
the origin of the anisotropic signal and a parameter, $\epsilon$, for
measuring it. An in-depth look at the fitting model and method we use
to extract the anisotropic signal is also given. We find that our model
parameters have different morphological structures in their derivatives
from $\epsilon$, although they can still be partially degenerate with
each other. These minor degeneracies appear to introduce a small bias
in $\epsilon$ at the $0.2\%$ level, far below our current level of
statistical precision.

We apply density field reconstruction and test the robustness of our
measured $\alpha$ and $\epsilon$ against changes in the reconstruction
parameters using 160 LasDamas mock catalogues. We find that reconstruction
appears to introduce some anisotropy into the quadrupole, however this
is adequately accounted for by our $A_2(r)$ nuisance parameters. We
then perform the same robustness checks on our fitting model using
the mock catalogues. Similar tests were also performed on the DR7 data
returning consistently robust results. We demonstrate that $\alpha$ and
$\epsilon$ have near-Gaussian posteriors. Hence estimating their errors
($\sigma_\alpha$ and $\sigma_\epsilon$) as the second moments of their
respective probability distributions is reasonable. The $\sigma_\alpha$,
$\sigma_\epsilon$ and $\rho_{\alpha\epsilon}$ values obtained from
the mocks and the DR7 data are mostly consistent with Fisher matrix
predictions.

We find that in the mocks and the DR7 data, our $\alpha$ error estimates
are slightly larger than those obtained when only the monopole is
fit. This small increase does not detract significantly from the
overall robustness of our measurements which we verify as discussed
above. About half of this increase is a result of including an FoG model
in our full monopole+quadrupole fits while the other half arises from
fitting for $\epsilon$. This first point suggests that our FoG model
does not match the data perfectly and may induce slight biases into
our measurements. Given the non-zero covariance between $\epsilon$ and
$\alpha$ the second point is not surprising. The behaviour of DR7 falls
completely within the locus of mock points and is therefore not unusual.

Our DR7 measurements of $D_A$ and $H$ before reconstruction are
consistent with those obtained by \citet{CW11} and \citet{CW12}
using the same dataset. The errors we measure, however, differ
slightly due to differences in our analysis techniques. The
errors on $\alpha$ and $\epsilon$ we measure from the DR7 data
are consistent with the scatter from the mocks. In addition,
our $\sigma_\epsilon/(1+\epsilon)$-to-$\sigma_\alpha/\alpha$
ratio agrees reasonably well with Fisher matrix predictions and our
$\sigma_H/H$-to-$\sigma_{D_A}/D_A$ ratio is $\sim2$ and in good agreement
with the predictions of \citet{SE07}.

Our post-reconstruction DR7 $\epsilon$ measurement agrees well with the
predictions from current datasets. Comparing the cosmological parameters
we obtain using our $\alpha$ and $\epsilon$ measurements with the
monopole-only measurements in \citet{Mehta12} yields consistent values
and errors. This suggests that although our errors on $\alpha$ are larger
than the monopole-only case, the $\epsilon$ measurement is providing
some additional constraint on the cosmological parameters. The validity
of previous monopole-only analyses is also confirmed by the similarity
between those cosmological results and the ones obtained here through
a more careful, fully-anisotropic analysis of the BAO.

We find that the anisotropic signal is stronger at higher redshifts which
suggests that its constraining power will become more apparent at higher
$z$. The recently obtained SDSS DR9 CMASS dataset has a higher number
density than the DR7 LRG sample, contains more galaxies and is at higher
redshift ($z=0.57$). The basic theory and methodology presented in this
work should serve as a foundation for obtaining a much better detection of
$\epsilon$, and subsequently $D_A$, $H$ and other cosmological parameters
from CMASS.

\section{Acknowledgments}
Funding for the Sloan Digital Sky Survey (SDSS) and SDSS-II has
been provided by the Alfred P. Sloan Foundation, the Participating
Institutions, the National Science Foundation, the U.S. Department of
Energy, the National Aeronautics and Space Administration, the Japanese
Monbukagakusho, the Max Planck Society and the Higher Education Funding
Council for England. The SDSS website is \texttt{http://www.sdss.org/.}

The SDSS is managed by the Astrophysical Research Consortium (ARC) for
the Participating Institutions. The Participating Institutions are the
American Museum of Natural History, Astrophysical Institute Potsdam,
University of Basel, University of Chicago, Drexel University, Fermilab,
the Institute for Advanced Study, the Japan Participation Group, the
Johns Hopkins University, the Joint Institute for Nuclear Astrophysics,
the Kavli Institute for Particle Astrophysics and Cosmology, the Korean
Scientist Group, the Chinese Academy of Sciences (LAMOST), Los Alamos
National Laboratory, the Max-Planck-Institute for Astronomy (MPIA), New
Mexico State University, Ohio State University, University of Pittsburgh,
University of Portsmouth, Princeton University, the United States Naval
Observatory and the University of Washington.

We thank the LasDamas collaboration for making their galaxy mock catalogs
public. XX thanks Kushal Mehta for useful conversations. XX and DJE
were supported by NSF grant AST-0707725 and NASA grant NNX07AH11G. NP
and AJC are partially supported by NASA grant NNX11AF43G. This work was
supported in part by the facilities and staff of the Yale University
Faculty of Arts and Sciences High Performance Computing Center.

\appendix

\section{Fisher matrix predictions} \label{app:fisher}

Using a Fisher matrix formalism, it is possible to derive theoretical
predictions for and expected correlations between the variances and
covariances of $\alpha$, $\epsilon$, $D_A$ and $H$. The values derived
here are utilized in the text as sanity checks for the values we measure.

We begin with the matrix equation
\begin{equation}
\Bigg(
\begin{matrix}
\sigma_\alpha^2 & \sigma_{\alpha\epsilon} \\
\sigma_{\alpha\epsilon} & \sigma_\epsilon^2
\end{matrix}
\Bigg)
= \Bigg(
\begin{matrix}
\dadd & \dadh \\
\dedd & \dedh 
\end{matrix}
\Bigg)
\Bigg(
\begin{matrix}
\sds & \sdh \\
\sdh & \shs
\end{matrix}
\Bigg)
\Bigg(
\begin{matrix}
\dadd & \dadh \\
\dedd & \dedh 
\end{matrix}
\Bigg)^T.
\end{equation}
Note that this is essentially the inverse process to Equation
(\ref{eqn:ermat}). Expanding we get
\begin{eqnarray}
\sas &=& \sds\left(\dadd\right)^2 + \shs\left(\dadh\right)^2 + 
2\sdh\dadd\dadh \nonumber \\
&& \\
\ses &=& \sds\left(\dedd\right)^2 + \shs\left(\dedh\right)^2 + 2\sdh\dedd\dedh.
\nonumber \\
&& \\ 
\sae &=& \sds\dadd\dedd + \sdh\left(\dadh\dedd + \dadd\dedh\right)  
\nonumber \\
&& + \shs\dadh\dedh 
\end{eqnarray}
Plugging in the relevant derivatives from Equations (\ref{eqn:ahd}) \&
(\ref{eqn:ehd}) we get
\begin{eqnarray}
\frac{\sigma_\alpha^2}{\alpha^2} &=& 
\frac{4}{9}\sigma_{\log D_A}^2 +
\frac{1}{9}\sigma_{\log H}^2 -
\frac{4}{9}\left(\frac{\sigma_{D_A H}}{D_A H}\right) \\
\frac{\sigma_\epsilon^2}{(1+\epsilon)^2} &=&
\frac{1}{9}\sigma_{\log D_A}^2 +
\frac{1}{9}\sigma_{\log H}^2 +
\frac{2}{9}\left(\frac{\sigma_{D_A H}}{D_A H}\right) \\
\frac{\sae}{\alpha(1+\epsilon)} &=& -\frac{2}{9}\sigma_{\log D_A}^2
+ \frac{1}{9}\sigma_{\log H}^2
- \frac{1}{9}\left(\frac{\sdh}{D_A H}\right)
\end{eqnarray}
where $\sigma^2_{\log y} = \frac{\sigma^2_y}{y^2}$.

The correlation coefficient between $D_A$ and $H$ is $\rho_{D_A H} =
\sigma_{D_A H}/\sigma_{D_A}\sigma_{H}$. If we write $f = \sigma_{\log
H}/\sigma_{\log D_A}$, then we have
\begin{eqnarray}
\frac{\sigma_\alpha^2}{\alpha^2} &=& \frac{1}{9}\sigma_{\log D_A}^2
(4 + f^2 - 4\rho_{D_A H}f) \\
\frac{\sigma_\epsilon^2}{(1+\epsilon)^2} &=& \frac{1}{9}\sigma_{\log D_A}^2
(1 + f^2 + 2\rho_{D_A H}f) \\
\frac{\sae}{\alpha(1+\epsilon)} &=& \frac{1}{9}\sigma_{\log D_A}^2
(-2 +f^2 - \rho_{D_A H}f).
\end{eqnarray}
Note that $f$ is just the ratio of $\sigma_H/H$-to-$\sigma_{D_A}/D_A$
which is typically $\sim2$ \citep{SE07}. The correlation coefficient
$\rho_{D_A H}$ is predicted to be $\sim0.4$. Hence, we have
\begin{eqnarray}
\frac{\sigma_\alpha}{\alpha} &=& \sigma_{\log{\alpha}} = 
0.73 \sigma_{\log D_A} 
\label{eqn:aerr} \\
\frac{\sigma_\epsilon}{1+\epsilon} &=& \sigma_{\log(1+\epsilon)} =
0.86 \sigma_{\log D_A},
\label{eqn:eerr}
\end{eqnarray}
which implies the ratio
\begin{equation}
\frac{\sigma_\epsilon}{1+\epsilon}\rm{-to-}\frac{\sigma_\alpha}{\alpha}
\sim 1.2.
\label{eqn:sigae}
\end{equation}

The correlation coefficient between $\alpha$ and $\epsilon$ is
\begin{eqnarray}
\rho_{\alpha\epsilon} &=& 
\frac{\sigma_{\alpha\epsilon}}{\sigma_\alpha \sigma_\epsilon} \\
&=& \frac{\alpha}{\sigma_\alpha}\frac{(1+\epsilon)}{\sigma_\epsilon}
\bigg(\frac{1}{9}\sigma_{\log D_A}^2\bigg)(-2 +f^2 - \rho_{D_A H}f).
\end{eqnarray}
Using Equations (\ref{eqn:aerr}) \& (\ref{eqn:eerr}) and
plugging in the assumed values of $f$ and $\rho_{D_A H}$ gives
\begin{equation}
\rho_{\alpha\epsilon}\sim0.21.
\label{eqn:rhoae}
\end{equation}

\end{document}